\documentclass[11pt,a4paper, titlepage]{article}
\usepackage{moreverb,url, dsfont, soul, booktabs, geometry}
\usepackage{moreverb, url, soul, subcaption, threeparttable, multicol, multirow, , makecell, bm}
\usepackage{enumitem}
\usepackage[title]{appendix}
\usepackage{mathtools}
	\title{Propensity score weighting under limited overlap and model misspecification}

\author{
Yunji Zhou\\Department of Biostatistics and Bioinformatics\\\& Duke Global Health Institute\\ Duke University, Durham, North Carolina, USA\\
	\and
Roland A. Matsouaka\\ Department of Biostatistics and Bioinformatics\\
\& Program for Comparative Effectiveness Methodology, \\Duke Clinical Research Institute\\Duke University, Durham, North Carolina, USA\\
 \and Laine Thomas\\ Department of Biostatistics and Bioinformatics\\
\& Program for Comparative Effectiveness Methodology, \\
Duke Clinical Research Institute\\Duke University, Durham, North Carolina, USA\\
}

\date{\today}
\begin{document}
	
	\maketitle
	\begin{abstract}
	Propensity score (PS) weighting methods are often used in non-randomized studies to  adjust for confounding and assess treatment effects. The most popular among them, the inverse probability weighting (IPW), assigns weights that are proportional to the inverse of the conditional probability of a specific treatment assignment, given observed covariates. 
	A key requirement for IPW estimation is the positivity assumption, i.e., the PS must be bounded away from 0 and 1. In practice, violations of the positivity assumption often manifest by  the presence of limited overlap in the PS distributions between treatment groups. When these practical violations occur, a small number of highly influential IPW weights may lead to unstable IPW estimators, with biased estimates and large variances. To mitigate these issues, a number of alternative methods have been proposed, including IPW trimming, overlap weights (OW), matching weights (MW), and entropy weights (EW). Because OW, MW, and EW target the population for whom there is equipoise (and with adequate overlap) and their estimands depend on the true PS, a common criticism is that these estimators may be more sensitive to misspecifications of the PS model. In this paper, we conduct extensive simulation studies to compare the performances of IPW and IPW trimming against those of OW, MW, and EW  under limited overlap and misspecified propensity score models. 
	Across the wide range of scenarios we considered, OW, MW, and EW consistently outperform IPW in terms of bias, root mean squared error, and coverage probability. \\
	\end{abstract}

		\section{Introduction}
	In non-randomized studies, propensity score (PS) weighting methods are often used to adjust for potential confounding when estimating treatment effects. The PS is the probability of receiving a treatment, conditional on the observed covariates. It can be used as a balancing score to compare treatment groups, i.e., given the PS, the distributions of observed covariates between treatment groups are similar \cite{rosenbaum1983central}.  Therefore, with the use of PSs, it is feasible to obtain, under certain conditions, results that mimic some aspects of a randomized trial study design and allows an appropriate estimation of the causal effects \cite{hernan2016using,hernan2011great}. 
	
	
	In this paper, we are concerned with the use of propensity score weights to estimate the average treatment effect (ATE).  The traditional inverse probability weighting (IPW) assigns weights to each observation from a study sample that are determined by the inverse probability of receiving the treatment that was actually received.  After weighting, both treatment groups will better reflect the distribution of patient characteristics in the population from which the sample was drawn \cite{hernan2018causal,horvitz1952generalization}. Under the assumptions of stable unit treatment value, conditional independence, and positivity (see Section \ref{sec:overview} for definitions)---along with a correctly specified PS model---IPW leads to consistent estimator of the ATE.\cite{kang2007demystifying} In practice, two important diagnostic evaluations are conducted when using IPW. One, to evaluate the covariate balance (before and after weighting) and ensure that weighting leads to comparable treatment groups, with respect to the measured covariates. The other, to assess the positivity assumption by looking at the overlap of the PS distributions between the treatment groups and their common support \cite{kang2016practice}. 
	Lack of sufficient overlap may be indicative of violation of the positivity assumption, which can result in extremely large IPW weights \cite{petersen2012diagnosing,austin2017performance,desai2019alternative}. Unfortunately, when present, IPW estimators can be unduly influenced by few observations with extreme weights, leading to both biased and unstable results \cite{ lee2011weight,petersen2012diagnosing,austin2017performance,ju2019adaptive,desai2019alternative}.
	To analyze data in the presence of extreme weights, we either truncate extreme weights by shrinking them to some smaller fixed value(s) (e.g., capping extreme weights to the 10th and 90th percentiles) \cite{cole2008constructing, ju2019adaptive} or exclude from our analysis subjects with extreme weights and  by restrict the analysis to observations with modest and less influential weights, potentially introducing bias and reducing efficiency.
	
	
	To mitigate these issues inherent to IPW estimators, several alternative methods (to handle or modify PS weights) have been proposed, including trimming IPW weights, overlap weights (OW), and matching (MW) weights \cite{crump2006moving,crump2009dealing,sturmer2010treatment,sturmer2014propensity, li2018addressing, mao2019propensity,glynn2019comparison}. 
	These weights are part of a class of balancing weights that target a judiciously chosen subpopulation of interest from which an estimand closely-related to the ATE can be estimated with better precision \cite{crump2009dealing, crump2006moving, li2018balancing, sturmer2010treatment,mao2019propensity,glynn2019comparison,yang2018asymptotic,li2018addressing}  
	While there are many versions of IPW trimming \cite{sturmer2010treatment,li2018addressing}, Crump et al. \cite{crump2009dealing} have shown that a good cutoff may be approximated by  0.1 and thus recommend, as rule-of-thumb, to exclude from statistical analysis observations with PS outside of the interval $[0.1, 0.9]$ for an improved estimation. 
	Both OW and MW emphasize a subpopulation that exhibits better overlap in the distribution of the (measured) covariates, similar to the target population of patients for whom there is clinical equipoise used in randomized clinical trials \cite{li2018balancing,mao2019propensity}.  
	These different weighting schemes offer substantial advantages compared to IPW, in terms of improved covariate balance, bias, and precision---assuming a correctly specified propensity score model \cite{hirano2003efficient,crump2006moving, crump2009dealing,li2013weighting,li2018balancing,mao2019propensity,li2018addressing,sturmer2010treatment,glynn2019comparison,yang2018asymptotic}. However, only IPW and trimming have been investigated under misspecifications of the propensity score model \cite{ lee2011weight,petersen2012diagnosing,ju2019adaptive,kang2016practice,austin2017performance,desai2019alternative}.
	
	In practice, true propensity scores are unknown in non-randomized studies and must be estimated. However, while OW and MW perform relatively well in the presence of extreme weights and under correctly specified PS models, little is known about their performances when the PS model is misspecified.   Because the estimands (i.e., the targets of inference) of OW and MW depend on the true PS (through their underlying target population), misspecifications of the PS model are of genuine concern; it appears that estimating these targets of inference may be tightly dependent on a correct specification of the PS model. Hence, it has been hypothesized that  OW and MW  estimators might be more sensitive to misspecifications of the propensity score model than IPW estimators. Nevertheless, the PS is a balancing score. As argued by Chakraborty and Moodie \cite{chakraborty2013statistical}, "'correct' specification of the PS model does not require complete knowledge" of the data generating process behind the treatment allocation. Of primary interest in most PS methods is whether the PS model includes adequately all the confounding variables of the treatment-outcome relationship
	and how well the proposed working model captures the inherent impact of these covariates on the treatment allocation. To the best of our knowledge, it is still unclear whether OW and MW estimators are robust to misspecifications of the PS model.
	
	Moreover, since the selection (or tilting) functions used to derive matching and overlap weights are part of a trio of functions commonly used in other fields, we also investigate the performance of the entropy weights (EW) that are obtained using the cross-entropy function. These three functions---usually referred to as misclassification error,  Gini index, and Shannon's  cross-entropy---are ubiquitous in classification and regression tree methods and in information theory \cite{breiman1984classification,hastie2009elements, mackay2003information}. 
	
	
	Therefore, in this paper, we investigate whether 
	OW, MW, and EW are better alternatives to IPW in the presence of limited PS overlap or misspecifications of the PS model. We use results from standard M-estimation theory \cite{stefanski2002calculus, lunceford2004stratification} to adjust for uncertainties in PS estimation  and derive in Appendix \ref{sec:sandwich_var}  empirical sandwich variance estimators for better precision. 
	
	
	The remaining of this paper is organized as follows. We begin Section \ref{sec:overview}  with our notations and a brief review of the class of balancing weights. We also provide a table that lists the different balancing weight methods, their estimands, and their corresponding weights. Our investigation is motivated by the example of estimating the effect of maternal smoking on offspring birth weight, in Section \ref{sec:motivation}. Section \ref{sec:simulation} details simulation settings and results. In Section \ref{sec:conclusion}, we discuss the findings and implications to analysis based on propensity score weighting.
	
	\section{Overview of the weighting methods}\label{sec:overview}
	\subsection{Notation and assumptions}
	Let $\boldsymbol X$ denote the vector of measured  pre-treatment covariates, $Y$ the observed outcome, and  $Z$ the  indicator of treatment options; we use the generic terms "treatment" $(Z=1)$   and  "control" $(Z=0)$. The  data consist of a sample $\{(\boldsymbol X_i, Z_i, Y_i), i=1, \dots, N\}$ of $N$ subjects from the population of interest to which the treatment will eventually be applied. The propensity score $e(\boldsymbol X) $ is the  probability  of treatment $Pr(Z = 1|\boldsymbol X)$, condition on the covariates $\boldsymbol X$ \cite{rosenbaum1983central}. 
	
	We consider the potential outcome framework of Rubin  \cite{rubin1997estimating}, where each subject has two potential outcomes $Y(0)$ and $Y(1)$, one of which would be observed under a given treatment assignment.  To infer causality, we make three untestable assumptions.  First, the stable unit treatment value assumption (SUTVA)\cite{rubin1980randomization}, i.e., for each subject, the observed outcome $Y_i=Z_iY_i(1)+ (1-Z_i)Y_i(0)$.
	Second, we assume the conditional independence of the treatment assignment, i.e., the set $\boldsymbol X$  is  sufficient enough to control for confounding, \cite{rosenbaum1983central} i.e,  $E\left[  Y(z) \mid Z=z, \boldsymbol X\right]=E\left[  Y(z) \mid \boldsymbol X\right], ~z=0, 1$.   Finally, a key requirement for PS methods is the positivity assumption, i.e.,  $P\left( \left\lbrace \boldsymbol x: \nu<e(\boldsymbol x)<1-\nu\right\rbrace \right) =1,$ for some $\nu  >0$.\cite{rosenbaum1983central} The positivity assumption states that each subject has a nonzero probability  to receive either treatment. Without positivity, groups can not be made comparable with the existing data, and the treatment effect of interest is not identifiable \cite{greenland1986identifiability}.
	
	
	
	\subsection{Practical violations of the positivity assumption}
	
	Practical violations of the positivity assumption occur when some subjects almost always (or almost never) receive treatment, $i=1, \dots, N$ i.e., when $\widehat e(\boldsymbol  x_i)\approx $ 0 or 1.  Such violations may arise for several reasons, including data limitations (where subjects with some specific covariates cannot possibly receive one of the treatment options), small sample size, and misspecifications of the PS model \cite{petersen2012diagnosing,westreich2010invited,austin2017performance}. When $\widehat e(\boldsymbol  x)\approx $ 0 or 1, which may be the result of a limited overlap in the covariate distributions of the two treatment groups, estimated IPW weights $\widehat w(\boldsymbol  x)$ can be  extreme for some participants. Unfortunately, extreme IPW weights often lead to biased and unstable ATE estimates, with large variances \cite{schafer2008average,kang2016practice,busso2014new,austin2017performance}.
	
	Moreover the occurrence of non-positivity sometimes suggests a conceptual problem with the study design; the sampling scheme includes people for whom the treatment decision is less ambiguous (or even vividly clear) that it is nearly deterministic. This is the case, for instance, when some patients have specific counter-indication for one the drugs under study, when frail and old patients are considered not good candidates for some invasive open-heart surgery such as a surgical aortic valve replacement \cite{rogers2018transcatheter}, or when some younger, pre-menopausal women with uterine fibroids forgo hysterectomy and opt for myomectomy, a uterine-sparing treatment option, to keep alive their dream to bear children \cite{wallace2020comparative}. The evidence for making treatment decisions in such patients is apparently already strong, and thus the need for treatment effect estimation is less clear.  This motivates the consideration of alternative target populations for whom the treatment effect may be more relevant and better estimated, with respect to bias and variance.
	
	\subsection{The class of balancing weights}
	
	The class of balancing weights includes traditional IPW along with alternatives (weight trimming, MW, OW and EW) that restrict estimation of the treatment effect to a region of reasonable positivity, bounded away from 0 and 1 (see Table \ref{tab:wgts_summary}) \cite{crump2006moving,crump2009dealing, schafer2008average,petersen2012diagnosing}. The essence of weighting is to use propensity scores to create a pseudopopulation in which treatment and control groups are balanced in covariates distributions. Let $f(\boldsymbol x)$ denote the marginal density of covariates $\boldsymbol X$ in the population that was sampled (combined over both treatments). Assume $f(\boldsymbol x)$ exists, then the density of the target population can be represented by $f(\boldsymbol  x)h(\boldsymbol  x)$, where $h(\boldsymbol  x)$ is a prespecified tilting function of $\boldsymbol x$ defining the target population. In other words, $h(\boldsymbol  x)$ serves to  re-distribute patient characteristics from distribution sampled, $f(\boldsymbol x)$, to another distribution that is more clinically relevant or statistically optimal. If $f(\boldsymbol x)$ is of primary interest, $h(\boldsymbol  x)=1$. Let $f_z(\boldsymbol x)=Pr(\boldsymbol X=\boldsymbol x\mid Z=z)$ be the density of $\boldsymbol X$ in the $z$ group, then $f_1(\boldsymbol x)\propto f(\boldsymbol x)e(\boldsymbol x)$ and $f_0(\boldsymbol x)\propto f(\boldsymbol x)(1-e(\boldsymbol x))$. For a given tilting function $h(\boldsymbol x)$, the corresponding weights for each treatment group $w_z(\boldsymbol x)$ are defined as follows.
	\begin{align}
	w_1(\boldsymbol  x)&\propto \frac{f(\boldsymbol x)h(\boldsymbol x)}{f(\boldsymbol x)e(\boldsymbol x)}=\frac{h(\boldsymbol x)}{e(\boldsymbol x)}, ~\text{for} ~z=1, \nonumber\\
	w_0(\boldsymbol  x)&\propto \frac{f(\boldsymbol  x)h(\boldsymbol  x)}{f(\boldsymbol  x)(1-e(\boldsymbol  x))}=\frac{h(\boldsymbol  x)}{(1-e(\boldsymbol  x))}, ~\text{for} ~z=0.
	\end{align}
	This class of weights $w_z(\boldsymbol  x)$ are called \textit{balancing weights} because of their property to balance the weighted covariates distributions:
	$f_1(\boldsymbol  x)w_1(\boldsymbol  x)=f_0(\boldsymbol  x)w_0(\boldsymbol  x)=f(\boldsymbol  x)h(\boldsymbol  x)$ \cite{li2018balancing}.

	\begin{table}[]
		\caption{Examples of tilting functions, targeted (sub)populations, causal estimands, and their weights} \label{tab:wgts_summary}
		\begin{center}
			\begin{threeparttable}[]\small
				\def\arraystretch{1.1}
				\begin{tabular*}{\textwidth}{c @{\extracolsep{\fill}} ccccccccccccccccccccc}
					\toprule
					Target  &   & &\multicolumn{2}{c}{Weights} & \\ \cmidrule(lr){4-5}
					population   & $h(\boldsymbol  x)$ & Estimand  &  $w_1(\boldsymbol  x)$& $ w_0(\boldsymbol  x)$& Name\\ \cmidrule(lr){1-6}
					overall &   $1$  &  ATE  &  $\displaystyle e(\boldsymbol  x)^{-1}$ & $(1-e(\boldsymbol  x))^{-1}$   
					& IPW         \\
					trimmed &   $I_{\alpha}(\boldsymbol  x)$     &  OSATE   &    $\displaystyle I_{\alpha}(\boldsymbol  x)e(\boldsymbol  x)^{-1}$ & $ I_{\alpha}(\boldsymbol  x)(1-e(\boldsymbol  x))^{-1}$   
					& Trimmed IPW\\ 
					treated &  $e(\boldsymbol  x)$   &  ATT  & $\displaystyle 1 $ & $ e(\boldsymbol  x)(1-e(\boldsymbol  x))^{-1}$   
					&  IPW for treated\\
					controls &  $1-e(\boldsymbol  x)$  &       ATC   &   $\displaystyle (1-e(\boldsymbol  x))e(\boldsymbol  x)^{-1}$ & $ 1$   
					& IPW for controls\\ \cmidrule(lr){1-6} 
					\multirow{3}{*}{equipoise} &  $e(\boldsymbol  x)\left(1-e(\boldsymbol  x)\right)$  &       ATO   &      $\displaystyle1-e(\boldsymbol  x) $ & $e(\boldsymbol  x)$& overlap weights\\
					&  $u(\boldsymbol  x) $  &          &       $\displaystyle u(\boldsymbol  x)e(\boldsymbol  x)^{-1}$ & $u(\boldsymbol  x)(1-e(\boldsymbol  x))^{-1}$   
					& matching weights\\
					&  $\xi(\boldsymbol  x)$  &          &     $\displaystyle \xi(\boldsymbol  x)e(\boldsymbol  x)^{-1}$ & $ \xi(\boldsymbol  x)(1-e(\boldsymbol  x))^{-1}$   
					&entropy	weights\\
					\bottomrule
				\end{tabular*}
				\begin{tablenotes}
					\footnotesize
					\item  $I_{\alpha}(\boldsymbol  x)=\mathds{1}( {\{\alpha\leq e(\boldsymbol  x)\leq 1-\alpha\}})$, where $0<\alpha<0.5$ and $\mathds{1}(.)$ is the standard indicator function;
					\item $u(\boldsymbol  x)=\min\{e(\boldsymbol  x), 1-e(\boldsymbol  x)\};$
					$\xi(\boldsymbol  x)=-\left\{e(\boldsymbol  x)\ln(e(\boldsymbol  x))+ \left(1-e(\boldsymbol  x)\right)\ln\left(1-e(\boldsymbol  x)\right)\right\}$.
					\item OSATE: Optimal sample average treatment effect; \cite{crump2009dealing}
					ATO: Average treatment effect for the overlap population. \cite{li2018balancing}
				\end{tablenotes}
			\end{threeparttable}
		\end{center}
	\end{table}

	\subsubsection{Target estimands}
	With a function $h(\boldsymbol x)$, our goal is to estimate the weighted average treatment effect \cite{hirano2003efficient,li2018balancing}
	\allowdisplaybreaks \begin{align}\label{eq:estimand}
	\Delta_h&=\displaystyle \frac{\displaystyle E[h(\boldsymbol X)(Y(1)-Y(0))]}{\displaystyle E(h(\boldsymbol X))}
	\end{align}
	When $h(\boldsymbol x)$ is equal to 1,  $e(\boldsymbol x),$ or $1-e(\boldsymbol x),$  the estimand ${\Delta}_h$ corresponds to the average treatment effect (ATE), the average treatment effect on the treated (ATT), or the average treatment effect on the controls (ATC), respectively \cite{hirano2003efficient,li2018addressing,li2018balancing}. Therefore, different tilting functions  $h(\boldsymbol  x)$ lead to different targets of inference, i.e., different ways to evaluate causal treatment effects (see Table \ref{tab:wgts_summary}).  The ideal target population (and tilting function) may vary across medical studies due to different clinical context, initial sampling scheme, or statistical considerations \cite{mao2019propensity,li2018addressing}. In addition, when the treatment effect is constant, the estimand will be the same for all $h$. It is only when the treatment effect is heterogeneous that different tilting functions can lead to different estimands.  Note also that $\Delta_h$ in equation \eqref{eq:estimand} is defined in term of the true propensity score, not on the estimated score. In the examples provided in Table 1, $h(X)\equiv h(e(X))$,  i.e., a function of the true (albeit unknown!) propensity score, except for the tilting function of ATE estimand via the IPW method. As such,  $\Delta_h$   is well-defined since we assume the existence of potential outcomes $Y(z)$,  $z=0, 1$, and of the true propensity score $e(X)$. Its identifiability is  guaranteed by the SUTVA assumption and the conditional independence of treatment assignment assumptions.  Now, whether we are able to correctly estimate such a quantity, based on  a data set at hand, is a different issue.  
	
	\begin{figure}[]\allowdisplaybreaks
		\begin{center}
			\hspace*{-1cm}
			\begin{subfigure}{.6\textwidth}
				\begin{center}
					\includegraphics[trim=45 32 135 70 clip, width=0.64\linewidth]{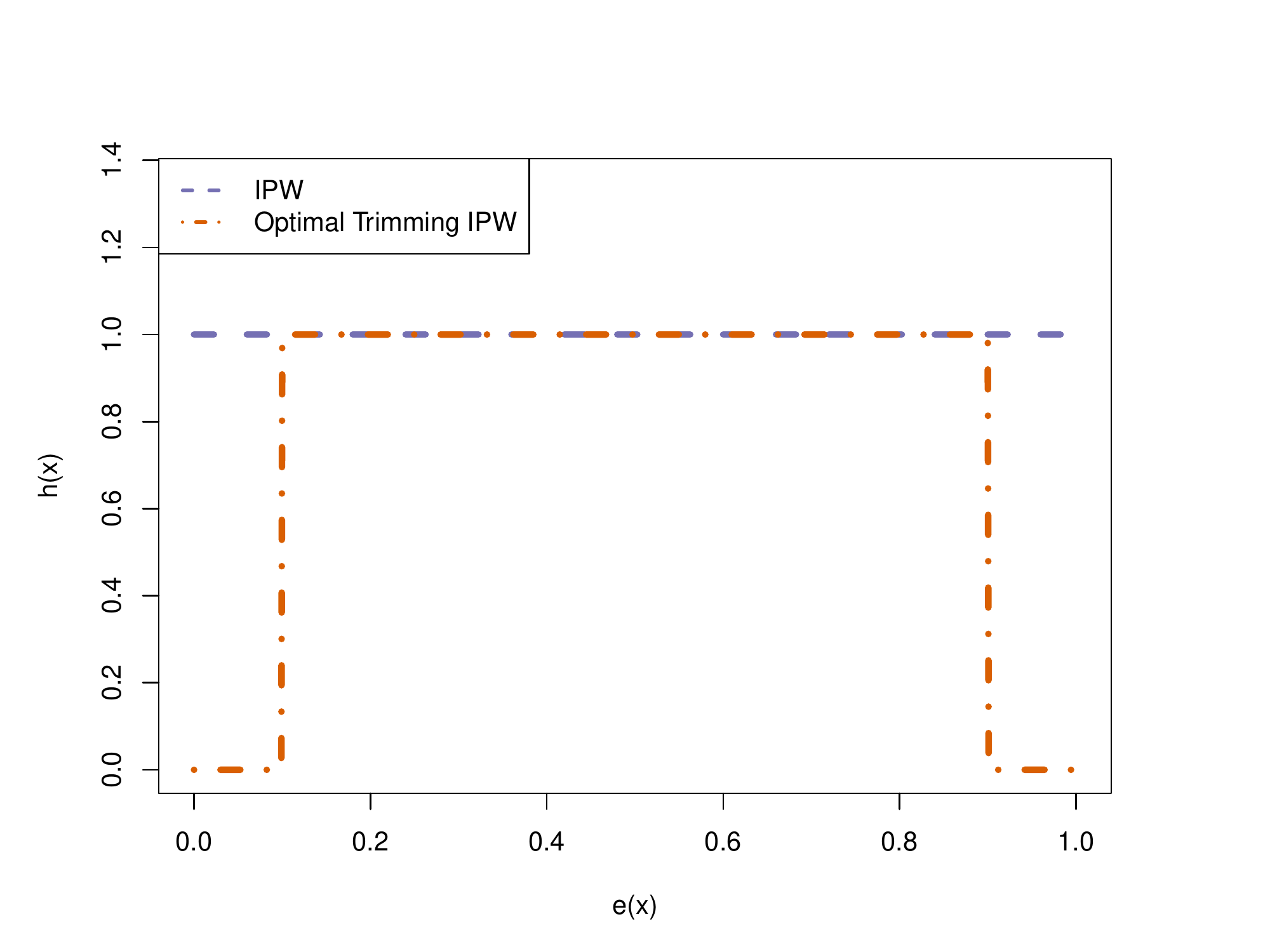}
					\label{fig:sub1}
				\end{center}
			\end{subfigure}%
			\hspace*{-1.5cm}
			\begin{subfigure}{.6\textwidth}	
				\begin{center}
					\includegraphics[trim=45 32 135 70  clip, width=0.64\linewidth]{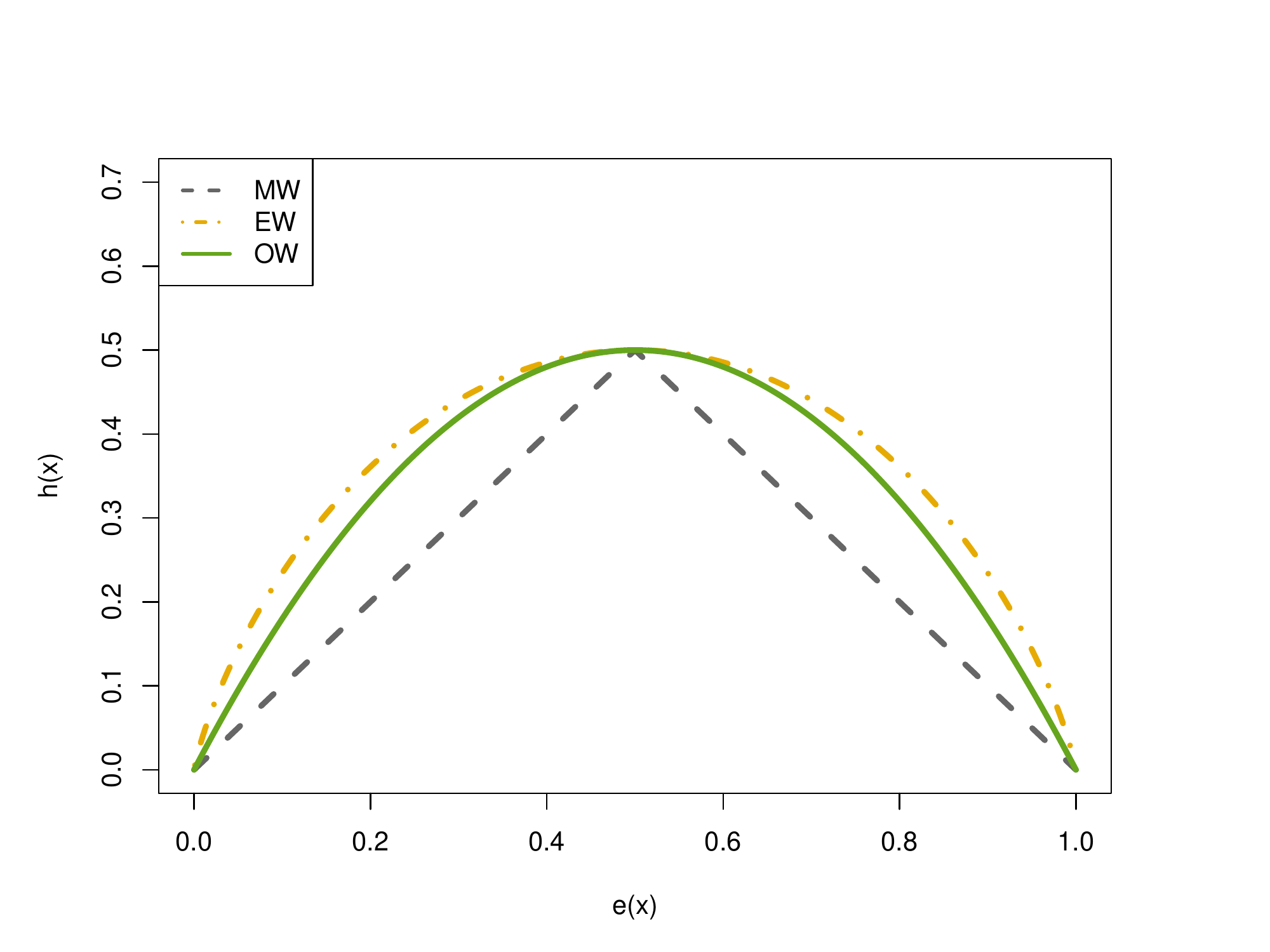}
					\label{fig:sub2}
				\end{center}
			\end{subfigure}
			\caption{Tilting functions for IPW and the optimal IPW trimming (left) and for OW,  MW, and EW (right)}\label{fig:4st_ps_bwgts}\label{fig:tilting_functions}
		\end{center}
	\end{figure}
	In Figure \ref{fig:tilting_functions} we compare some of the tilting functions; the functions were scaled (by multiplying each function by some constant) in a way that they all reach 0.5 as maximum whenever $e(\boldsymbol  x)=0.5$. Multiplying a tilting function by a constant is not essential and doesn't change the nature of the estimand of interest.
	This figure shows how different tilting functions re-weight the target population.  IPW does not change the distribution at all.  IPW with trimming excludes all patients with PS values beyond a threshold (e.g., keep only observations with PS inside the [0.1, 0.9]).  In contrast, OW, MW, and EW keep the entire sample size and tilt $f(\boldsymbol x)$ towards 0.50 and away from 0 or 1. This circumvents the need to discard observations with extreme weights since their relative weights and influence dwindle the closer the PS approaches 0 or 1. In fact, these estimators counter the unduly influence of observations at the tails of the PS distributions might have exerted on the treatment effect estimation by smoothly down-weighting them and thus strategically target the subpopulation of subjects for whom there is most clinical equipoise  \cite{li2018addressing,mao2019propensity}. The difference between OW, MW and EW is in how sharply the tails are down-weighted (Figure \ref{fig:tilting_functions}).
	

	\subsubsection{Estimation}
	
	The estimator $\widehat{\Delta}_h$  of the causal target of inference $\Delta_h$ can be determined as
	\allowdisplaybreaks\begin{align}\label{eq.estimand}
	\widehat{\Delta}_h=\frac{\displaystyle \sum_{i=1}^{N}Z_iY_i\widehat w_i(\boldsymbol  x)}{\displaystyle\sum_{i=1}^{N}Z_i\widehat w_i(\boldsymbol  x)}-\frac{\displaystyle\sum_{i=1}^{N}(1-Z_i)Y_i\widehat w_i(\boldsymbol  x)}{\displaystyle\sum_{i=1}^{N}(1-Z_i)\widehat w_i(\boldsymbol  x)}
	\end{align}
	where $\widehat w(\boldsymbol  x)=Z\widehat w_1(\boldsymbol  x)+(1-Z)\widehat w_0(\boldsymbol  x)$ are the estimated weights.
	For non-randomized studies, the PS is unknown and $e(\boldsymbol x)$ usually estimated by logistic regression.
	
	The statistical properties of the estimator $\widehat{\Delta}_h$ can be found in the literature \cite{hirano2003efficient,li2013weighting,li2018addressing,li2018balancing,mao2019propensity,yang2018asymptotic}.  When $h(\boldsymbol  x)=1,$ the above estimator  $\widehat{\Delta}_h$ correspond to the modified Horvitz-Thompson IPW estimator, also known as the Hajek's estimator  \cite{hajek1971comment}. It is more efficient than the standard Horvitz-Thompson IPW estimator \cite{robins2007comment,lunceford2004stratification,busso2014new} and correspond to a version of stabilized IPW estimator of Hernan and Robins\cite{hernan2006estimating} with weights $\displaystyle \frac{Zg(Z)}{e_i(X)}$ and  $\displaystyle \frac{(1-Z)g(Z)}{1-e_i(X)}$, if we define  
	$$g(Z)=\left[\displaystyle{Z}\left( {\displaystyle\sum_{i=1}^{N}Z_ie_i(X)^{-1}}\right) ^{-1} + \displaystyle{(1-Z)}\left( {\displaystyle\sum_{i=1}^{N}(1-Z_i)(1-e_i(X_i))^{-1}}\right) ^{-1} \right]^{-1} $$
	However, we did not considered the version of stabilized-weight estimator for which $g(Z)=ZP(Z=1)+(1-Z)P(Z=0)$ \cite{hernan2018causal, hernan2006estimating} since we decided to use  normalized weights for all estimators we compared, for consistency. Nevertheless, we have provided the sandwich variance estimators of  these latter stabilized weights in  the Appendix \ref{sec:stabilizedIPW_var}.
	
	In summary, whenever the PS model is correctly specified, but there are extreme IPW weights, the above alternatives (OW, MW, and EW) are more reliable than the IPW and even IPW trimming: no exclusion, no reduction in sample size, improved covariate balance, better precision, and better point estimation \cite{crump2009dealing,crump2006moving,li2018addressing,li2018balancing,mao2019propensity,li2013weighting}.
	When extreme weights are not present and correspondingly the PS distribution has good overlap between treatment groups, balancing weights perform similarly.
	However, little is known about the behaviors of OW, MW, and EW when the PS model is misspecified. To fill this gap in our knowledge, we first compare the asymptotic statistical  properties of IPW, OW, MW, and EW. Later, we run simulation studies to evaluate their performance under PS model misspecification, in finite samples.

	\subsection{Asymptotic behaviors when the positivity assumption is violated or the propensity score model is misspecified }
	The asymptotic behaviors of IPW, OW, MW, and EW estimators can be assessed using results from the semi-parametric theory (see, for instance, Tsiatis \cite{tsiatis2007semiparametric}).
	Given  $h(\boldsymbol  x)$, the estimator $\widehat \Delta_h=\widehat \Delta_{1h}-\widehat \Delta_{0h}$ can be derived via the solutions $\left( \widehat \Delta_{1h}, \widehat \Delta_{0h}\right)$ to the estimating equation $\displaystyle\sum_{i=1}^{N}\displaystyle \begin{pmatrix}
	\displaystyle \frac{Z_ih(\boldsymbol  X)}{e(\boldsymbol  X_i)}(Y_i-\Delta_{1h}), ~
	\displaystyle \frac{(1-Z_i)h(\boldsymbol  X)}{1-e(\boldsymbol  X_i)}(Y_i-\Delta_{0h}))
	\end{pmatrix}'=0, $ with respect to $(\Delta_{1h}, \Delta_{0h})$, i.e., $$\left( \widehat \Delta_{1h}, \widehat \Delta_{0h}\right) =\left(\frac{\displaystyle \sum_{i=1}^{N}Z_iY_i\widehat w_i(\boldsymbol  x)}{\displaystyle\sum_{i=1}^{N}Z_i\widehat w_i(\boldsymbol  x)},~ 
	\displaystyle \frac{\displaystyle\sum_{i=1}^{N}(1-Z_i)Y_i\widehat w_i(\boldsymbol  x)}{\displaystyle\sum_{i=1}^{N}(1-Z_i)\widehat w_i(\boldsymbol  x)} \right). $$
	Consider, $m_z(\boldsymbol X)=E(Y|Z=z, \boldsymbol X)$ for $z=0, 1$ and $\sigma_z^2(\boldsymbol X)=Var(Y|Z=z, \boldsymbol X)$. As shown by Hirano et al. \cite{hirano2003efficient},
	the efficient influence function of $\widehat \Delta_h$  is given by 
	\begin{align*}\allowdisplaybreaks
	&\text{EIF}_{h}=\displaystyle \displaystyle \frac{h(\boldsymbol  X)}{E\left[h(\boldsymbol  X)\right]}
	\left[ \frac{ZY}{e(\boldsymbol  X)}-\frac{(1-Z)Y}{1-e(\boldsymbol  X)} - \{Z-e(\boldsymbol  X)\} \left\lbrace m_1(\boldsymbol  X)+ m_0(\boldsymbol  X)\right\rbrace  -\Delta_{h}  \right]
	\end{align*}
	
	The asymptotic variance of $\widehat \Delta_h$ is the variance of its influence function $\text{EIF}_{h}$,\cite{tsiatis2007semiparametric} i.e., 
	\begin{align}\allowdisplaybreaks\label{eq:asympt_variance}
	&\text{AV}(\widehat \Delta_{h})=\displaystyle \displaystyle E\left[h(\boldsymbol  X)\right]^{-2} E\left[h(\boldsymbol X)^2 \left\lbrace\sigma_1^2(\boldsymbol X)e(\boldsymbol X)^{-1} + \sigma_0^2(\boldsymbol X)(1-e(\boldsymbol X))^{-1}\right\rbrace \right]. 
	\end{align}
	This last equation \eqref{eq:asympt_variance} can be used to show the stark contrast between the asymptotic variance of the IPW estimator (i.e., $h(\boldsymbol X)=1$) and those of the  OW, MW, and EW estimators when the propensity scores $e(\boldsymbol X)$ are near 0 or 1. 
	
	 Whenever $e(\boldsymbol X)\approx$ 0 or 1 for some observations, their contributions to the  asymptotic variance $\text{AV}(\widehat \Delta_{h})$ of either of the last 3 estimators are negligible, since  $\allowdisplaybreaks\displaystyle {h(\boldsymbol X)}\approx 0 $ and $h(\boldsymbol X)^2 \left\lbrace\sigma_1^2(\boldsymbol X)e(\boldsymbol X)^{-1} + \sigma_0^2(\boldsymbol X)(1-e(\boldsymbol X))^{-1}\right\rbrace\approx 0 $ in these regions of the PS spectrum. However,  for the IPW estimator,   $\text{AV}(\widehat \Delta_{h})=\displaystyle E\left[\sigma_1^2(\boldsymbol X)e(\boldsymbol X)^{-1} + \sigma_0^2(\boldsymbol X)(1-e(\boldsymbol X))^{-1}\right]$ 
	can take large values when  $e(\boldsymbol X)\approx$ 0 or 1, even if the treatment effect is constant.
	
	Furthermore, when the PS model is misspecified, $\widehat e(\boldsymbol X)$ converges to $\widetilde e(\boldsymbol X)\neq e(\boldsymbol X)$ and thus yields a biased estimator $\widetilde \Delta_h$. The asymptotic bias of  $ \widetilde \Delta_h$  is equal to 
	\allowdisplaybreaks \begin{align}\label{eq:asympot_bias}
	\text{ABias} (\widehat \Delta_{h})= & \frac{E\left[\displaystyle\frac{e(\boldsymbol X)}{
			\widetilde e(\boldsymbol X)} \widetilde h(\boldsymbol X) m_1(\boldsymbol  X)  \right]}{E\left[\displaystyle\frac{e(\boldsymbol X)}{
			\widetilde e(\boldsymbol X)} \widetilde h(\boldsymbol X)
		\right]}
	-\frac{E\left[\displaystyle\frac{ 1-e(\boldsymbol X)}{1-
			\widetilde e(\boldsymbol X)} \widetilde h(\boldsymbol X)
		m_0(\boldsymbol  X)\right]}{E\left[\displaystyle\frac{ 1-e(\boldsymbol X)}{1-
			\widetilde e(\boldsymbol X)} \widetilde h(\boldsymbol X) \right]}  -\Delta_{h}
	\end{align}
	based on the results from Appendix \ref{sec:asymptotic_results}.
	
	Therefore, if the treatment is constant, then the difference $m_1(\boldsymbol  X)-m_0(\boldsymbol  X)$ is constant for $z=0,1$ and  $\text{ABias} (\widehat \Delta_{h})\approx 0$. However, when the treatment effect is heterogeneous and the PS model is misspecified, i.e., $\widetilde e(\boldsymbol X)\neq  e(\boldsymbol X)$, the impact of such misspecification in the asymptotic bias of the IPW estimator $\widehat \Delta_{h=1}$ can be exacerbated by the presence of observations with $ \widetilde e(\boldsymbol X)\approx$ 0 or 1, since its asymptotic bias \eqref{eq:asympot_bias} is
	\begin{align*}\allowdisplaybreaks
	&E\left[\displaystyle\frac{e(\boldsymbol X)}{
		\widetilde e(\boldsymbol X)}\right]^{-1}\!\!\!E\!\left[\displaystyle\frac{e(\boldsymbol X)}{
		\widetilde e(\boldsymbol X)}m_1(\boldsymbol  X)\right]-E\left[\displaystyle\frac{ 1-e(\boldsymbol X)}{1-
		\widetilde e(\boldsymbol X)}\right]^{-1}\!\!\!E\!\left[\displaystyle\frac{ 1-e(\boldsymbol X)}{1-
		\widetilde e(\boldsymbol X)}m_0(\boldsymbol  X)\right] -\Delta_{h}.
	\end{align*}
	For the OW, MW, or EW estimator,  $\displaystyle\frac{e(\boldsymbol X)\widetilde h(\boldsymbol X) }{
		\widetilde e(\boldsymbol X)}\approx 0$ and $\displaystyle\frac{ 1-e(\boldsymbol X) \widetilde h(\boldsymbol X) }{1-
		\widetilde e(\boldsymbol X)}\approx 0$ in \eqref{eq:asympot_bias}, when $\widetilde e(\boldsymbol X)\approx 0$ or 1, and thus do not contribute much to the asymptotic bias.
	
	Overall, OW, MW, and EW have better asymptotic behaviors compared to IPW weights. While we expect these behaviors we have uncovered here to carry over in finite sample size, we run  simulation studies, in Section \ref{sec:simulation}, to find out.
	\section{Motivating example} \label{sec:motivation}
	
	To set the stage, we consider data from the North Carolina vital statistics recorded between 1998 and 2002 by the North Carolina State Center Health Services, accessible through the Odum Institute at the University of North Carolina in Chapel Hill. The goal is to estimate the effects of maternal smoking on infant birth weights (measured in grams) from first-time black mothers ($N=157,988$). 
	
\begin{figure}[h]
		\setlength{\fboxsep}{0pt}%
		\setlength{\fboxrule}{0pt}%
		\begin{center}
			\includegraphics[trim=37 15  30 57, clip, width=0.7\linewidth]{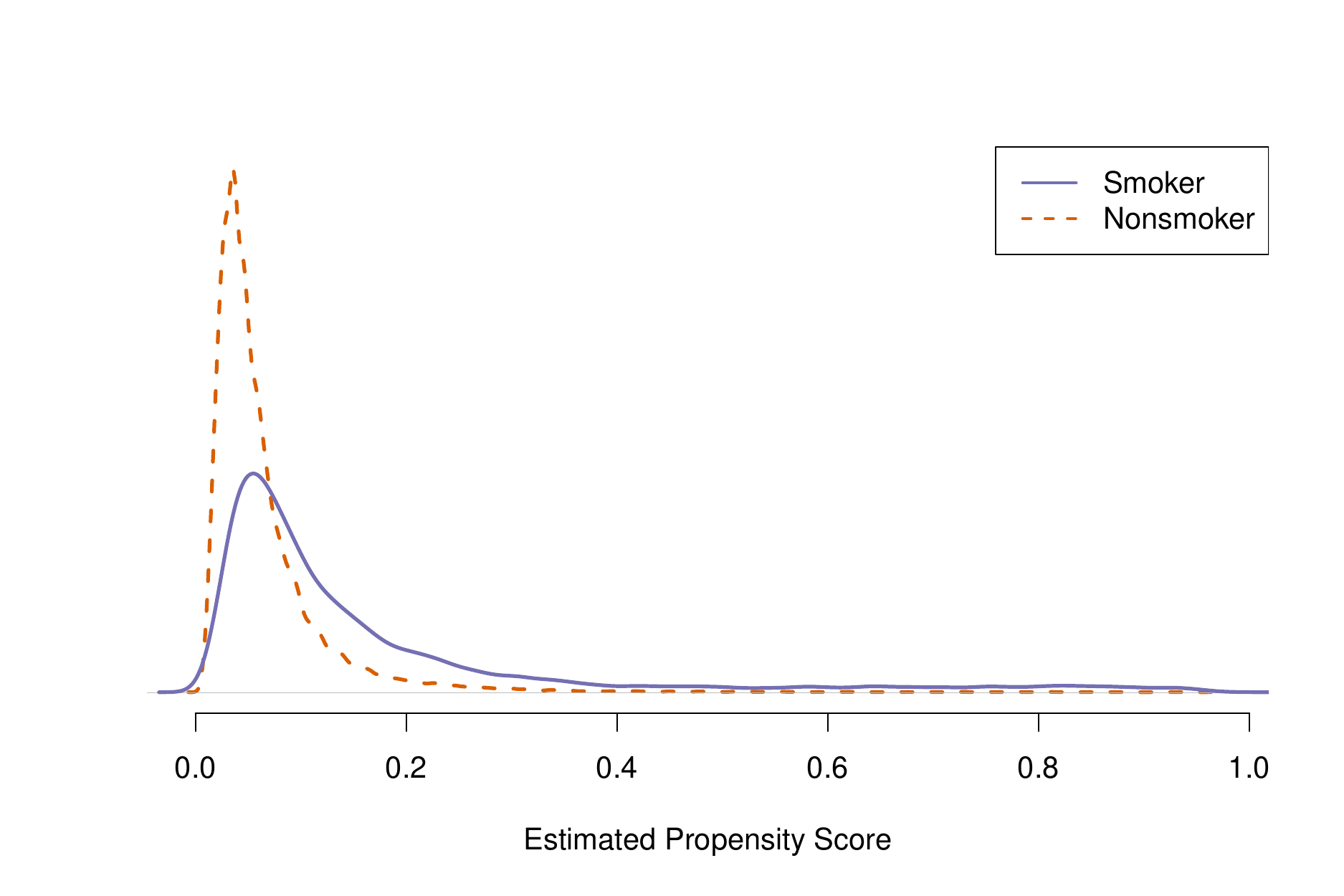}
			\footnotesize	\begin{tabular}{cccccccccccccccccccccc}
				\toprule
				&   Min.  & 1st Qu.  &  Median   &   Mean  & 3rd Qu.   &   Max. \\ \cmidrule(lr){1-7}
				Smoker     &  0.0044 & 0.06 & 0.10 & 0.17 & 0.18 & 0.98 \\
				Nonsmoker & 0.0014 & 0.03 & 0.05 & 0.07 &  0.08 & 0.97 \\
				
				\bottomrule
			\end{tabular}
			\caption{ NC birth weights: Distribution of estimated propensity scores\label{PS_NCBW}}
		\end{center}
	\end{figure}

\begin{figure*}
		\setlength{\fboxsep}{0pt}%
		\setlength{\fboxrule}{0pt}%
		\begin{center}
			\includegraphics[trim=18 8 12 20, clip, width=0.55\linewidth]{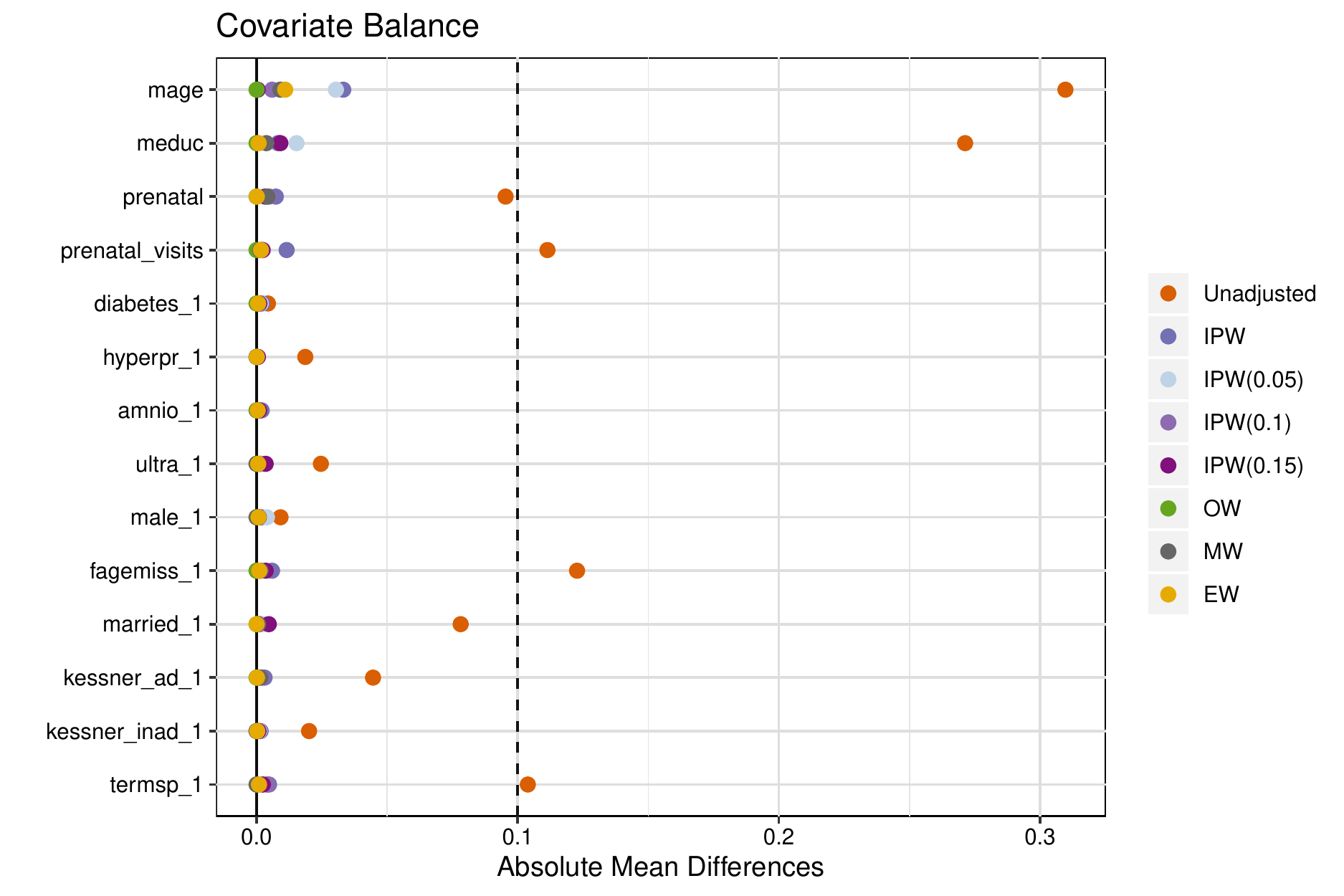}
		\end{center}
		\caption{NC birth weights: Covariate balance (smokers vs. nonsmokers)\label{Balance_NCBW}}
	\end{figure*}

	Given that maternal smoking is non-randomized, propensity scores are used to adjust for differences in women who smoke and those who do not. The propensity score model, based on a logistic regression, includes 4 continuous variables (mother's age, years of education, month of first prenatal visit, and number of prenatal visits), as well as 11 binary variables (indicators of  baby's gender, mother's marital status, gestational diabetes, hypertension, amniocentesis, ultrasound exam and previous terms where newborn died, adequate Kessner Index, inadequate Kessner Index,  alcohol consumption during pregnancy, and whether father's age is missing). The distributions of estimated propensity scores (Figure \ref{PS_NCBW}), show a good overlap of the propensity scores of mothers who smoked during their pregnancy and those who did not. However, a large number of propensity score values are near 0, suggesting potential problems from extreme weights.  This motivates potential gains from balancing weights (trimming, OW, MW, EW) beyond standard IPW.  Applying these methods we find that good covariate balance is achieved between the two exposure groups for all balancing weights (Figure~\ref{Balance_NCBW}).
	
	\begin{figure*}
		\setlength{\fboxsep}{0pt}%
		\setlength{\fboxrule}{0pt}%
		\begin{center}
			\includegraphics[trim=15 15 8 35, clip, width=0.55\linewidth]{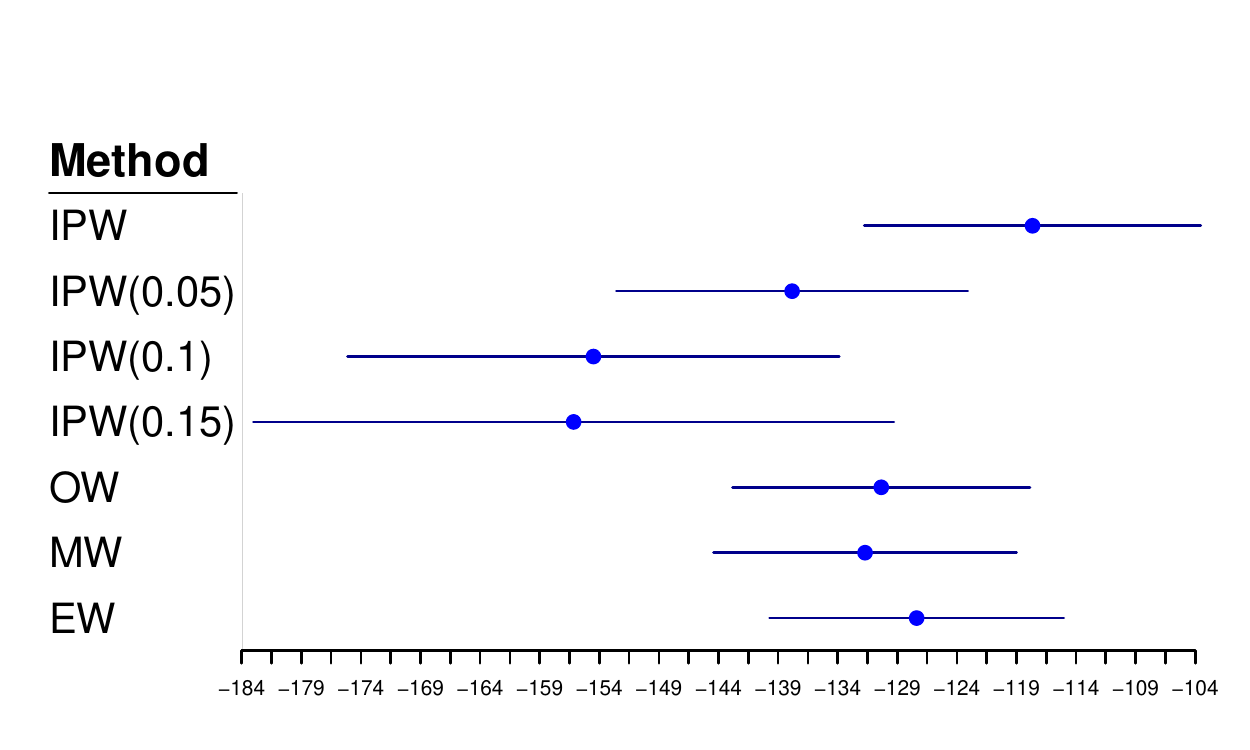}
		\end{center}
		\caption{NC birth weight: Forest plot\label{forest_NCBW}}
	\end{figure*}

	
	Next, each of the balancing weights is used to estimate the corresponding weighted-average treatment effect defined by equation \eqref{eq:estimand}.
	Assuming that the propensity score model is correctly specified, the point estimates and confidence intervals are displayed in Figure~\ref{forest_NCBW}.
	The conclusion is the same across the different analyses: birth weights are lower if the mothers smoked than not.  However, the estimated magnitude ranges from -118 grams (IPW) to -156 grams (IPW with trimming at 0.15), and the confidence intervals differ substantially.  As expected,  OW is the most efficient (the smallest standard error) \cite{li2018addressing,li2018balancing,li2019propensity}, while IPW is the least efficient.
	
	Note that trimming further inflates the standard error of IPW estimates by excluding  from the original sample a large proportion of mothers with estimated propensity scores near 0. This can be seen in the resulting standard errors, but also prior to analysis of outcome through a ``design effect'' approximation of Kish \cite{li2018balancing, kish1985survey}. This approximate variance inflation is reported in Table~\ref{NCBW}.
	\begin{align}
	\text{VI}=\displaystyle \frac{N_1(N-N_1)}{N} \left[ \frac{\displaystyle\sum_{i=1}^{N} Z_iw(\boldsymbol  x_i)^2}{\left(\displaystyle\sum_{i=1}^{N}Z_iw(\boldsymbol  x_i)\right)^2} + \frac{\displaystyle\sum_{i=1}^{N} (1-Z_i)w(\boldsymbol  x_i)^2}{\left(\displaystyle\sum_{i=1}^{N}(1-Z_i)w(\boldsymbol  x_i)\right)^2}\right], ~\text{for}~N_1=\sum^{N}_{i=1} Z_i.
	\end{align}
	Out of $157,988$ mothers, $130,570$ (82.6\%) of them  would be trimmed with the cutoff of $0.10$. Therefore, excluding these mothers from the analysis using Crump's optimal threshold of 0.10 \cite{crump2009dealing} leads to an estimate of the ATE equal to -154.49 grams (SE = 10.58 grams). In this case, optimal trimming does not reduce standard error of IPW estimate, suggesting that the performance of trimming is inconsistent. In addition, $79,125$ (50\%) and $144,983$ (91.8\%) mothers would be excluded using, respectively, the cutoffs of $0.05$ and $0.15$. The last two options result both in considerable reductions of sample size and large variation in the causal effect estimate. The estimate goes from -137.82 (SE = 7.56 grams) to  -156.15 (SE = 13.78 grams) when the threshold increases from 0.05 to 0.15.

	\begin{table}[ht]
		\def\arraystretch{1.1}
		\centering\sf\caption{Sample size, variance inflation, estimate and standard error of NC birth weight data} \label{NCBW}
		\begin{tabular}{rcccccccccccccccccccccccccccccccccccccccc}
			\toprule
			& IPW & IPW(0.05) & IPW(0.1) & IPW(0.15) & OW & MW & EW \\
			\cmidrule(lr){1-8}
			$n$ & 157,988 & 78,863 & 27,418 & 13,005 & 157,988 & 157,988 & 157,988 \\ 
			$\%$ & 100 & 49.92 & 17.35 & 8.23 & 100 & 100 & 100 \\ 
			\text{VI} & 1.66 & 1.65 & 2.85 & 4.81 & 1.12 & 1.14 & 1.14 \\\midrule \addlinespace 
			$\widehat\Delta_h$ & -117.65 & -137.82 & -154.49 & -156.15 & -130.34 & -131.70 & -127.37 \\ 
			SE & 7.24 & 7.56 & 10.58 & 13.78 & 6.40 & 6.52 & 6.33 \\
			
			\bottomrule
		\end{tabular}
		\begin{tablenotes}
			\tiny
			\item $n$: trimmed sample size (when applicable); $\%$: percentage of data used (out of $N=157,988$); SE: standard error.
			\item VI: variance inflation; $\widehat\Delta_h$: point estimate of causal treatment effect; SE: standard error.
		\end{tablenotes}
		
	\end{table}

	As demonstrated in the example, the balancing weights that target patients for whom there is treatment equipoise (OW, MW, EW) exhibit preferable statistical properties in terms of good covariate balances and superior precision. These would seem to be superior methods to handle tails in the propensity score distribution. However, we can not rule out mis-specification in the propensity score model.  Are the apparent gains associated with OW, MW and EW achieved by sacrificing robustness to mis-specification? Even if the propensity score model is misspecified, we hope the appealing statistical properties of these balancing weights continue to exist and our choice of propensity score weighting method is robust. To investigate how robust are the balancing weights (OW, MW, EW) under various propensity score model misspecifications, different degrees of propensity score overlap, and different sample sizes.

	\section{Simulation studies} \label{sec:simulation}
	\subsection{Our objectives}
	Several simulation studies have been  already  conducted to assess the performance of IPW under propensity score misspecification \cite{kang2016practice, kang2007demystifying,mao2019propensity,kreif2013regression,austin2017performance} or practical violations of the positivity assumption   \cite{petersen2012diagnosing,kang2016practice,pirracchio2014improving,austin2017performance}.  Building upon these investigations, our objective is to 
	investigate how robust are the OW, MW, and EW to PS model misspecifications and under 3 degrees of PS overlap (good, moderate, or poor), for $N=500$, $1000$, and $2000$. 
	\subsection{Simulation setup}
	
	We considered both medium and low  prevalence of treatment (see Table \ref{trt_prevalence}) and carried out extensive simulation studies using two different data generating processes (DGPs), for each of the intended PS model misspecifications. For each DGP, we generated a total of 1,000 simulated datasets.  
	
	Each method considered, targets a different estimand of interest. Therefore,    throughout the simulations, we calculated the true estimands for IPW, OW, MW, and EW under heterogeneous treatment effect  using "super-populations" of size $10^7$ units, based on the true parameter coefficients, covariates, and models.
	
	\subsubsection{Variable omission.}
	We  generated $\boldsymbol X=(X_1, \dots, X_6)$ and $Z$ following the DGP of Li and Greene \cite{li2013weighting}, with $X_4 \sim Ber(0.5)$, $X_3\sim Ber(0.4+0.2X_4)$,
	$X_5=X_{1}^{2},$  $X_6=X_{2}X_{4},$
	\begin{flalign*}
	\text{where}~&\begin{pmatrix} X_{1}\\
	X_{2}
	\end{pmatrix}\! \sim  N\!\begin{bmatrix}\begin{pmatrix}
	X_{4}-X_{3}+0.5X_{3} X_{4}\\
	-X_{4}+X_{3}+X_{3} X_{4}
	\end{pmatrix}, \begin{pmatrix}
	2-X_{3} & 0.25(1+X_{3})\\
	0.25(1+X_{3}) & 2-X_{3}				
	\end{pmatrix} 	\end{bmatrix},
	\end{flalign*}
	and $Z\sim Bernoulli(e(\boldsymbol X))$, 	where $e(\boldsymbol X)=[1+\exp\{-(\beta_{0}+\beta_{1}X_{1}+\dots +\beta_{6}X_{6})\}]^{-1}.$
	
\begin{figure*}[h]
		\begin{center}
			\includegraphics[trim=2 20 19 35, clip, width=0.75\linewidth]{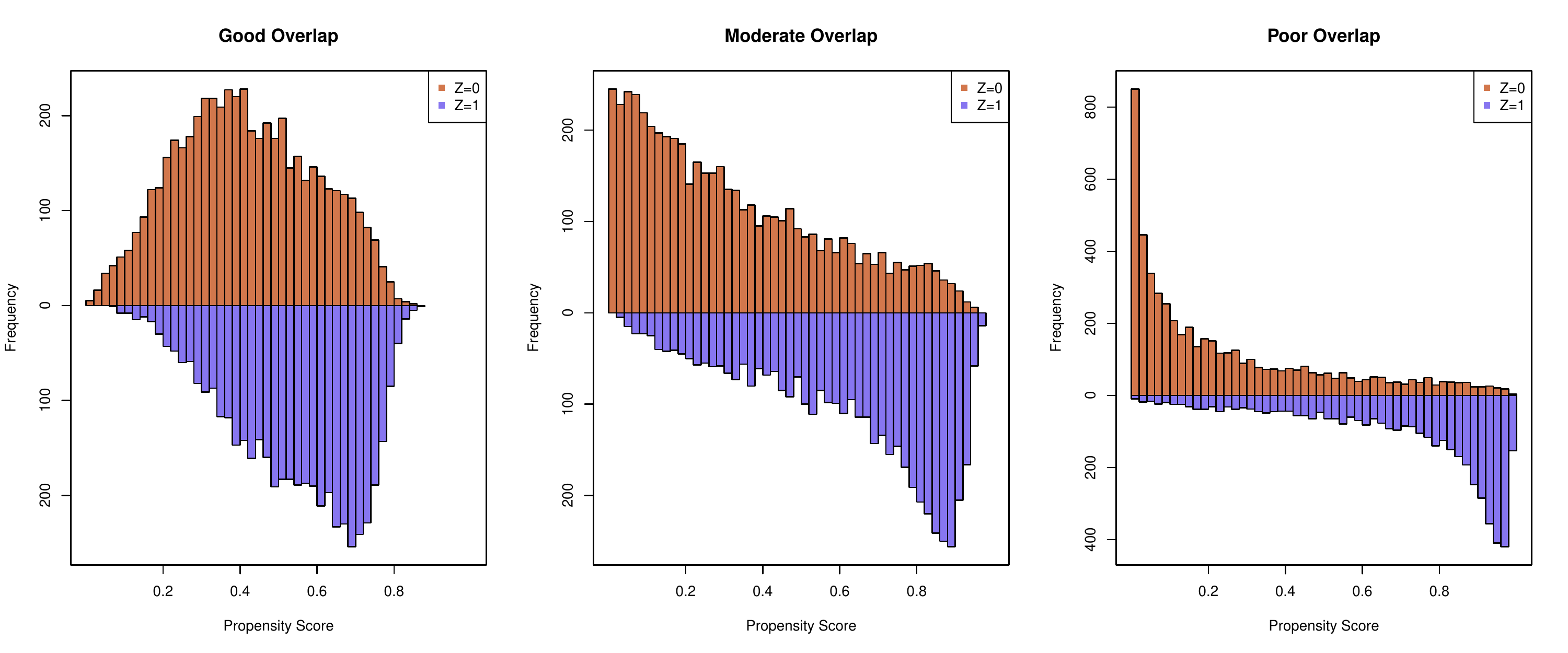}
			\includegraphics[trim=2 20 19 35, clip, width=0.75\linewidth]{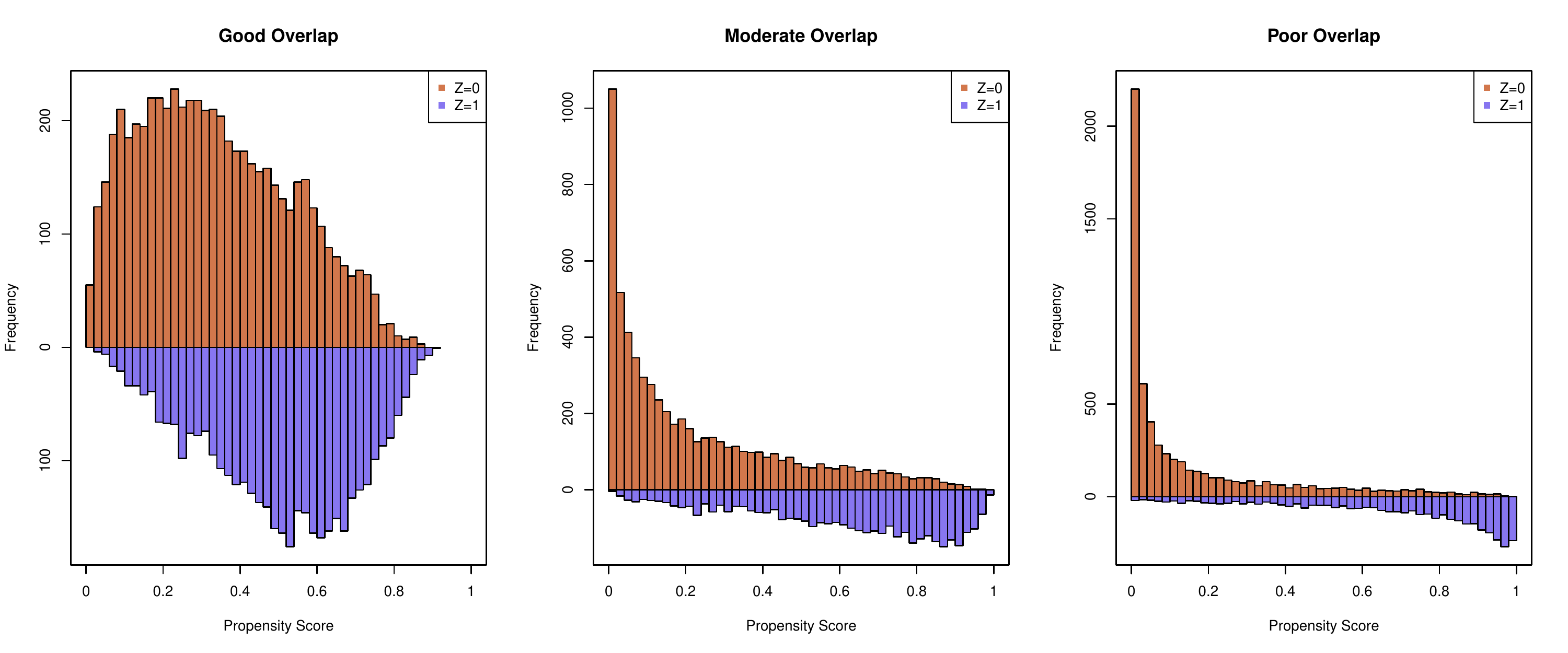}
		\end{center}
		\caption{
			Histograms of the propensity scores from simulated data, with good (left), moderate (middle), and poor (right) overlap. Variable omission (top panel) and variable transformation (bottom panel).
			\label{Mao_Overlap}}
	\end{figure*}

	We considered, respectively, $(\beta_{0},\beta_{1}, \dots, $ $\beta_{6})=$ $\allowdisplaybreaks (-0.5,$ $0.3,$ $0.4,$ $0.4,$ $ 0.4,$ $-0.1,$ $-0.1)$, $(-1,$ $0.6,$ $0.8,$ $0.8,$ $0.8,$ $-0.2,$ $-0.2)$, and $(-1.5,$ $0.9,$ $1.2,$ $1.2,$ $1.2,$ $-0.3,$ $-0.3)$ for good, moderate, and poor overlap of PS distributions, as shown in Figure \ref{Mao_Overlap}. For low prevalence of treatment, we simply changed the intercept $\beta_{0}$ to $-1.5$,
	$-3$, and $-4$ for good, moderate, and poor PS overlaps (see Figure \ref{Mao_Overlap_Asy}).
	
	We chose $\Delta=1$  (resp. $\Delta=-4e(\boldsymbol X)^2+3.94e(\boldsymbol X)+0.69$) for a homogeneous  (resp. heterogeneous) treatment effect. Then, we generated the outcome  $Y\sim 0.5+\Delta Z+X_{1}+0.6X_{2}+2.2X_{3}+1.2X_{4}+0.1{X_{5}}+X_{6}+N(0 ,1)$.
	Finally,  to misspecify the true PS model, we omitted (one at a time)
	the  variables $X_2$,  $X_5=X_1^2$, and  $X_6=X_{2}X_{4}$.

	\subsubsection{Variable transformation.}
	We generated  $\boldsymbol X=(X_1, \dots, X_5)$ and $Z$ such that
	$$
	\begin{pmatrix} X_{1}\\
	X_{2}
	\end{pmatrix} \sim  N\begin{bmatrix}\begin{pmatrix}
	2\\
	4
	\end{pmatrix}, \begin{pmatrix}
	1 & 0.2\\
	0.2 & 1				
	\end{pmatrix} 	\end{bmatrix}, ~ \begin{pmatrix} X_{3}\\
	X_{4}
	\end{pmatrix} \sim  N\begin{bmatrix}\begin{pmatrix}
	2\\
	4
	\end{pmatrix}, \begin{pmatrix}
	1 & 0.2\\
	0.2 & 1				
	\end{pmatrix} 	\end{bmatrix},	~ X_5=X_{1}^{2},
	$$
	and $Z\sim Bernoulli(e(\boldsymbol X))$, 	where $e(\boldsymbol X)=[1+\exp\{-(\beta_{0}+\beta_{1}X_{1}+\dots +\beta_{5}X_{5})\}]^{-1}.$
	
	Under medium treatment prevalence (see Table \ref{trt_prevalence}), we chose $(\beta_{0},$ $\beta_{1},$ $\dots,$ $ \beta_{5})=$ $(-0.2,$ $0.3,$ $0.15,$ $0.22,$ $0.15,$ $-0.15)$, $(-0.4,$ $0.6,$ $0.3,$ $0.44,$ $0.3,$ $-0.3)$, and $(-0.6,$ $ 0.9,$ $0.45,$ $0.66,$ $0.45,$ $-0.45)$, respectively, to obtain good, moderate, and poor PS distributions overlap (Figure~\ref{Mao_Overlap}). For low prevalence of treatment, we change the intercept $\beta_{0}$ into $-1.5$, $-2.5$, and $-3.5$, respectively, for good, moderate, and poor overlap (see Figure \ref{Mao_Overlap_Asy}).
	
	We generated the outcome $Y\sim 9 $ + $\Delta Z + $  $0.1X_1$$ -0.05(X_2 - X_3 + X_4)$$ +  N(0,1)$, with $\Delta=1$  and $\Delta=-4e(\boldsymbol X)^2+3.94e(\boldsymbol X)+0.69$, respectively, for homogeneous and heterogeneous treatment effects.	
	Finally, we considered the misspecified propensity score model, $\allowdisplaybreaks\displaystyle \widetilde e(\boldsymbol X))=[1+\exp\{-(\beta_{0}+\beta_{1}M_{1}+\dots+\beta_{4}M_{4})\}]^{-1}$ where $M_2=X_2(1+X_1)+10$ and $M_3=\left( 0.04X_3+0.6\right)^2$, following the DGPs of Kang and Schafer and Kreif et al. \cite{kang2007demystifying, kreif2013regression}.
	We used, respectively, $\displaystyle (M_1, M_4)=\left(\exp( 0.1X_1), (X_4+20)^2\right)$  and $\left(\exp\left(0.33X_1\right), \left(0.1X_2+X_4+20\right)^2\right)$ for mild and major PS model misspecifications.
	
	\begin{table}[]
		\centering
		\begin{threeparttable}[]
			
			\caption{Average treatment prevalence for different levels of PS overlap.\label{trt_prevalence} { }}
			\centering\small\sf
			\begin{tabular}{p{3.5cm}p{1cm}p{1cm}p{0.7cm}p{0.1cm}p{1cm}p{1cm}p{0.7cm}  }
				\toprule
				& \multicolumn{3}{c}{Medium prevalence} && \multicolumn{3}{c}{Low prevalence} \\ \cmidrule(lr){2-4}\cmidrule(lr){6-8}
				PS misspecification   & Good & Moderate & Poor  &&Good & Moderate & Poor  \\
				\cmidrule(lr){1-8}

				Variable omission & 47.42\% & 47.34\% & 47.12\% & & 	27.7\% & 	18.07\% & 	17.99\% \\
				Variable transformation  & 40.76\% & 37.63\% &	37.00\% &&  18.51\% & 	12.79\% & 	10.87\% \\
				\bottomrule
			\end{tabular}
		\end{threeparttable}
	\end{table}			
	\subsection{Simulation results}
	For each method, we provide its large sample size "true" estimand based on  the DGP considered and 	
	summarize the simulation results in terms of relative bias, root mean squared error (RMSE), empirical standard deviation (SD), average estimated standard error (SE), and  coverage probability (CP) of the $95\%$ confidence interval. 
	
	To be concise, we only present here the results for $N=2000$ under heterogeneous treatment effect (see Tables \ref{varomi_HE}--\ref{T2}). 
	We report, in the Appendix, the Cleveland plots for covariate balance and the boxplots of the relative  biases when the treatment effect is heterogeneous (Section \ref{sec:boxplots}). We also provide the results for $N=2000$ under homogeneous treatment effect (Section \ref{sec:homo200_results}), the corresponding boxplots of relative biases, and figures of covariate balance under different degrees of PS model misspecification (Section \ref{sec:boxplots}) as well as the additional results for $N=500$ and $N=1000$ (Section \ref{sec:additional_tables}) and under low prevalence of treatment for $N=2000$ (Section \ref{sec:lowpre}). 
	
	\subsubsection{Variable omission.}	
	The results under variable omission are shown in Table \ref{varomi_HE}.
	\begin{table}[]
		\centering
		\begin{threeparttable}[]
			\centering\scriptsize\sf
			\caption{Variable omission: heterogeneous treatment effect ($N=2000$).\label{varomi_HE} { }}
			\begin{tabular}{p{1.5cm}p{0.6cm}p{0.4cm}p{0.9cm}p{0.55cm}p{0.55cm}p{0.55cm}p{0.55cm}   p{0.75cm}p{0.55cm}p{0.55cm}p{0.55cm}p{0.55cm} }
				\toprule
				&  &  & \multicolumn{10}{c}{Propensity score misspecification} \\\cmidrule(lr){4-13}
				&  &  & \multicolumn{5}{c}{None }& \multicolumn{5}{c}{Missing $X_2$} \\ \cmidrule(lr){4-8}\cmidrule(lr){9-13}
				Weight & Overlap & True & Bias & RMSE & SD & SE & CP & Bias & RMSE & SD & SE & CP \\
				\cmidrule(lr){1-13}

				IPW & Good & 1.52 & -0.20 & 8.42 & 8.42 & 7.72 & 0.94 & 12.12 & 19.96 & 7.61 & 7.16 & 0.25 \\
				IPW(0.05) & Good & 1.53 & -0.09 & 6.63 & 6.63 & 6.71 & 0.95 & 12.05 & 19.55 & 6.63 & 6.63 & 0.23 \\
				IPW(0.1) & Good & 1.53 & -0.10 & 5.82 & 5.82 & 5.95 & 0.96 & 11.80 & 19.09 & 6.08 & 6.18 & 0.17 \\
				IPW(0.15) & Good & 1.54 & -0.10 & 5.38 & 5.38 & 5.54 & 0.95 & 11.29 & 18.38 & 5.82 & 5.92 & 0.17 \\
				OW & Good & 1.55 & -0.08 & 4.77 & 4.77 & 4.83 & 0.95 & 11.90 & 19.20 & 5.37 & 5.43 & 0.08 \\
				MW & Good & 1.56 & -0.06 & 5.01 & 5.01 & 5.06 & 0.95 & 12.18 & 19.87 & 5.65 & 5.70 & 0.08 \\
				EW & Good & 1.54 & -0.10 & 4.88 & 4.88 & 4.95 & 0.95 & 11.91 & 19.17 & 5.41 & 5.48 & 0.09 \\ \addlinespace
				
				IPW & Mod & 1.32 & -0.13 & 31.09 & 31.11 & 21.66 & 0.84 & 24.78 & 38.18 & 19.58 & 15.72 & 0.34 \\
				IPW(0.05) & Mod & 1.36 & 0.38 & 10.79 & 10.78 & 10.51 & 0.93 & 21.92 & 31.67 & 10.47 & 10.26 & 0.19 \\
				IPW(0.1) & Mod & 1.42 & 0.25 & 7.56 & 7.56 & 7.64 & 0.95 & 20.59 & 30.15 & 7.68 & 7.76 & 0.04 \\
				IPW(0.15) & Mod & 1.47 & 0.01 & 6.85 & 6.85 & 6.70 & 0.94 & 20.18 & 30.62 & 7.19 & 6.96 & 0.02 \\
				OW & Mod & 1.44 & -0.14 & 5.61 & 5.61 & 5.58 & 0.95 & 23.00 & 33.63 & 6.17 & 6.07 & 0.00 \\
				MW & Mod & 1.47 & -0.26 & 5.90 & 5.89 & 5.88 & 0.95 & 23.29 & 34.88 & 6.52 & 6.44 & 0.00 \\
				EW & Mod & 1.42 & -0.13 & 6.03 & 6.03 & 6.01 & 0.94 & 22.98 & 33.17 & 6.38 & 6.29 & 0.00 \\ \addlinespace
				
				IPW & Poor & 1.17 & 9.16 & 53.33 & 52.27 & 35.56 & 0.74 & 39.82 & 58.20 & 34.89 & 26.91 & 0.42 \\
				IPW(0.05) & Poor & 1.29 & 0.85 & 11.55 & 11.50 & 11.49 & 0.94 & 29.66 & 39.99 & 11.16 & 11.17 & 0.11 \\
				IPW(0.1) & Poor & 1.39 & 0.57 & 8.65 & 8.62 & 8.35 & 0.94 & 28.77 & 40.95 & 8.30 & 8.24 & 0.00 \\
				IPW(0.15) & Poor & 1.47 & 0.23 & 7.76 & 7.76 & 7.80 & 0.95 & 27.36 & 40.91 & 7.79 & 7.74 & 0.00 \\
				OW & Poor & 1.38 & 0.29 & 6.32 & 6.31 & 6.40 & 0.95 & 32.25 & 44.94 & 6.41 & 6.66 & 0.00 \\
				MW & Poor & 1.42 & 0.22 & 6.60 & 6.60 & 6.75 & 0.96 & 31.98 & 46.05 & 6.78 & 7.03 & 0.00 \\
				EW & Poor & 1.34 & 0.40 & 7.02 & 7.01 & 7.10 & 0.95 & 32.43 & 44.12 & 6.79 & 7.07 & 0.00 \\  \addlinespace

				&  &  &  \multicolumn{5}{c}{Missing $X_1^2$} &   \multicolumn{5}{c}{Missing $X_2X_4$}  \\ \cmidrule(lr){4-8}\cmidrule(lr){9-13}
				Weight & Overlap & True & Bias & RMSE & SD & SE & CP & Bias & RMSE & SD & SE & CP\\
				\cmidrule(lr){1-13}

				IPW & Good & 1.52 & -4.27 & 9.38 & 6.76 & 6.76 & 0.86 & -2.85 & 9.79 & 8.77 & 8.07 & 0.94 \\
				IPW(0.05) & Good & 1.53 & -4.52 & 9.63 & 6.73 & 6.71 & 0.85 & -2.91 & 8.74 & 7.53 & 7.40 & 0.92 \\
				IPW(0.1) & Good & 1.53 & -4.94 & 9.85 & 6.29 & 6.32 & 0.78 & -3.05 & 8.44 & 7.03 & 6.88 & 0.90 \\
				IPW(0.15) & Good & 1.54 & -4.87 & 9.44 & 5.70 & 5.79 & 0.76 & -3.09 & 8.15 & 6.62 & 6.54 & 0.89 \\
				OW & Good & 1.55 & -4.08 & 7.99 & 4.89 & 4.93 & 0.75 & -2.77 & 7.24 & 5.82 & 5.70 & 0.88 \\
				MW & Good & 1.56 & -3.69 & 7.68 & 5.07 & 5.11 & 0.81 & -2.46 & 6.97 & 5.81 & 5.75 & 0.90 \\
				EW & Good & 1.54 & -4.19 & 8.17 & 4.99 & 5.04 & 0.75 & -2.84 & 7.38 & 5.94 & 5.82 & 0.89 \\ \addlinespace
				
				IPW & Mod & 1.32 & -14.10 & 29.30 & 22.60 & 20.11 & 0.92 & -5.90 & 28.22 & 27.14 & 21.18 & 0.94 \\
				IPW(0.05) & Mod & 1.36 & -11.28 & 19.41 & 11.85 & 11.34 & 0.78 & -6.43 & 14.74 & 11.85 & 11.80 & 0.90 \\
				IPW(0.1) & Mod & 1.42 & -8.86 & 14.84 & 7.94 & 7.80 & 0.65 & -4.46 & 10.63 & 8.56 & 8.50 & 0.89 \\
				IPW(0.15) & Mod & 1.47 & -8.25 & 13.95 & 6.84 & 6.74 & 0.55 & -3.57 & 9.19 & 7.54 & 7.41 & 0.89 \\
				OW & Mod & 1.44 & -7.65 & 12.33 & 5.59 & 5.62 & 0.49 & -4.42 & 9.06 & 6.46 & 6.31 & 0.81 \\
				MW & Mod & 1.47 & -6.97 & 11.74 & 5.72 & 5.81 & 0.57 & -3.81 & 8.60 & 6.53 & 6.38 & 0.85 \\
				EW & Mod & 1.42 & -8.60 & 13.65 & 6.17 & 6.15 & 0.48 & -4.93 & 9.86 & 6.96 & 6.86 & 0.82 \\ \addlinespace
				
				IPW & Poor & 1.17 & -32.06 & 71.18 & 60.52 & 45.66 & 0.92 & -5.26 & 51.92 & 51.58 & 38.64 & 0.89 \\
				IPW(0.05) & Poor & 1.29 & -14.56 & 22.25 & 11.83 & 11.98 & 0.68 & -6.52 & 15.36 & 12.84 & 12.30 & 0.91 \\
				IPW(0.1) & Poor & 1.39 & -12.51 & 19.36 & 8.42 & 8.46 & 0.46 & -4.25 & 10.83 & 9.07 & 8.96 & 0.90 \\
				IPW(0.15) & Poor & 1.47 & -11.73 & 18.89 & 7.75 & 7.70 & 0.39 & -3.43 & 9.63 & 8.21 & 8.30 & 0.91 \\
				OW & Poor & 1.38 & -11.03 & 16.53 & 6.46 & 6.36 & 0.33 & -4.91 & 9.68 & 6.93 & 7.05 & 0.85 \\
				MW & Poor & 1.42 & -10.06 & 15.76 & 6.57 & 6.53 & 0.41 & -4.06 & 9.05 & 6.97 & 7.17 & 0.89 \\
				EW & Poor & 1.34 & -12.95 & 19.03 & 7.68 & 7.34 & 0.35 & -6.04 & 11.36 & 7.95 & 8.03 & 0.85 \\
				
				\bottomrule
			\end{tabular}
			\begin{tablenotes}
				\tiny
				\item  Mod: Moderate; IPW($\alpha$): trimmed IPW with $I_{\alpha}(x)=\mathds{1}({\{\alpha\leq e(x)\leq 1-\alpha\}})$, for   $\alpha=0.05, 0.10,$ and $0.15.$
				\item Bias: relative bias in percentage$\times 100$; RMSE: root mean-squared error$\times 100$; SD: empirical standard deviation$\times 100$; SE: average estimated standard error$\times 100$; CP: 95\%  coverage probability. The results are based on 1000 simulated data sets.
			\end{tablenotes}
		\end{threeparttable}
	\end{table}		
	Among all the methods and in almost all the situations we considered, IPW is more sensitive to poorer overlap and PS model misspecification.
	When overlap is good and the model is correctly specified, IPW, as well as the other methods, provides satisfactory estimation with small bias and variance. However, when overlap worsens, even in the scenarios where the propensity scores are estimated appropriately, the relative bias and standard error of IPW estimates increase rapidly and can be much higher than those of the other methods. IPW is also shown to be less accurate and efficient when the propensity score model is misspecified. This is true especially under moderate and poor overlap. In a few number of cases, the difference in bias among IPW estimand and the other methods can be minute given the fact that the relative bias in Table \ref{varomi_HE} is presented on $10^{-2}$ scale; however, the difference in standard error is substantial and not to be neglected.

	While IPW performs poorly, the weighting methods that target the clinical equipoise (OW, MW, and EW) maintain their ability to estimate causal effects with relatively small bias and small variance when $X_1^2$ and $X_2X_4$ are omitted. IPW trimming reduces the relative bias and standard error of IPW estimates, but is often slightly less accurate and less efficient compared to OW, MW, and EW. For example, the standard error of optimal trimming estimate is roughly $1.3$ times higher than that of the OW estimate across the majority of the scenarios.
	Of all the IPW trimming methods we considered, the optimal IPW trimming (with the cutoff $\alpha=0.1$) was the usually most reliable and the most efficient. In general, OW, MW, and EW exhibit stronger stability and higher efficiency against propensity score model misspecifications compared to IPW.
	
	These findings are true except when we omit the main variable $X_2$.  As expected, none of the methods we considered provide satisfactory results and the relative advantage of IPW over the other methods is trivial, considering the magnitudes of the bias and standard error. In fact because $X_2$ is a strong confounder of the treatment--outcome relationship, excluding it from a postulated PS model lead to unreliable results, regardless of the method considered and the degree of the overlap of the PS distributions between the treatment groups.
	
	Moreover, some of the coverage probabilities of $95\%$ confidence intervals can be misleading. Confusion can arise since the coverage probability of IPW is generally better than OW, MW, EW, and trimming. However, the variance of IPW is tremendous so that it maintains acceptable coverage even when the bias of IPW is substantial.
	
	Finally, as indicated in Table \ref{varomi_HE}, the range of "true" values of the estimands under treatment heterogeneity is 1.52--1.56 under good overlap, 1.32--1.47 under moderate overlap, and 1.17--1.47 under poor overlap. In most scenarios, the true estimands of the optimal IPW trimming (i.e., with the cutoff $\alpha=0.1$) and OW are particularly close in values, rendering similar target of inference for these two methods.

	\subsubsection{Variable transformation.}
	
\begin{center}
	\begin{table}

		\begin{threeparttable}
			\centering\scriptsize\sf
			\caption{Variable transformation: heterogeneous treatment effect  ($N=2000$).\label{T2} { }}
			\begin{tabular}{p{1.1cm}p{0.6cm} p{0.35cm}   p{0.65cm} p{0.55cm}p{0.35cm}p{0.35cm}p{0.45cm}p{0.75cm}p{0.65cm}p{0.45cm}p{0.35cm}p{0.35cm}  p{0.7cm}p{0.45cm}p{0.35cm}p{0.35cm}p{0.35cm}}
				\toprule
				&  &  & \multicolumn{15}{c}{Propensity score misspecification} \\\cmidrule(lr){4-18}
				&  &  & \multicolumn{5}{c}{None }& \multicolumn{5}{c}{Mild} &   \multicolumn{5}{c}{Major}   \\ \cmidrule(lr){4-8}\cmidrule(lr){9-13}\cmidrule(lr){14-18}
				Weight & Overlap & True &Bias & RMSE & SD & SE & CP & Bias & RMSE & SD & SE  & CP & Bias & RMSE & SD & SE & CP\\
				\cmidrule(lr){1-18}					
				
				IPW 		& Good & 1.46 & 0.30 & 6.12 & 6.11 & 5.82 &  0.95 & 2.17 & 6.75 & 5.96 & 5.65 &  0.91 & 2.56 & 7.66 & 6.69 & 6.01 &  0.88 \\
				IPW(0.05) 	& Good & 1.47 & 0.29 & 5.55 & 5.54 & 5.36 &  0.95 & 1.47 & 5.99 & 5.59 & 5.34 &  0.93 & 1.93 & 6.18 & 5.49 & 5.30 &  0.91 \\
				IPW(0.1) 	& Good & 1.50 & 0.29 & 5.43 & 5.42 & 5.24 &  0.95 & 0.77 & 5.44 & 5.32 & 5.23 &  0.95 & 1.12 & 5.61 & 5.35 & 5.19 &  0.94 \\
				IPW(0.15) 	& Good & 1.53 & 0.25 & 5.45 & 5.44 & 5.25 &  0.95 & 0.43 & 5.30 & 5.26 & 5.24 &  0.96 & 0.76 & 5.38 & 5.25 & 5.19 &  0.95 \\
				OW 			& Good & 1.52 & 0.22 & 5.19 & 5.18 & 5.06 &  0.95 & 0.91 & 5.32 & 5.14 & 5.03 &  0.94 & 1.42 & 5.53 & 5.09 & 5.00 &  0.93 \\
				MW 			& Good & 1.54 & 0.20 & 5.25 & 5.25 & 5.12 &  0.95 & 0.66 & 5.27 & 5.18 & 5.09 &  0.95 & 1.16 & 5.43 & 5.14 & 5.05 &  0.93 \\
				EW 			& Good & 1.51 & 0.24 & 5.20 & 5.19 & 5.07 &  0.95 & 1.10 & 5.42 & 5.16 & 5.05 &  0.93 & 1.61 & 5.67 & 5.12 & 5.01 &  0.92 \\ \addlinespace
				
				IPW 		& Mod & 1.24 & 0.10 & 15.00& 15.01& 10.53 & 0.91 & 3.54 & 16.89& 16.32& 12.47&  0.85 & 2.94 & 27.44 & 27.21 & 18.24 &  0.84 \\
				IPW(0.05) 	& Mod & 1.35 & 0.13 & 6.85 & 6.86 & 6.66 &  0.94 & 1.88 & 7.26 & 6.81 & 6.75 &  0.93 & 2.39 & 7.35 & 6.60 & 6.59 &  0.92 \\
				IPW(0.1) 	& Mod & 1.42 & 0.15 & 6.40 & 6.40 & 6.44 &  0.95 & 1.00 & 6.45 & 6.30 & 6.48 &  0.96 & 1.25 & 6.35 & 6.10 & 6.31 &  0.96 \\
				IPW(0.15)	& Mod & 1.48 & 0.08 & 6.29 & 6.30 & 6.56 &  0.97 & 0.29 & 6.46 & 6.44 & 6.58 &  0.96 & 0.35 & 6.10 & 6.08 & 6.39 &  0.97 \\
				OW 			& Mod & 1.43 & 0.02 & 5.72 & 5.73 & 5.98 &  0.97 & 1.27 & 5.94 & 5.66 & 5.91 &  0.95 & 2.01 & 6.20 & 5.50 & 5.77 &  0.93 \\
				MW 			& Mod & 1.47 & -0.02& 5.74 & 5.74 & 6.07 &  0.96 & 0.80 & 5.82 & 5.70 & 6.00 &  0.96 & 1.44 & 5.92 & 5.54 & 5.86 &  0.95 \\
				EW 			& Mod & 1.40 & 0.01 & 5.82 & 5.83 & 6.04 &  0.96 & 1.64 & 6.23 & 5.79 & 5.99 &  0.94 & 2.44 & 6.61 & 5.66 & 5.86 &  0.92 \\ \addlinespace
				
				IPW 		& Poor & 1.10 & 4.01 & 18.92& 18.40& 12.64&  0.80 & 5.76 & 29.94 & 29.28 & 22.12& 0.82 & 5.84 & 40.87 & 40.38 & 41.34 &  0.84 \\
				IPW(0.05)	& Poor & 1.31 & 0.26 & 8.21 & 8.21 & 7.77 &  0.93 & 1.89 & 8.70 & 8.34 & 7.89 &  0.92 & 1.77 & 8.14 & 7.81 & 7.57 &  0.93 \\
				IPW(0.1) 	& Poor & 1.40 & 0.01 & 7.88 & 7.88 & 7.58 &  0.94 & 0.20 & 7.61 & 7.60 & 7.61 &  0.95 & -0.09 & 7.24 & 7.24 & 7.27 &  0.94 \\
				IPW(0.15) 	& Poor & 1.47 & -0.11& 8.04 & 8.05 & 7.81 &  0.95 &-0.57 & 7.84 & 7.80 & 7.79 &  0.95 & -1.16 & 7.56 & 7.37& 7.39&  0.94 \\
				OW 			& Poor & 1.38 & -0.07& 7.00 & 7.01 & 6.90 &  0.95 & 1.20 & 7.03 & 6.83 & 6.75 &  0.94 & 1.78 & 6.95 & 6.51 & 6.46 &  0.93 \\
				MW 			& Poor & 1.43 & -0.15& 7.08 & 7.08 & 7.01 &  0.95 & 0.55 & 6.89 & 6.85 & 6.88 &  0.95 & 0.98 & 6.68 & 6.54 & 6.59 &  0.95 \\
				EW 			& Poor & 1.35 & 0.01 & 7.15 & 7.15 & 6.99 &  0.94 & 1.78 & 7.44 & 7.05 & 6.90 &  0.93 & 2.51 & 7.49 & 6.69 & 6.59 &  0.91 \\

				\bottomrule
			\end{tabular}
			\begin{tablenotes}
				\tiny
				\item  Mod: Moderate; IPW($\alpha$): trimmed IPW with $I_{\alpha}(x)=\mathds{1}({\{\alpha\leq e(x)\leq 1-\alpha\}})$, for   $\alpha=0.05, 0.10,$ and $0.15.$
				\item Bias: relative bias in percentage$\times 100$; RMSE: root mean-squared error $\times 100$; SD: empirical standard deviation$\times 100$; SE: average estimated standard error$\times 100$; CP: 95\%  coverage probability. The results are based on 1000 simulated data sets.
			\end{tablenotes}
		\end{threeparttable}
	\end{table}
\end{center}	
	
	The range of the "true" values of the estimands, under treatment heterogeneity, widens as the degree of overlap of the propensity score distributions worsens, going from 1.46--1.54 (good overlap) to 1.24--1.48 (moderate overlap) and 1.10--1.47 (poor overlap). Yet again, we observe that the true estimands (i.e., the target of inference) of OW and optimal IPW trimming ($\alpha=0.1$) are close to each other. This result provides the opportunity to further examine and compare the performances of OW and such an optimal IPW trimming.

	Congruent with the results under variable omission, the results obtained when we transformed the variables used in the true PS model also lead to the same conclusion as indicated in Table \ref{T2}. Indeed, IPW is sensitive to the specification of propensity score model and the degree of overlap. Under good overlap of propensity score distributions and correctly specified propensity score models, all the methods lead to similar causal effect estimates. The biases of different methods are small, corresponding RMSE and standard errors are also satisfactory and close in values. However, when the overlap of the distributions of the PS between the treatment groups becomes more and more limited, going from good to poor,  the performance of IPW become very unstable, leading to more biased estimates and large variability under both minor and major PS model misspecifications. The root mean squared error of IPW estimates increases from $6.12\times10^-2$ to $40.87\times10^-2$ in presence of poor overlap and major propensity score misspecification. This trend can also be observed in other measures in Table \ref{T2}, including standard error and coverage probability.

	Again, the results in Table \ref{T2} demonstrate the robustness of the class of balancing weights that target clinical equipoise (OW, MW, and EW) against poor overlap and propensity score model misspecification. For this class of weights, the inflation in bias and variance is negligible and the ability to maintain satisfactory causal effect estimation is stable across all scenarios. Again, IPW trimming improves the performance of IPW. Furthermore, the optimal IPW trimming (with $\alpha=0.1$) yields very good results, with even smaller biases than OW, MW, and EW in most of the scenarios; however, the standard errors of OW, MW, and EW are consistently smaller. It is worth noting that in this set of simulations, overall the biases from all methods (IPW, IPW trimming, OW, MW, and EW) are relatively small in scale and, thus, the variance continues to play a preeminent  role in comparing them.

	The appendices \ref{sec:boxplots}, \ref{sec:additional_tables}, and \ref{sec:lowpre} include a number of additional simulation settings: (1) homogenous treatment effects; (2) smaller samples sizes ($N=500$ and $N=1000$); (3) low treatment prevalence. Results in these settings were similar to those in Table \ref{varomi_HE} and Table \ref{T2}.  That is IPW yielded some bias and severely inflated standard errors when the propensity score distribution had moderate to poor overlap.  OW, MW and EW all performed similarly to each other and moderately better than trimming. All methods performed poorly when an important confounder ($X_2$) was omitted form the analysis.  In smaller sample sizes and low treatment prevalence, the problems with IPW were even more pronounced and not always corrected by trimming.  The benefit of OW, MW and EW was more evident in these settings.  \\

	\par 
	\section{Discussion}\label{sec:conclusion}
	We investigate the impacts of both propensity score misclassification and practical violations of positivity on OW, MW, and EW compared to IPW and IPW trimming.  In extensive simulations we find that IPW is sensitive to both misspecification of the PS model and the degrees of overlap of the PS distributions. Its relative bias and standard error increase rapidly as the PS model becomes erroneous and as the impact of misspecification enlarges. On the contrary, the changes in relative bias and standard for OW, MW, and EW, were modest and usually less poignant compared to IPW.
	Two interesting facts might explain such a seemingly robustness of these methods. The estimand of OW, MW, and EW is not defined on the estimated but true PS. Therefore, even when the PS model is misspecified, the estimand is still well-defined. In fact, this estimand equals the average treatment effect whenever the treatment effect is constant. Besides, it is commonly known that PS misspecification usually matters most in the area of limited PS distributions overlap or when the propensity scores are near 0 or 1 \cite{li2018addressing,li2018balancing,li2018propensity,ju2019adaptive,kang2016practice,khan2010irregular}. Our simulation studies also demonstrated that the weighted average treatment effect difference is more likely to be affected by a PS model misspecification when treatment effect is heterogeneous, rather than homogeneous. Because overlap, matching, and entropy weighting smoothly down-weight the influence of observations at both end of PS spectrum and focus on regions of the distributions the (measured) covariates with the most overlap, their robustness is expected as it captures the treatment effect where it is more informative \cite{li2018addressing}.
	
	We also found that 
	by targeting the middle range, IPW trimming can considerably improve the performance of IPW. Our simulations seem to indicate that trimming at 0.10 is better than trimming at 0.05 or 0.15. However, as it was the case with our motivating example, it is difficult to optimally determine which cutoff $\alpha$ can provide the best bias-variance trade-off, despite the cutoff $\alpha=0.10$ recommended by Crump et al.\cite{crump2006moving, crump2009dealing} 	as a good rule-of-thumb.
	Depending on the data at hand, the sample size, the degree of PS distributions overlap, as well as the type of misspecification, trimming at 0.10 is not always better than trimming at 0.15 (or other higher cutpoint).  While increasing the trimming cutpoint seems to decrease the variance of the estimators, there is no control whatsoever on the amount of bias such a process introduce into the point estimate.
	
	In practice, various cutoffs $\alpha$ are still being used  without a clear rationale as to why (1) we need to choose a specific cuttoff and (2) we must rather discard observations outside of the interval $[\alpha, 1-\alpha]$ instead of outside of the $\alpha$-th and $(1-\alpha)$-th percentiles or vice versa 	\cite{kang2016practice,sturmer2014propensity,sturmer2010treatment,lee2011weight,cole2008constructing}.
	In reality, an efficient and optimal trimming algorithm requires a clear bias-variance trade-off and cannot be fixed in advance as it is likely to be data-driven: excessive trimming may improve the variance by removing participants with extreme weights, but introduce bias and not enough trimming may somehow help reduce bias, but not improve the variance significantly (or vice-versa).   Unfortunately, such a data-adaptive nature of IPW trimming is also its main drawback as it makes the related conclusions from IPW trimming subjective susceptible to the arbitrary choice of cutoffs and mode of trimming (i.e., interval range vs. percentiles), which result in unclear and unstable causal effect estimation.
	
	In light of these results, for our motivating example on maternal smoking and its effect on infants' birth weight, we can confidently rely on the results obtain with OW, MW, and EW if we factor in  the large number of participants with a propensity score near 0
	and extremely large proportions of participants who would be excluded had we decide to trim at 0.05 (50\% will be excluded), at 0.10 (with 82.6\% excluded) or at 0.15 (with 91.8\% excluded). In this example, using OW (which has the smallest standard error), we will conclude that, in our data set, infants born to mothers who smoked while pregnant weigh, on average, 132.77 grams less than  infants born to mothers who did not smoke.
	
	As illustrated in our motivating example, in practice, only a PS model that passes overlap and balance checks will be ultimately used to evaluate  causal effects. Besides, it has become ubiquitous that the choice of variables to include in a PS model needs to be examined carefully. It is recommended that we should only include true confounders, variables that are directly related to the treatment assignment and the outcome of interest, as well as variables are  predictors of the outcome (but not necessarily of the treatment). Variables that are mediators of the treatment--outcome relationship, only predictors of the treatment,  post-treatment, or time-varying, or treatment dependent must not be included \cite{brookhart2006variable, brookhart2013propensity, wooldridge2002inverse,sturmer2014propensity,westreich2011role}. As a result, glaring and severe problems of misspecification are less likely to appear in real-world data analysis. Nevertheless, if unmeasured confounding is an issue, a sensitivity analysis is highly recommended.

	In general, overlap of the PS distributions, balance of the covariates, and inference can be improved by adding in the PS model higher order and interaction terms \cite{rubin2004principles,rosenbaum1983central,austin2011introduction}. Furthermore, more flexible models or approaches, including generalized additive model and random forest, may be leveraged to efficiently handle with higher order  and interactions terms in a model and provide valuable information for PS estimation. If after these preemptive measures, we are still confronted with the limited overlap or the presence of observations with PSs near 0 or 1 that may jeopardize the validity of our IPW estimation, our simulation studies clearly highlight situations where alternative methods such as OW, MW, and EW may be advantageous. In these situations, we recommend using the inverse weighting methods that target the population of patients for whom there is a clinical equipoise, i.e., OW, MW, or EW.
	
	In conclusion, one of the key questions that motivated this paper was whether IPW is better than OW at reducing relative bias while improving efficiency. 	 Overall, our simulation studies indicate that OW, MW, and EW outperformed IPW, with OW providing a most efficient estimate.

		\bibliographystyle{acm}
		\bibliography{propensity_misspecification_arxiv}
		
		\vspace*{.5cm}
		\appendix{\sf{\bf \Large Appendix}}
		\newcounter{Appendix}[section]
		\numberwithin{equation}{section}
		\renewcommand\theequation{\Alph{section}.\arabic{equation}}
		\numberwithin{table}{section}
		\numberwithin{figure}{section}
		%

		\section{Sandwich variance for the estimators}\label{sec:sandwich_var}
		\subsection{Sandwich variance for the balancing-weight estimators}\label{sec:balancedwgt_est_var}
		The large-sample variance estimator of $\widehat \Delta_h$ can be derived using M-theory. In this paper, we determine the empirical sandwich variance estimator,  assuming regularity conditions hold \cite{stefanski2002calculus,lunceford2004stratification}.
		
		Given the function  $h(\boldsymbol  x)$, the estimator $\widehat \Delta_h$ can be written as  $\widehat \Delta_h=\widehat \Delta_{1h}-\widehat \Delta_{0h}$, where  $\left( \widehat \Delta_{1h}, \widehat \Delta_{0h}\right)$ are the solutions to the estimating equations $$\displaystyle\sum_{i=1}^{N}\displaystyle \begin{pmatrix}
		\displaystyle \frac{Z_ih(\boldsymbol  X_i)}{e(\boldsymbol  X_i)}(Y_i-\Delta_{1h}), ~
		\displaystyle \frac{(1-Z_i)h(\boldsymbol  X_i)}{1-e(\boldsymbol  X_i)}(Y_i-\Delta_{0h}))
		\end{pmatrix}'=0, $$ with respect to $(\Delta_{1h}, \Delta_{0h})$, along with the estimating equation  for the logistic regression  parameters $\beta$, 
		\begin{equation} \label{eq:logistic.esteq}
		\displaystyle\psi_{{\beta}}(\beta, \boldsymbol{X})=\displaystyle\sum_{i=1}^{N}[Z_i-e({X_i})]\boldsymbol X_i=0
		\end{equation} where $e(\boldsymbol {X})=P(Z=1|\boldsymbol {X})=\left\{ 1+\exp(-{\boldsymbol  X}{\beta})\right\}^{-1}$.
		
		For the treatment group, i.e., when $Z=1$, we have  $$\displaystyle U(\widehat \Delta_{1h},\hat{e}(\boldsymbol X_i)) =\sum_{i=1}^{N}\frac{Z_i\widehat  h(\boldsymbol  X_i)}{\hat{e}(\boldsymbol  X_i)}(Y_i-\widehat \Delta_{1h})=0$$ via the first  equation. 
		We consider the Taylor's expansions, 
		\begin{flalign*}
		&\begin{cases}
		0=\displaystyle U(\widehat \Delta_{1h},\hat{e}(\boldsymbol  X_i))=U(\Delta_{1h},e(\boldsymbol  X_i))+\frac{\partial U( \Delta_{1h},  e(\boldsymbol  X_i))}{\partial {
				\Delta_{1h}}}(\widehat \Delta_{1h}-\Delta_{1h}) +  \frac{\partial U(\Delta_{1h}, e(\boldsymbol  X_i))}{\partial {
				\beta'}} (\hat{\beta}-\beta)+o_p(1)\\
		0=\displaystyle \psi_{{\beta}} (\boldsymbol  X_i, \hat{\beta})= \displaystyle \psi_{{\beta}} (\boldsymbol  X_i,{\beta}) +  \frac{\partial \displaystyle \psi_{{\beta}} (\boldsymbol  X_i, \hat{\beta})}{\partial {
				\beta'}} (\hat{\beta}-\beta)+o_p(1).
		\end{cases}\\
		&\Rightarrow  \displaystyle\begin{cases}
		0=\displaystyle\sum_{i=1}^{N}\frac{Z_ih(\boldsymbol  X_i)}{e(\boldsymbol  X_i)}(Y_i-\Delta_{1h})-\sum_{i=1}^{N}\frac{Z_ih(\boldsymbol  X_i)}{e(\boldsymbol  X_i)}(\hat\Delta_{1h}-\Delta_{1h})+\displaystyle  U_{\beta} (\hat{\beta}-\beta)+o_p(1)\\
		0=\displaystyle\sum_{i=1}^{N}[Z_i-e(\boldsymbol {X_i})]\boldsymbol X_i - \displaystyle\sum_{i=1}^{N}\left[ e({\boldsymbol X_i})(1-e({\boldsymbol X_i}))\boldsymbol X_i \boldsymbol X_i'\right]   (\hat{\beta}-\beta)+o_p(1)
		\end{cases}\\
		&\text{with}~U_{\beta}=\displaystyle \frac{\partial U(\Delta_{1h}, e(\boldsymbol  X_i))}{\partial {
				\beta'}}  =\displaystyle \sum_{i=1}^{N}\displaystyle \frac{Z_i\displaystyle \left[ h_{\beta}(\boldsymbol  X_i)-(1-e(\boldsymbol  X_i))h(\boldsymbol  X_i)\boldsymbol  X_i'\right] }{e(\boldsymbol  X_i)}(Y_i-\Delta_{1h}) \\
		&	\text{and} ~~ \displaystyle h_{\beta}(\boldsymbol  X_i)=\frac{\partial h(\boldsymbol  X_i)}{\partial {
				\beta'}} \\ 
		&\Rightarrow  \displaystyle\begin{cases} (\hat\Delta_{1h} -\Delta_{1h})\displaystyle\frac{1}{N}\displaystyle\sum_{i=1}^{N}\frac{Z_ih(\boldsymbol  X_i)}{e(\boldsymbol  X_i)}= \displaystyle\frac{1}{N}\left[ \displaystyle \sum_{i=1}^{N}\frac{Z_ih(\boldsymbol  X_i)}{e(\boldsymbol  X_i)}(Y_i-\Delta_{1h})+U_{\beta} (\hat{\beta}-\beta)\right]  \\
		(\hat{\beta}-\beta)\displaystyle\frac{1}{N}\displaystyle\sum_{i=1}^{N}\left[ e({\boldsymbol X_i})(1-e({\boldsymbol X_i}))\boldsymbol X_i \boldsymbol X_i'\right]  =\displaystyle\frac{1}{N}
		\displaystyle\sum_{i=1}^{N}[Z_i-e(\boldsymbol{X_i})]\boldsymbol X_i 
		\end{cases}	\\
		&\Rightarrow  \displaystyle\begin{cases} 	\hat\Delta_{1h} -\Delta_{1h}= \displaystyle\frac{\hat E_{1h}^{-1}}{{N}}\left[ \displaystyle \sum_{i=1}^{n}\frac{Z_ih(\boldsymbol  X_i)}{e(\boldsymbol  X_i)}(Y_i-\Delta_{1h})+U_{\beta} (\hat{\beta}-\beta)\right]; ~ ~\hat E_{1h}= \displaystyle\frac{1}{N}\displaystyle\sum_{i=1}^{N}\frac{Z_ih(\boldsymbol  X_i)}{e(\boldsymbol  X_i)}
		\\
		\hat{\beta}-\beta  =  \displaystyle\frac{\hat E_{\beta\beta}^{-1}} {{N}}\displaystyle\sum_{i=1}^{N}[Z_i-e(\boldsymbol{X_i})]\boldsymbol X_i; ~~~ \hat E_{\beta\beta} = \displaystyle\frac{1}{N}\displaystyle\sum_{i=1}^{N}\left[ e({\boldsymbol X_i})(1-e({\boldsymbol X_i}))\boldsymbol X_i \boldsymbol X_i'\right]
		\end{cases}	
		\end{flalign*}
		
		$$\Rightarrow\displaystyle
		\hat\Delta_{1h} -\Delta_{1h}= \displaystyle\frac{\hat E_{1h}^{-1}}{N} \sum_{i=1}^{N} \left[ \frac{Z_ih(\boldsymbol  X_i)}{e(\boldsymbol  X_i)}(Y_i-\Delta_{1h}) +[Z_i-e(\boldsymbol{X_i})]\displaystyle\frac{1} {{N}} U_{\beta} \hat E_{\beta\beta}^{-1}\boldsymbol X_i\right]. $$
		Similarly, for $Z=0$, we have $\displaystyle V(\widehat \Delta_{0h},\hat{e}(\boldsymbol X_i)) =\sum_{i=1}^{N}\frac{(1-Z_i)\widehat  h(\boldsymbol  X_i)}{1-\hat{e}(\boldsymbol  X_i)}(Y_i-\widehat \Delta_{0h})=0$. We can show that
		\begin{align*}
		 \displaystyle
		\hat\Delta_{0h} -\Delta_{0h}&= \displaystyle\frac{\hat E_{0h}^{-1}}{N} \sum_{i=1}^{N} \left[ \frac{(1-Z_i)h(\boldsymbol  X_i)}{1-e(\boldsymbol  X_i)}(Y_i-\Delta_{0h})  +[Z_i-e(\boldsymbol{X_i})]\displaystyle\frac{1} {{N}} V_{\beta} \hat E_{\beta\beta}^{-1}\boldsymbol X_i\right];	
		\\
		\hat E_{0h}&= \displaystyle\frac{1}{N}\displaystyle\sum_{i=1}^{N}\frac{(1-Z_i)h(\boldsymbol  X_i)}{1-e(\boldsymbol  X_i)}	
		~~~\text{and}\\
		 V_{\beta} &= \displaystyle \frac{\partial V(\Delta_{0h}, e(\boldsymbol  X_i))}{\partial {
				\beta'}}  =\displaystyle \sum_{i=1}^{N}\displaystyle \frac{(1-Z_i)\left[ h_{\beta}(\boldsymbol  X_i)+e(\boldsymbol  X_i)h(\boldsymbol  X_i)\boldsymbol  X_i'\right] }{\left( 1-e(\boldsymbol  X_i)\right)}(Y_i-\Delta_{0h}).
		\end{align*}
		Thus,
		\allowdisplaybreaks\begin{align}\label{eq:delta_diff1}
		\displaystyle 
		\sqrt{N}(\hat\Delta_{h}-\Delta_{h})&=\sqrt{N}\left[ (\hat\Delta_{1h} -\Delta_{1h})-(\hat\Delta_{0h} -\Delta_{0h})\right] \\
		&= \displaystyle\frac{\hat E_{1h}^{-1}}{\sqrt{N}} \sum_{i=1}^{N} \left[ \frac{Z_ih(\boldsymbol  X_i)}{e(\boldsymbol  X_i)}(Y_i-\Delta_{1h})\right] -\displaystyle\frac{\hat E_{0h}^{-1}}{\sqrt{N}} \sum_{i=1}^{N} \left[ \frac{(1-Z_i)h(\boldsymbol  X_i)}{1-e(\boldsymbol  X_i)}(Y_i-\Delta_{0h}) \right]  \nonumber\\
		&+\displaystyle\frac{\hat E_{1h}^{-1}}{\sqrt{N}} \sum_{i=1}^{N} \left[ [Z_i-e(\boldsymbol{X_i})]\displaystyle\frac{1} {{N}} U_{\beta} \hat E_{\beta\beta}^{-1}\boldsymbol X_i\right]- \displaystyle\frac{\hat E_{0h}^{-1}}{N} \sum_{i=1}^{N} \left[ [Z_i-e(\boldsymbol{X_i})]\displaystyle\frac{1} {{N}} V_{\beta} \hat E_{\beta\beta}^{-1}\boldsymbol X_i\right] 	\nonumber
		\end{align}
		Note that  both $\hat E_{1h}= \displaystyle\frac{1}{N}\displaystyle\sum_{i=1}^{N}\frac{Z_ih(\boldsymbol  X_i)}{e(\boldsymbol  X_i)}$ and $\hat E_{0h}= \displaystyle\frac{1}{N}\displaystyle\sum_{i=1}^{N}\frac{(1-Z_i)h(\boldsymbol  X_i)}{1-e(\boldsymbol  X_i)}$ converge in probability to $ E\left[h(\boldsymbol  X)\right]$ , by the weak law of large numbers. 
		Indeed,  $$\hat E_{1h}\longrightarrow \displaystyle E\left[ \frac{Zh(\boldsymbol X)}{e(\boldsymbol  X)}\right] = \displaystyle E\left[ E\left[ \frac{Zh(\boldsymbol  X)}{e(\boldsymbol  X)}\Big|X\right]\right] =\displaystyle E\left[  \frac{h(\boldsymbol  X)}{e(\boldsymbol  X)} E\left[Z |X\right]\right] = E\left[h(\boldsymbol  X)\right].$$
		Likewise, $$\hat E_{0h}= \displaystyle\frac{1}{N}\displaystyle\sum_{i=1}^{N}\frac{(1-Z_i)h(\boldsymbol  X_i)}{1-e(\boldsymbol  X_i)}\longrightarrow E\left[h(\boldsymbol  X)\right]$$ in probability.	
		We can replace both $\hat E_{1h}$ and $\hat E_{0h}$ by $\hat E_{h}=\displaystyle\frac{1}{N}\displaystyle\sum_{i=1}^{N}h(\boldsymbol  X_i)$ in \eqref{eq:delta_diff1}, i.e., $\sqrt{N}(\hat\Delta_{h}-\Delta_{h})$ equals
		
		\begin{align*}
		\displaystyle 
		&\displaystyle\frac{\hat E_{h}^{-1}}{\sqrt{N}} \sum_{i=1}^{N} \left[ \frac{Z_ih(\boldsymbol  X_i)}{e(\boldsymbol  X_i)}(Y_i-\Delta_{1h})- \frac{(1-Z_i)h(\boldsymbol  X_i)}{1-e(\boldsymbol  X_i)}(Y_i-\Delta_{0h}) 
		- [Z_i-e(\boldsymbol{X_i})] \displaystyle\frac{1} {{N}} \left(  V_{\beta} - U_{\beta}\right)  \hat E_{\beta\beta}^{-1}\boldsymbol X_i	\right] 
		\end{align*}
		Therefore,  an approximate sampling  variance of $\hat\Delta_{h}$ can be computed as $ \displaystyle\frac{1}{N^2}  \sum_{i=1}^{N} \hat I_h^2 $ where
		$$
		\displaystyle 
		\hat I_h = \hat E_{h}^{-1}\left[ \frac{Z_i\hat h(\boldsymbol  X_i)}{\hat e(\boldsymbol  X_i)}(Y_i-\hat \Delta_{1h})- \frac{(1-Z_i)\hat h(\boldsymbol  X_i)}{1- \hat e(\boldsymbol  X_i)}(Y_i-\hat \Delta_{0h}) 
		- [Z_i- \hat e(\boldsymbol{X_i})] \hat H_{\beta} \hat E_{\beta\beta}^{-1}\boldsymbol X_i	\right]\!\!; ~\hat H_{\beta}= \displaystyle\frac{1} {{N}} \left( \hat  V_{\beta} - \hat U_{\beta}\right). 
		$$
		The normalized IPW is a special case of the balanced weights where the tilting function $h(\boldsymbol  x)= 1$.  The sampling  variance of $\hat\Delta_{\{h=1\}}$ is given by $ \displaystyle\frac{1}{N^2}  \sum_{i=1}^{N} \hat I_{\{h=1\}}^2 $ with,
		
		$$
		\displaystyle 
		\hat I_{\{h=1\}} =  \frac{Z_i}{\hat e(\boldsymbol  X_i)}(Y_i-\hat \Delta_{1\{h=1\}})- \frac{(1-Z_i)}{1- \hat e(\boldsymbol  X_i)}(Y_i-\hat \Delta_{0\{h=1\}}) 
		- [Z_i- \hat e(\boldsymbol{X_i})] \hat H_{\beta} \hat E_{\beta\beta}^{-1}\boldsymbol X_i	 
		$$
		where $\hat H_{\beta}= \displaystyle\frac{1} {{N}} (\hat  V_{\beta} - \hat U_{\beta}) = \displaystyle\frac{1}{N}\displaystyle \sum_{i=1}^{N}\left[ \displaystyle \frac{(1-Z_i)\hat   e(\boldsymbol  X_i)}{\left( 1- \hat  e(\boldsymbol  X_i)\right)}(Y_i-\hat  \Delta_{0\{h=1\}})+\displaystyle \frac{Z_i(1-\hat  e(\boldsymbol  X_i)) }{\hat  e(\boldsymbol  X_i)}(Y_i-\hat  \Delta_{1\{h=1\}}) \right] \boldsymbol  X_i'$
		
		\subsection{Sandwich variance for the stabilized IPW estimator}\label{sec:stabilizedIPW_var}
		The stabilized IPW estimator, based on  Hernan and Robin's definition \cite{hernan2006estimating}, is given by  
		\begin{align}
		\hat\Delta_{s} =  \frac{1}{N}\displaystyle\sum_{i=1}^{N}\left[ \displaystyle \frac{Z_iP(Z=1)Y_i}{\hat e(\boldsymbol  X_i)} -\displaystyle \frac{(1-Z_i)\left[ 1-P(Z=1)\right]Y_i }{1-\hat e(\boldsymbol  X_i)}\right]. 
		\end{align}
		Thus, the estimator $\hat\Delta_{s} $ can be written as $\hat\Delta_{s} = \hat\Delta_{1s}-\hat\Delta_{0s}$, where $\hat\Delta_{1s}= \displaystyle\frac{1}{N}\displaystyle\sum_{i=1}^{N} \displaystyle \frac{Z_iP(Z=1)Y_i}{\hat e(\boldsymbol  X_i)}$ and  $\hat\Delta_{0s}=\displaystyle \frac{(1-Z_i)\left[ 1-P(Z=1)\right]Y_i }{1-\hat e(\boldsymbol  X_i)}$ are solutions  to the (system of) estimating equations
		\allowdisplaybreaks $$\displaystyle\sum_{i=1}^{N}\displaystyle \begin{pmatrix}
		\displaystyle \frac{Z_iP(Z=1)}{e(\boldsymbol  X_i)}Y_i-\Delta_{1s}, ~
		\displaystyle \frac{(1-Z_i)\left[ 1-P(Z=1)\right] }{1-e(\boldsymbol  X_i)}Y_i-\Delta_{0s})
		\end{pmatrix}'=0, $$ with respect to $(\Delta_{1s}, \Delta_{0s})$, along with the logistic estimating equation   \eqref{eq:logistic.esteq}  of  $\beta$.%
		
		For  $Z=1$, the equation $U^s(\hat \Delta_{1s}, \hat e(\boldsymbol  X_i))=\displaystyle\sum_{i=1}^{N}
		\left[ \displaystyle \frac{Z_iP(Z=1)}{\hat e(\boldsymbol  X_i)}Y_i-\hat \Delta_{1s} \right] =0$ leads to 
		\begin{flalign*}
		&
		0=\displaystyle\sum_{i=1}^{N}\left[ \frac{Z_iP(Z=1)}{e(\boldsymbol  X_i)}Y_i-\Delta_{1s}\right] -N(\hat\Delta_{1s}-\Delta_{1s})+\displaystyle  U_{\beta}^s (\hat{\beta}-\beta)+o_p(1)\\
		&\text{via a Taylor's expansion, with} ~ U_{\beta}^s=\displaystyle \frac{\partial U^s(\Delta_{1s}, e(\boldsymbol  X_i))}{\partial {
				\beta'}}  =-\displaystyle \sum_{i=1}^{N}\displaystyle \frac{Z_iP(Z=1)\displaystyle (1-e(\boldsymbol  X_i))}{e(\boldsymbol  X_i)}Y_i\boldsymbol  X_i'.\\ 
		&\Rightarrow  \hat\Delta_{1s} -\Delta_{1s}= \displaystyle\frac{1}{N}\left[ \displaystyle \sum_{i=1}^{N}\left( \frac{Z_iP(Z=1)}{e(\boldsymbol  X_i)}Y_i-\Delta_{1s}\right) +U_{\beta}^s (\hat{\beta}-\beta)\right] \\
		&\Rightarrow\sqrt{N}(\hat\Delta_{1s} -\Delta_{1s})= \displaystyle\frac{1}{\sqrt{N}}\displaystyle \sum_{i=1}^{N}\left[ \left( \frac{Z_iP(Z=1)}{e(\boldsymbol  X_i)}Y_i-\Delta_{1s}\right) +[Z_i-e(\boldsymbol{X_i})]\displaystyle\frac{1} {{N}} U_{\beta}^s \hat E_{\beta\beta}^{-1}\boldsymbol X_i\right] \\
		&\text{since}~ \hat{\beta}-\beta  =  \displaystyle\frac{\hat E_{\beta\beta}^{-1}} {{N}}\displaystyle\sum_{i=1}^{N}[Z_i-e(\boldsymbol{X_i})]\boldsymbol X_i. 
		\end{flalign*}
		Similarly, for the control group where $Z=0$, the estimating equation $$ V^s(\hat \Delta_{0s}, \hat e(\boldsymbol  X_i))=\displaystyle\sum_{i=1}^{N}
		\left[ \displaystyle \frac{(1-Z_i)\left[ 1-P(Z=1)\right] }{1-\hat e(\boldsymbol  X_i)}Y_i-\hat \Delta_{0s} \right] =0$$ leads to  
		$$ \displaystyle\sqrt{N}(\hat\Delta_{0s} -\Delta_{0s})= \displaystyle\frac{1}{\sqrt{N}}\left[ \displaystyle \sum_{i=1}^{N}\left( \frac{(1-Z_i)\left[ 1-P(Z=1)\right] }{1-e(\boldsymbol  X_i)}Y_i-\Delta_{0s}\right) +[Z_i-e(\boldsymbol{X_i})]\displaystyle\frac{1} {{N}} V_{\beta}^s \hat E_{\beta\beta}^{-1}\boldsymbol X_i\right] $$ where $V_{\beta}^s=\displaystyle \frac{\partial V^s(\Delta_{0s}, e(\boldsymbol  X_i))}{\partial {
				\beta'}}  =\displaystyle \sum_{i=1}^{N}\displaystyle \frac{(1-Z_i)\left[ 1-P(Z=1)\right] \displaystyle e(\boldsymbol  X_i)}{(1-e(\boldsymbol  X_i))}Y_i\boldsymbol  X_i'.$
		\allowdisplaybreaks\begin{align*}
		&\Rightarrow\sqrt{N}(\hat\Delta_{s}-\Delta_{s})=\displaystyle\frac{1}{\sqrt{N}} \sum_{i=1}^{N} \left[ \frac{Z_iP(Z=1)}{e(\boldsymbol  X_i)}Y_i- \frac{(1-Z_i)\left[ 1-P(Z=1)\right] }{1-e(\boldsymbol  X_i)}Y_i-\Delta_{s} 
		-[Z_i-e(\boldsymbol{X_i})] H_\beta^s  \hat E_{\beta\beta}^{-1}\boldsymbol X_i\right]
		\end{align*}
		where  $\displaystyle H_\beta^s = \displaystyle\frac{1} {{N}} (V_{\beta}^s -U_{\beta}^s) =\displaystyle\frac{1} {{N}}\displaystyle \sum_{i=1}^{N}\left[ \displaystyle \frac{Z_iP(Z=1)\displaystyle (1-e(\boldsymbol  X_i))}{e(\boldsymbol  X_i)} +\displaystyle \frac{(1-Z_i)\left[ 1-P(Z=1)\right] \displaystyle e(\boldsymbol  X_i)}{(1-e(\boldsymbol  X_i))}\right] Y_i\boldsymbol  X_i'.$
		
		Therefore, the empirical sample variance of the estimator $\hat \Delta_{s}$ can be estimated by $ \displaystyle\frac{1}{N^2}  \sum_{i=1}^{N} \hat I_s^2 $ where
		$$
		\displaystyle 
		\hat I_s =  \frac{Z_i\hat P(Z=1)}{\hat e(\boldsymbol  X_i)}Y_i- \frac{(1-Z_i)\left[ 1-P(Z=1)\right] }{1- \hat e(\boldsymbol  X_i)}Y_i-\hat \Delta_{s} 
		- [Z_i- \hat e(\boldsymbol{X_i})] \hat H_{\beta}^s \hat E_{\beta\beta}^{-1}\boldsymbol X_i
		$$
		with $\displaystyle\hat  H_\beta^s  =\displaystyle\frac{1} {{N}}\displaystyle \sum_{i=1}^{N}\left[ \displaystyle \frac{Z_iP(Z=1)\displaystyle (1-\hat  e(\boldsymbol  X_i))}{\hat  e(\boldsymbol  X_i)} +\displaystyle \frac{(1-Z_i)\left[ 1-P(Z=1)\right] \displaystyle \hat  e(\boldsymbol  X_i)}{(1-\hat  e(\boldsymbol  X_i))}\right] Y_i\boldsymbol  X_i'.$

		\section{Asymptotic bias of a regular asymptotic linear (RAL) estimator}\label{sec:asymptotic_results}
		Consider the RAL $\widehat\theta$, solution to the estimating equation $U_n(\theta)=\displaystyle\sum_{i=1}^{N}U_{ij}(\boldsymbol X_i, Z_i, Y_i;\theta)=0,$ with respect to $\theta$. We use the Taylor's expansion
		$$
		0=\displaystyle U_{n}(\widehat\theta)=U_{n}(\theta_0) + \left[ \displaystyle \partial U_{n}(\theta)/\partial\theta\right]_{\theta_0} (\widehat\theta-\theta_0)+o_p(1),
		$$ to derive the asymptotic bias of the estimator $\widehat\theta$ of  $\theta_0$, 
		i.e.,  $$\widehat\theta-\theta_0=-\left[ \displaystyle \frac{1}{N}\displaystyle \partial U_{n}(\theta)/\partial\theta'\right]^{-1}_{\theta_0}\displaystyle \frac{1}{N}\left\lbrace \displaystyle U_{n}(\theta_0) \right\rbrace +o_p(1), $$ under some regularity conditions. Therefore,  the asymptotic bias of the estimator $\widehat\theta$ is approximately equal to
		\begin{align}
		-E\left[\partial U(\boldsymbol X, Z, Y;\theta_0)/\partial\theta'\right]^{-1}\displaystyle E\left\lbrace \displaystyle U(\boldsymbol X, Z, Y;\theta_0) \right\rbrace
		\end{align}
		\section{Simulation results,  Part I:\\ 
			Medium treatment prevalence and N = 2000.}\label{sec:boxplots}
		\subsection{Homogeneous treatment effect}\label{sec:homo200_results}
		\subsubsection{Variable omission and variable transformation}
\begin{figure}[]
			\begin{center}
				\includegraphics[trim=0 0 0 0, clip, width= 1\linewidth]{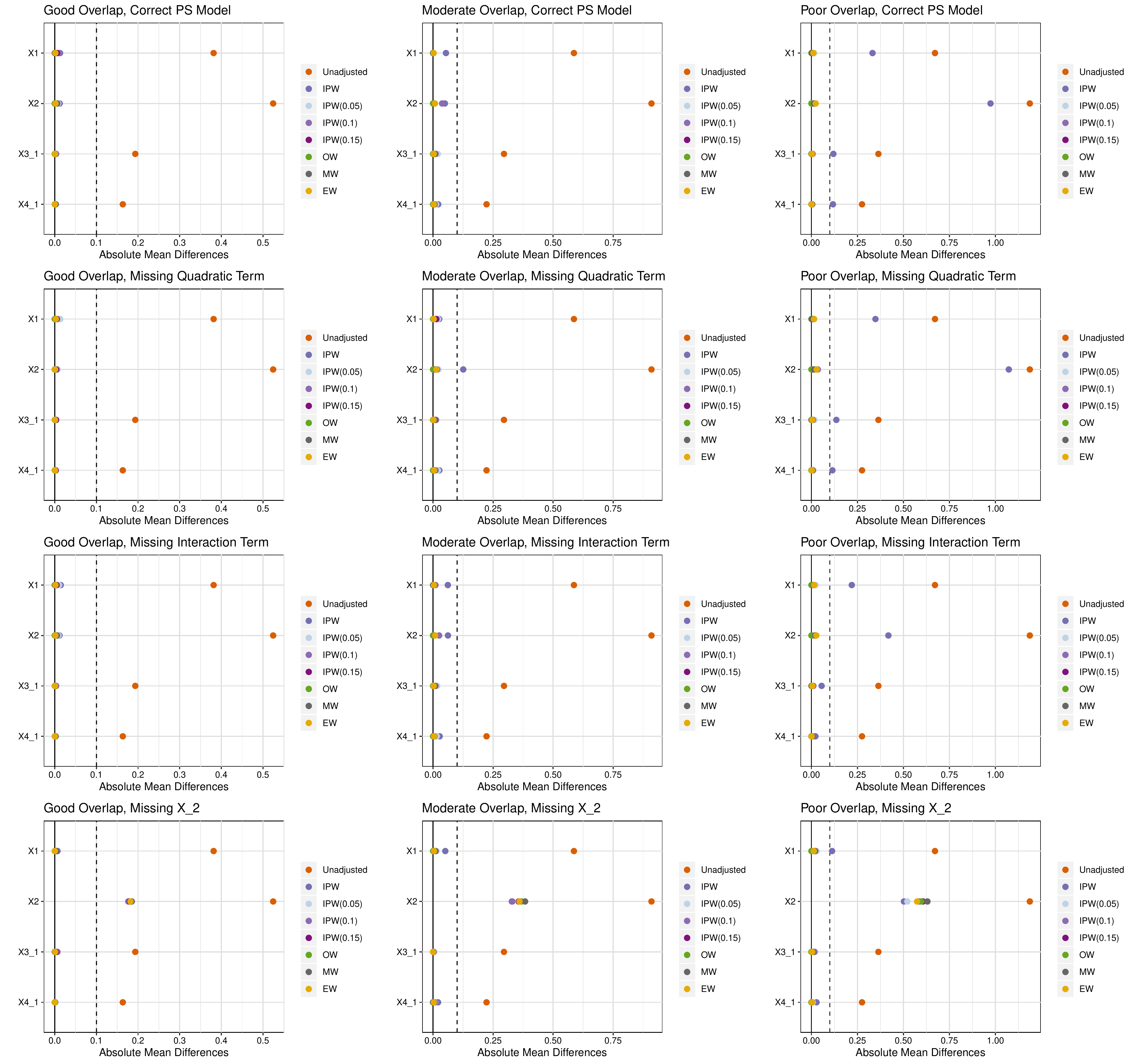}
				\caption{Variable Omission: Covariate balance with $N=2000$. \label{Balance_Mao} }
			\end{center}
			
		\end{figure}

\begin{figure}[]
			\begin{center}
				\includegraphics[trim=0 0 0 0, clip, width= 1\linewidth]{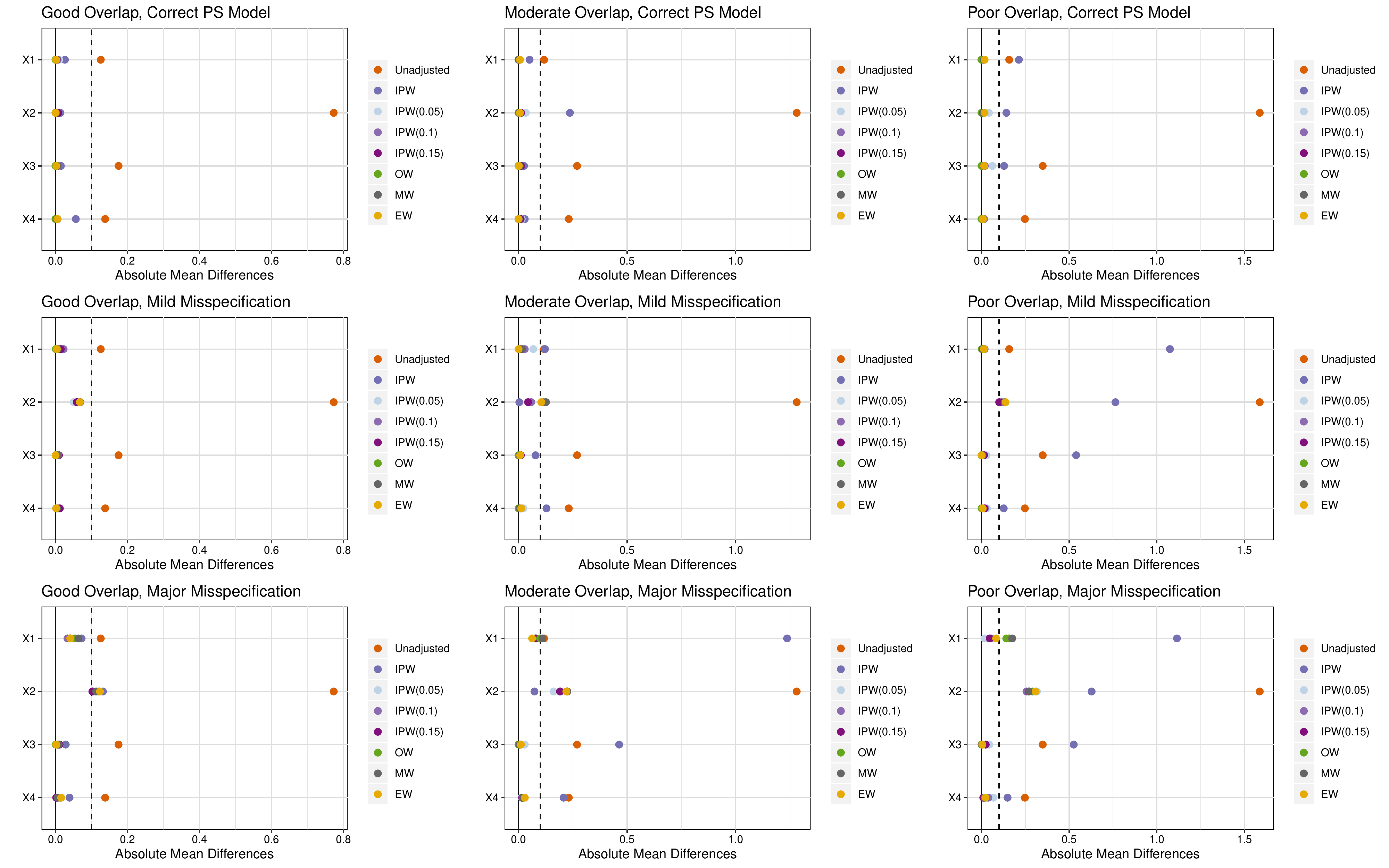}
				\caption{Variable Transformation: Covariate balance with $N=2000$. \label{Balance_KS} }
			\end{center}
			
		\end{figure}

		Figures \ref{Balance_Mao} and  \ref{Balance_KS} summarize the covariate balance induced by different weighting methods by different degrees of overlap and different  PS model misspecifications under variable omission and variable transformation simulation settings. 

		Both Tables  \ref{T1} and \ref{varomi} show the results under misspecified PS model using a constant treatment effect and a simple size of 2,000 observations.
		
\begin{figure*}[]
			\begin{center}
				\includegraphics[trim=5 20 10 5, clip, width=1\linewidth]{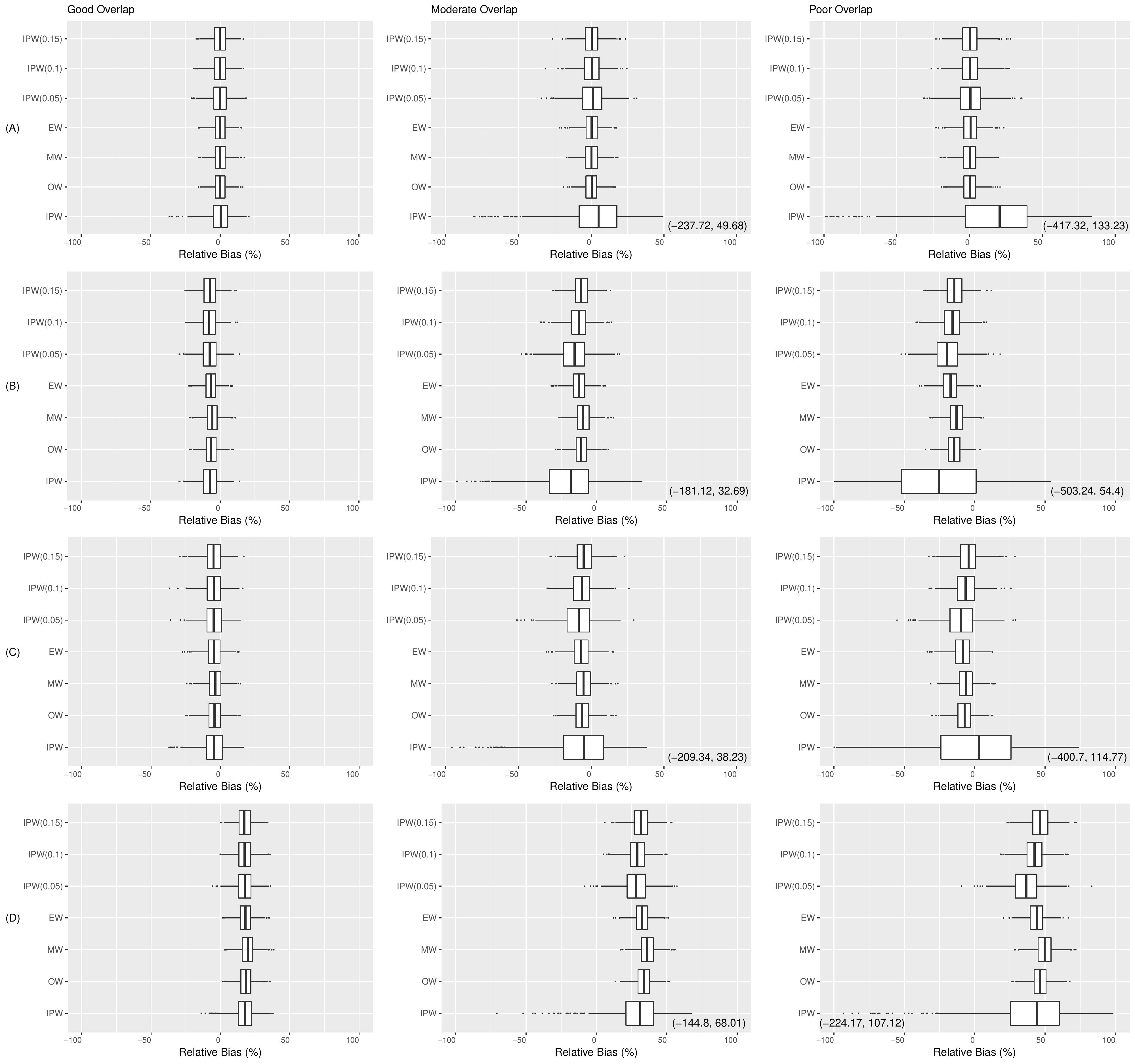}
			\end{center}
			\caption{Variable omission: Relative bias  (homogeneous treatment effect), with $N=2000$. \label{Boxplot_Mao} 
			}
			\subcaption*{{\bf Legend:} A: Correct PS Model; B: Missing $X_1^2$; C: Missing $X_2X_4$; D: Missing $X_2$.}
		\end{figure*}

\begin{figure*}[]
			\begin{center}
				\includegraphics[trim=5 20 10 5, clip, width=1\linewidth]{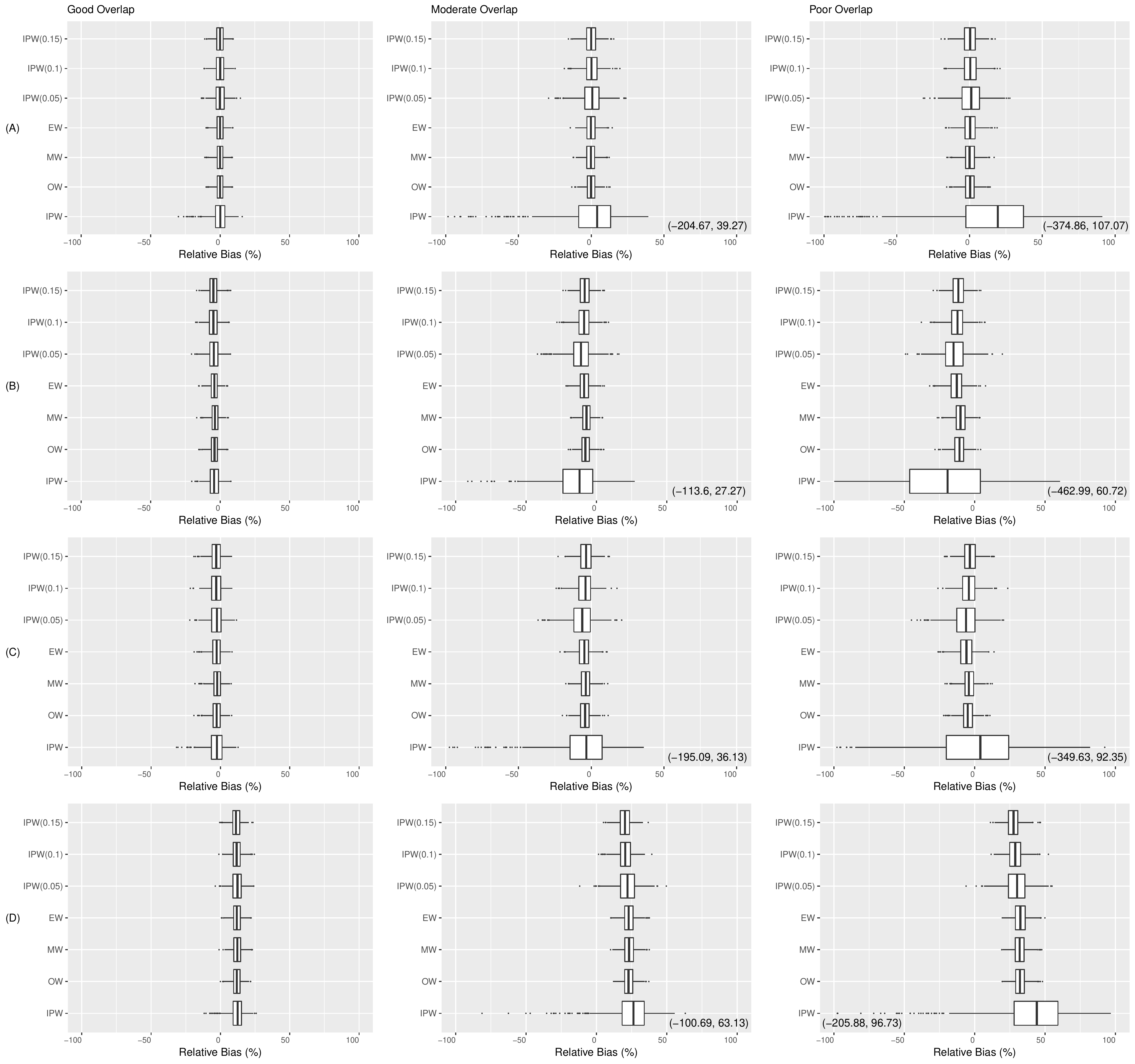}
			\end{center}
			\caption{Variable omission: Relative bias (heterogeneous treatment effect), with $N=2000$.\label{Boxplot_HE_Mao} 
			}
			\subcaption*{{\bf Legend:} A: Correct PS Model; B: Missing $X_1^2$; C: Missing $X_2X_4$; D: Missing $X_2$.}
		\end{figure*}

\begin{figure}[t]
			
			\begin{center}
				\includegraphics[trim=5 20 5 5, clip, width=0.8\linewidth]{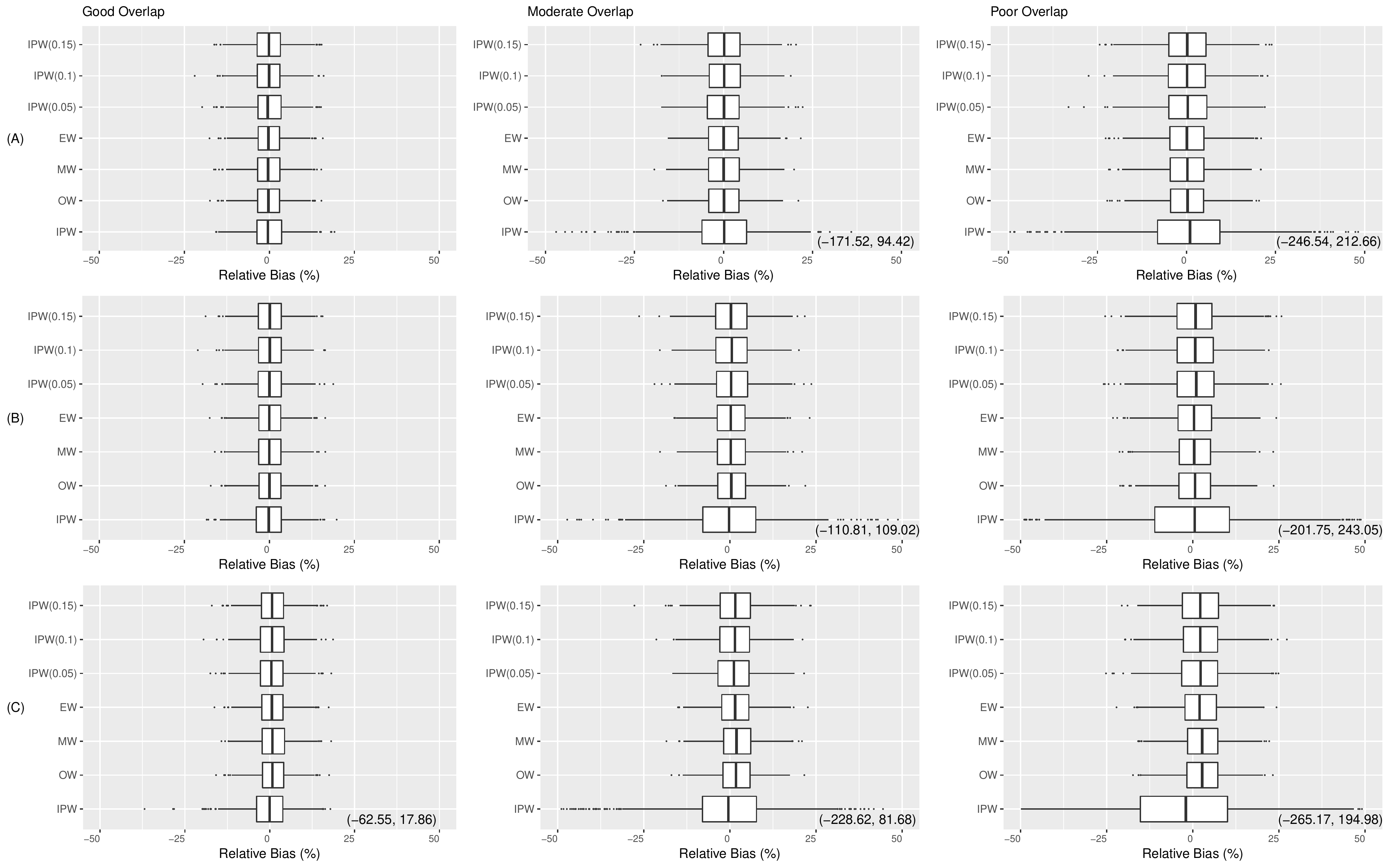}
				\caption{Variable Transformation: Relative bias (homogeneous treatment effect), with  $N=2000$. \label{Boxplot}}
			\end{center}
		\end{figure}

\begin{figure}[]
			\begin{center}
				\includegraphics[trim=5 20 5 5, clip, width= 0.8\linewidth]{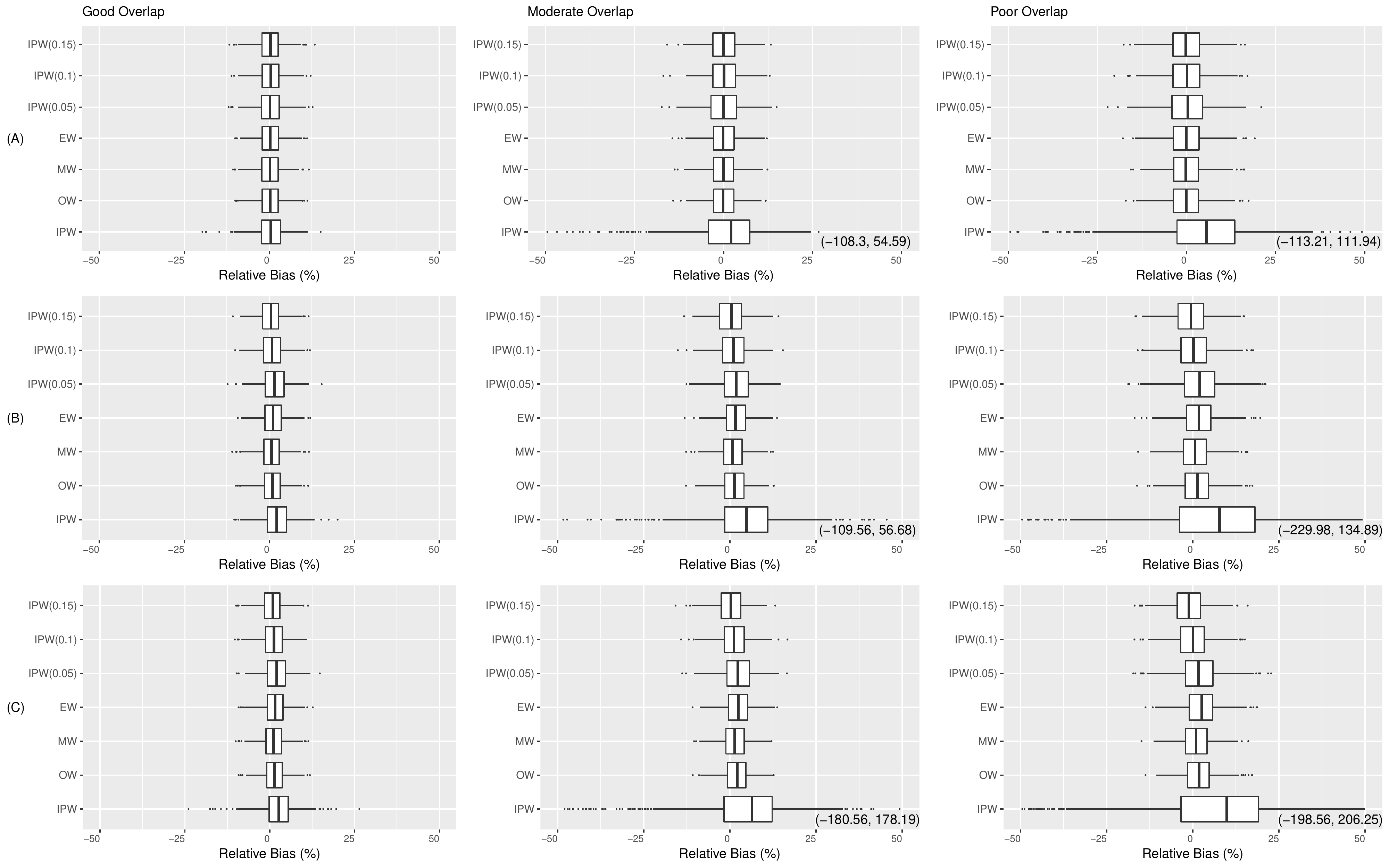}
				\caption{Variable Transformation: Relative bias (Heterogeneous treatment effect) , with  $N=2000$. \label{BoxplotHE} }
			\end{center}
			\subcaption*{{\bf Legend:} A: Correct PS Model; B: Mild PS Model Misspecification; C: Major PS Model Misspecification.}
			
		\end{figure}

		Figures \ref{Boxplot_Mao} and \ref{Boxplot_HE_Mao} show the boxplots of the relative bias percentages for the different PS weighting methods, under variable omissions.  Similarly, both  \ref{Boxplot} and \ref{BoxplotHE}, represent the boxplots of relative bias percentages, when true variables are transformed as specified in the main text. For better visualization, we trimmed some outlying IPW relative bias percentage estimates beyond 50\%. The intervals indicate the ranges of the IPW relative bias estimates in these situations where some of the outlying values did not fit on the graphs.
		
		
		\begin{table}[]
			\centering
			\begin{threeparttable}
				\centering\scriptsize\sf
				\caption{Variable transformation: homogeneous treatment effect ($N=2000$).\label{T1} { }}
				\begin{tabular}{p{1.1cm}p{0.6cm}p{0.3cm}p{0.8cm}p{0.4cm}p{0.3cm}p{0.3cm}p{0.35cm}  p{0.7cm}p{0.4cm}p{0.3cm}p{0.3cm}p{0.2cm}  p{0.7cm}p{0.4cm}p{0.3cm}p{0.3cm}p{0.2cm}}
					\toprule
					&  &  & \multicolumn{15}{c}{Propensity score misspecification} \\\cmidrule(lr){4-18}
					&  &  & \multicolumn{5}{c}{None }& \multicolumn{5}{c}{Mild} &   \multicolumn{5}{c}{Major}   \\ \cmidrule(lr){4-8}\cmidrule(lr){9-13}\cmidrule(lr){14-18}
					Weight & Overlap & True &Bias & RMSE & SD & SE & CP & Bias & RMSE & SD & SE  & CP & Bias & RMSE & SD & SE & CP\\
					\cmidrule(lr){1-18}				
					
					IPW 	  & Good & 1 & -0.12 & 5.52 & 5.52 & 5.61 & 95 & -0.16 & 5.46 & 5.46 & 5.53 &  95 & -0.41 & 6.70 & 6.69 & 5.99 &  96 \\
					IPW(0.05) & Good & 1 & -0.14 & 5.11 & 5.11 & 5.28 & 96 & 0.11 & 5.12 & 5.13 & 5.26 &  96 & 0.55 & 5.05 & 5.03 & 5.20 &  95 \\
					IPW(0.1)  & Good & 1 & -0.17 & 5.07 & 5.07 & 5.20 & 95 & 0.23 & 5.10 & 5.10 & 5.17 &  95 & 0.72 & 5.04 & 4.99 & 5.13 &  95 \\
					IPW(0.15) & Good & 1 & -0.15 & 5.07 & 5.07 & 5.22 & 95 & 0.24 & 5.01 & 5.01 & 5.19 &  96 & 0.77 & 5.03 & 4.98 & 5.14 &  95 \\
					OW		  & Good & 1 & -0.19 & 4.87 & 4.87 & 5.01 & 96 & 0.13 & 4.83 & 4.83 & 4.98 &  96 & 0.89 & 4.84 & 4.76 & 4.94 &  95 \\
					MW		  & Good & 1 & -0.20 & 4.93 & 4.93 & 5.08 & 96 & 0.07 & 4.88 & 4.88 & 5.04 &  96 & 0.98 & 4.92 & 4.83 & 5.01 &  95 \\
					EW 		  & Good & 1 & -0.18 & 4.87 & 4.87 & 5.02 & 96 & 0.12 & 4.84 & 4.84 & 4.99 &  96 & 0.79 & 4.84 & 4.78 & 4.96 &  95 \\ \addlinespace
					
					IPW & Mod & 1 & 0.07 & 14.45 & 14.46 & 9.78 &  93 & -0.47 & 15.27 & 15.27 & 11.96 &  95 & -4.69 & 27.60 & 27.21 & 18.29 &  96 \\
					IPW(0.05) & Mod & 1 & 0.10 & 6.62 & 6.63 & 6.54 &  94 & 0.57 & 6.81 & 6.79 & 6.63 &  94 & 1.13 & 6.68 & 6.59 & 6.46 &  94 \\
					IPW(0.1)  & Mod & 1 & 0.24 & 6.40 & 6.40 & 6.38 &  95 & 0.52 & 6.50 & 6.48 & 6.39 &  95 & 1.47 & 6.50 & 6.33 & 6.23 &  94 \\
					IPW(0.15) & Mod & 1 & 0.11 & 6.60 & 6.61 & 6.53 & 95 & 0.42 & 6.67 & 6.66 & 6.52 &  95 & 1.54 & 6.57 & 6.39 & 6.33 &  95 \\
					OW        & Mod & 1 & 0.09 & 5.92 & 5.92 & 5.88 &  95 & 0.45 & 5.88 & 5.86 & 5.80 &  95 & 1.89 & 6.02 & 5.72 & 5.67 &  94 \\
					MW 		  & Mod & 1 & 0.09 & 5.98 & 5.98 & 6.00 &  95 & 0.35 & 5.91 & 5.90 & 5.92 &  96 & 2.08 & 6.13 & 5.77 & 5.78 &  94 \\
					EW 		  & Mod & 1 & 0.08 & 6.00 & 6.00 & 5.93 &  95 & 0.44 & 5.98 & 5.96 & 5.87 &  95 & 1.63 & 6.03 & 5.81 & 5.74 &  94 \\ \addlinespace
					
					IPW       & Poor & 1 & 0.44 & 20.56& 20.56 & 12.55& 90 &-2.42 & 33.03 & 32.96 & 23.37 &  93 & -8.45 & 44.12 & 43.32 & 43.19 &  96 \\
					IPW(0.05) & Poor & 1 & 0.11 & 7.86 & 7.87 & 7.66 &  94 & 0.69 & 7.95 & 7.92 & 7.76 &  94 & 1.95 & 8.02 & 7.79 & 7.45 &  93 \\
					IPW(0.1)  & Poor & 1 & 0.12 & 7.73 & 7.74 & 7.51 &  94 & 0.65 & 7.76 & 7.73 & 7.53 &  94 & 2.00 & 7.70 & 7.44 & 7.17 &  93 \\
					IPW(0.15) & Poor & 1 & 0.10 & 8.06 & 8.07 & 7.77 &  95 & 0.54 & 8.00 & 7.98 & 7.72 &  94 & 2.15 & 7.88 & 7.58 & 7.32 &  93 \\
					OW        & Poor & 1 & 0.10 & 6.95 & 6.95 & 6.75 &  94 & 0.59 & 6.89 & 6.87 & 6.61 &  94 & 2.63 & 7.11 & 6.61 & 6.32 &  91 \\
					MW        & Poor & 1 & 0.11 & 7.06 & 7.06 & 6.89 &  94 & 0.51 & 7.00 & 6.99 & 6.76 &  94 & 2.84 & 7.30 & 6.72 & 6.47 &  91 \\
					EW        & Poor & 1 & 0.08 & 7.05 & 7.06 & 6.82 &  95 & 0.54 & 7.04 & 7.03 & 6.75 &  94 & 2.19 & 7.10 & 6.76 & 6.46 &  93 \\
					\bottomrule
				\end{tabular}
				\begin{tablenotes}
					\tiny
					\item  Mod: Moderate; IPW($\alpha$): trimmed IPW with $I_{\alpha}(x)=\mathds{1}({\{\alpha\leq e(x)\leq 1-\alpha\}})$, for   $\alpha=0.05, 0.10,$ and $0.15.$
					\item Bias: relative bias in percentage$\times 100$; RMSE: root mean-squared error$\times 100$; SD: empirical standard deviation$\times 100$; SE: average estimated standard error$\times 100$; CP: 95\%  coverage probability$\times 100$. The results are based on 1000 simulated data sets.
				\end{tablenotes}
			\end{threeparttable}
		\end{table}		
		
		\begin{table}[]
			\centering
			\begin{threeparttable}[]
				\centering\scriptsize\sf
				\caption{Variable omission: homogeneous treatment effect ($N=2000$).\label{varomi} { }}
				\begin{tabular}{p{1.2cm}p{0.7cm}p{0.5cm}  p{0.8cm}p{0.55cm}p{0.55cm}p{0.55cm}p{0.65cm}   p{0.7cm}p{0.55cm}p{0.75cm}p{0.55cm}p{0.55cm}}
					\toprule
					&  &  & \multicolumn{10}{c}{Propensity score misspecification} \\\cmidrule(lr){4-13}
					&  &  & \multicolumn{5}{c}{None }& \multicolumn{5}{c}{Missing $X_2$} \\ \cmidrule(lr){4-8}\cmidrule(lr){9-13}
					Weight & Overlap & True & Bias & RMSE & SD & SE & CP & Bias & RMSE & SD & SE & CP \\
					\cmidrule(lr){1-13}	
					
					IPW 		& Good & 1 & -0.18 & 8.03 & 8.04 & 7.20 &  93 & 17.35 & 18.85 & 7.36 & 6.81 &  27 \\
					IPW(0.05) 	& Good & 1 & -0.14 & 6.61 & 6.61 & 6.40 &  94 & 17.37 & 18.56 & 6.53 & 6.39 &  23 \\
					IPW(0.1) 	& Good & 1 & -0.19 & 5.93 & 5.93 & 5.79 &  95 & 17.38 & 18.42 & 6.09 & 6.04 &  19 \\
					IPW(0.15) 	& Good & 1 & -0.17 & 5.63 & 5.63 & 5.47 &  95 & 17.23 & 18.18 & 5.80 & 5.83 &  16 \\
					OW 			& Good & 1 & -0.04 & 4.91 & 4.91 & 4.81 &  95 & 18.31 & 19.09 & 5.38 & 5.37 &  8.0 \\
					MW 			& Good & 1 & 0.08  & 5.11 & 5.11 & 5.06 &  95 & 19.44 & 20.24 & 5.63 & 5.65 &  7.0 \\
					EW 			& Good & 1 & -0.08 & 5.03 & 5.04 & 4.88 &  95 & 18.06 & 18.86 & 5.44 & 5.40 &  8.0 \\ \addlinespace
					
					IPW 		& Mod & 1 & 0.55 & 28.01& 28.02& 19.59 &  86 & 29.02 & 34.24 & 18.20 & 14.62 &  37 \\
					IPW(0.05) 	& Mod & 1 & 0.65 & 10.09& 10.08& 10.12 &  95 & 28.05 & 29.74 & 9.88 & 9.96 &  22 \\
					IPW(0.1) 	& Mod & 1 & 0.36 & 7.24 & 7.23 & 7.49 &  96 & 28.95 & 29.84 & 7.23 & 7.59 &  4.0 \\
					IPW(0.15) 	& Mod & 1 & 0.34 & 6.52 & 6.52 & 6.62 &  96 & 31.53 & 32.27 & 6.84 & 6.83 &  00 \\
					OW 			& Mod & 1 & 0.24 & 5.47 & 5.47 & 5.48 &  95 & 33.62 & 34.13 & 5.90 & 5.94 &  00 \\
					MW 			& Mod & 1 & 0.14 & 5.89 & 5.89 & 5.84 &  96 & 36.19 & 36.74 & 6.35 & 6.34 &  00 \\
					EW 			& Mod & 1 & 0.24 & 5.77 & 5.77 & 5.80 &  95 & 32.31 & 32.87 & 6.05 & 6.10 &  00 \\ \addlinespace
					
					IPW & Poor & 1 & 10.11 & 49.47 & 48.45 & 33.16 &  76 & 39.74 & 50.43 & 31.06 & 25.53 &  48 \\
					IPW(0.05) & Poor & 1 & 0.54 & 10.75 & 10.74 & 11.12 &  95 & 36.40 & 38.00 & 10.91 & 10.91 &  11 \\
					IPW(0.1) & Poor & 1 & 0.36 & 7.91 & 7.91 & 8.18 &  96 & 42.27 & 43.01 & 7.92 & 8.11 &  00 \\
					IPW(0.15) & Poor & 1 & 0.18 & 7.55 & 7.56 & 7.72 &  96 & 46.53 & 47.17 & 7.77 & 7.70 &  00 \\
					OW & Poor & 1 & 0.14 & 6.07 & 6.07 & 6.25 &  95 & 46.39 & 46.83 & 6.35 & 6.53 &  00 \\
					MW & Poor & 1 & 0.11 & 6.50 & 6.50 & 6.65 &  96 & 49.69 & 50.15 & 6.76 & 6.96 &  00 \\
					EW & Poor & 1 & 0.27 & 6.56 & 6.56 & 6.83 &  96 & 43.89 & 44.38 & 6.56 & 6.88 &  00 \\  \addlinespace

					&  &  &  \multicolumn{5}{c}{Missing $X_1^2$} &   \multicolumn{5}{c}{Missing $X_2X_4$}  \\ \cmidrule(lr){4-8}\cmidrule(lr){9-13}
					Weight & Overlap & True & Bias & RMSE & SD & SE & CP & Bias & RMSE & SD & SE & CP\\
					\cmidrule(lr){1-13}

					IPW 		& Good & 1 & -7.49 & 10.14 & 6.84 & 6.58 &  80 & -4.54 & 9.76 & 8.64 & 7.75 &  91 \\
					IPW(0.05) 	& Good & 1 & -7.60 & 10.20 & 6.81 & 6.55 &  78 & -4.67 & 9.00 & 7.70 & 7.23 &  90 \\
					IPW(0.1) 	& Good & 1 & -7.88 & 10.17 & 6.43 & 6.22 &  75 & -4.82 & 8.72 & 7.27 & 6.83 &  88 \\
					IPW(0.15) 	& Good & 1 & -7.55 & 9.60 & 5.94 & 5.74 &  73 & -4.81 & 8.42 & 6.92 & 6.55 &  88 \\
					OW 			& Good & 1 & -6.45 & 8.18 & 5.04 & 4.94 &  74 & -4.33 & 7.45 & 6.07 & 5.75 &  87 \\
					MW 			& Good & 1 & -5.67 & 7.67 & 5.17 & 5.13 &  79 & -3.78 & 7.08 & 6.00 & 5.80 &  89 \\
					EW 			& Good & 1 & -6.75 & 8.49 & 5.16 & 5.02 &  73 & -4.45 & 7.65 & 6.22 & 5.85 &  87 \\ \addlinespace
					
					IPW 		& Mod & 1 & -20.89 & 30.11 & 21.69& 19.29&  89 & -8.09 & 25.91& 24.63 & 19.59 &  95 \\
					IPW(0.05) 	& Mod & 1 & -16.24 & 19.64 & 11.06& 10.97&  72 & -8.96 & 14.43& 11.31 & 11.56 &  92 \\
					IPW(0.1) 	& Mod & 1 & -12.61 & 14.69 & 7.55 & 7.66 &  64 & -6.64 & 10.64& 8.32 & 8.43 &  89 \\
					IPW(0.15) 	& Mod & 1 & -10.82 & 12.67 & 6.59 & 6.69 &  64 & -5.00 & 8.97 & 7.44 & 7.39 &  90 \\
					OW 			& Mod & 1 & -10.66 & 12.03 & 5.58 & 5.55 &  52 & -6.25 & 9.00 & 6.48 & 6.31 &  82 \\
					MW 			& Mod & 1 & -9.39  & 11.07 & 5.87 & 5.79 &  61 & -5.37 & 8.51 & 6.61 & 6.39 &  85 \\
					EW 			& Mod & 1 & -12.23 & 13.63 & 6.02 & 6.00 &  46 & -6.97 & 9.82 & 6.92 & 6.80 &  82 \\ \addlinespace
					
					IPW 		& Poor & 1 & -41.09 & 70.73 & 57.60 & 44.13 &  93 & -7.73 & 48.79 & 48.20 & 36.74 &  90 \\
					IPW(0.05) 	& Poor & 1 & -19.65 & 22.47 & 10.90 & 11.60 &  63 & -9.76 & 15.50 & 12.05 & 12.07 &  89 \\
					IPW(0.1) 	& Poor & 1 & -16.14 & 18.01 & 7.99 & 8.29 &  52 & -6.37 & 10.80 & 8.72 & 8.88 &  90 \\
					IPW(0.15) 	& Poor & 1 & -14.44 & 16.29 & 7.54 & 7.63 &  53 & -4.78 & 9.58 & 8.31 & 8.26 &  90 \\
					OW 			& Poor & 1 & -14.65 & 15.86 & 6.10 & 6.23 &  34 & -7.41 & 10.10 & 6.87 & 6.96 &  82 \\
					MW 			& Poor & 1 & -12.98 & 14.47 & 6.39 & 6.47 &  48 & -6.25 & 9.41 & 7.04 & 7.12 &  86 \\
					EW 			& Poor & 1 & -17.49 & 18.80 & 6.90 & 7.09 &  30 & -8.88 & 11.76 & 7.72 & 7.87 &  81 \\
					
					\bottomrule
				\end{tabular}
				\begin{tablenotes}
					\tiny
					\item  Mod: Moderate; IPW($\alpha$): trimmed IPW with $I_{\alpha}(x)=\mathds{1}({\{\alpha\leq e(x)\leq 1-\alpha\}})$, for   $\alpha=0.05, 0.10,$ and $0.15.$
					\item Bias: relative bias in percentage$\times 100$; RMSE: root mean-squared error$\times 100$; SD: empirical standard deviation$\times 100$; SE: average estimated standard error$\times 100$; CP: 95\%  coverage probability$\times 100$. The results are based on 1000 simulated data sets.
				\end{tablenotes}
			\end{threeparttable}
		\end{table}		
		
		
		
		
		\section{Simulation results,  Part II: \\Medium prevalence, N = 500 and N = 1000.}\label{sec:additional_tables}
		
		\subsection{Under variable omission}
		Tables \ref{varomi500} and \ref{varomi_HE500} (resp. Tables \ref{varomi1000} and \ref{varomi_HE1000}) provide the results under variable omission for $N=500$ (resp.  $N=1000$) for homogeneous as well as heterogeneous treatment effects.
		\begin{table}[]
			\centering
			\begin{threeparttable}[]
				\centering\scriptsize\sf
				\caption{Variable omission: homogeneous treatment effect ($N=500$).\label{varomi500} { }}
				\begin{tabular}{p{1cm}p{0.61cm}p{0.75cm}  p{0.8cm}p{0.8cm}p{0.8cm}p{0.7cm}p{0.7cm}   p{0.7cm}p{0.7cm}p{0.7cm}p{0.7cm}p{0.7cm}}
					\toprule
					&  &  & \multicolumn{10}{c}{Propensity score misspecification} \\\cmidrule(lr){4-13}
					&  &  & \multicolumn{5}{c}{None }& \multicolumn{5}{c}{Missing $X_2$} \\ \cmidrule(lr){4-8}\cmidrule(lr){9-13}
					Weight & Overlap & True & Bias & RMSE & SD & SE & CP & Bias & RMSE & SD & SE & CP \\
					\cmidrule(lr){1-13}	
					
					IPW & Good & 1 & 1.15 & 15.02 & 14.98 & 13.97 & 0.93 & 18.51 & 23.26 & 14.10 & 13.38 & 0.65 \\
					
					IPW(0.05) & Good & 1 & 0.58 & 13.02 & 13.01 & 12.95 & 0.94 & 18.20 & 22.40 & 13.07 & 12.88 & 0.67 \\
					
					IPW(0.1) & Good & 1 & 0.51 & 11.48 & 11.47 & 11.81 & 0.96 & 17.91 & 21.58 & 12.05 & 12.28 & 0.67 \\
					
					IPW(0.15) & Good & 1 & 0.26 & 10.66 & 10.66 & 11.12 & 0.95 & 17.59 & 21.00 & 11.48 & 11.80 & 0.67 \\
					
					OW & Good & 1 & 0.12 & 9.58 & 9.59 & 9.60 & 0.95 & 18.33 & 21.20 & 10.66 & 10.73 & 0.58 \\
					
					MW & Good & 1 & 0.06 & 10.14 & 10.15 & 10.30 & 0.95 & 19.23 & 22.28 & 11.24 & 11.43 & 0.61 \\
					
					EW & Good & 1 & 0.23 & 9.69 & 9.69 & 9.73 & 0.95 & 18.23 & 21.16 & 10.75 & 10.77 & 0.58 \\ \addlinespace

					IPW & Mod & 1 & 6.76 & 41.11 & 40.57 & 29.41 & 0.82 & 32.51 & 43.49 & 28.91 & 24.93 & 0.56 \\
					
					IPW(0.05) & Mod & 1 & 1.22 & 19.76 & 19.73 & 19.65 & 0.94 & 29.44 & 34.96 & 18.87 & 19.34 & 0.61 \\
					
					IPW(0.1) & Mod & 1 & 0.07 & 14.59 & 14.59 & 15.24 & 0.95 & 29.74 & 33.11 & 14.56 & 15.39 & 0.49 \\
					
					IPW(0.15) & Mod & 1 & -0.44 & 13.21 & 13.21 & 13.67 & 0.96 & 31.55 & 34.44 & 13.83 & 14.11 & 0.40 \\
					
					OW & Mod & 1 & -0.32 & 11.03 & 11.03 & 10.97 & 0.95 & 33.63 & 35.64 & 11.80 & 11.90 & 0.19 \\
					
					MW & Mod & 1 & -0.38 & 11.77 & 11.77 & 11.89 & 0.95 & 36.05 & 38.19 & 12.61 & 12.83 & 0.19 \\
					
					EW & Mod & 1 & -0.07 & 11.58 & 11.59 & 11.51 & 0.95 & 32.54 & 34.69 & 12.02 & 12.16 & 0.23 \\ \addlinespace

					IPW & Poor & 1 & 22.03 & 73.18 & 69.81 & 46.42 & 0.70 & 45.14 & 73.19 & 57.64 & 41.55 & 0.56 \\
					
					IPW(0.05) & Poor & 1 & 0.41 & 21.97 & 21.98 & 22.09 & 0.94 & 37.38 & 43.34 & 21.95 & 21.65 & 0.53 \\
					
					IPW(0.1) & Poor & 1 & 0.20 & 16.24 & 16.24 & 16.84 & 0.95 & 43.02 & 46.06 & 16.48 & 16.54 & 0.26 \\
					
					IPW(0.15) & Poor & 1 & 0.21 & 16.69 & 16.70 & 15.99 & 0.94 & 47.05 & 49.67 & 15.94 & 15.77 & 0.16 \\
					
					OW & Poor & 1 & -0.38 & 12.73 & 12.73 & 12.47 & 0.94 & 46.41 & 48.25 & 13.20 & 13.08 & 0.06 \\
					
					MW & Poor & 1 & -0.24 & 13.38 & 13.38 & 13.54 & 0.95 & 49.72 & 51.64 & 13.96 & 14.08 & 0.06 \\
					
					EW & Poor & 1 & -0.14 & 14.10 & 14.11 & 13.48 & 0.93 & 43.96 & 46.18 & 14.14 & 13.67 & 0.12 \\ \addlinespace

					&  &  &  \multicolumn{5}{c}{Missing $X_1^2$} &   \multicolumn{5}{c}{Missing $X_2X_4$}  \\ \cmidrule(lr){4-8}\cmidrule(lr){9-13}
					Weight & Overlap & True & Bias & RMSE & SD & SE & CP & Bias & RMSE & SD & SE & CP\\
					\cmidrule(lr){1-13}	
					
					IPW & Good & 1 & -6.79 & 14.91 & 13.28 & 13.55 & 0.94 & -3.24 & 16.02 & 15.70 & 15.10 & 0.94 \\
					
					IPW(0.05) & Good & 1 & -6.86 & 14.78 & 13.10 & 13.33 & 0.94 & -3.88 & 15.11 & 14.61 & 14.48 & 0.94 \\
					
					IPW(0.1) & Good & 1 & -7.05 & 13.97 & 12.06 & 12.49 & 0.93 & -4.27 & 14.18 & 13.53 & 13.72 & 0.94 \\
					
					IPW(0.15) & Good & 1 & -6.82 & 13.05 & 11.14 & 11.54 & 0.92 & -4.59 & 13.63 & 12.84 & 13.10 & 0.94 \\
					
					OW & Good & 1 & -6.23 & 11.55 & 9.73 & 9.85 & 0.91 & -4.27 & 12.30 & 11.54 & 11.49 & 0.94 \\
					
					MW & Good & 1 & -5.69 & 11.67 & 10.20 & 10.34 & 0.92 & -3.89 & 12.37 & 11.75 & 11.77 & 0.95 \\
					
					EW & Good & 1 & -6.42 & 11.72 & 9.82 & 10.03 & 0.90 & -4.22 & 12.41 & 11.68 & 11.65 & 0.93 \\ \addlinespace

					IPW & Mod & 1 & -17.49 & 46.50 & 43.11 & 33.13 & 0.94 & -3.73 & 40.73 & 40.58 & 31.99 & 0.89 \\
					
					IPW(0.05) & Mod & 1 & -15.26 & 26.15 & 21.25 & 21.23 & 0.94 & -9.29 & 24.33 & 22.50 & 22.54 & 0.95 \\
					
					IPW(0.1) & Mod & 1 & -13.08 & 19.87 & 14.97 & 15.41 & 0.90 & -7.44 & 18.61 & 17.07 & 17.02 & 0.93 \\
					
					IPW(0.15) & Mod & 1 & -11.66 & 18.00 & 13.72 & 13.67 & 0.86 & -6.07 & 16.33 & 15.17 & 15.02 & 0.93 \\
					
					OW & Mod & 1 & -11.22 & 15.81 & 11.14 & 11.10 & 0.81 & -6.97 & 14.42 & 12.63 & 12.61 & 0.90 \\
					
					MW & Mod & 1 & -9.93 & 15.32 & 11.68 & 11.71 & 0.86 & -6.01 & 14.09 & 12.75 & 12.97 & 0.93 \\
					
					EW & Mod & 1 & -12.53 & 17.38 & 12.04 & 11.88 & 0.82 & -7.51 & 15.55 & 13.62 & 13.50 & 0.90 \\ \addlinespace

					IPW & Poor & 1 & -24.21 & 91.54 & 88.32 & 59.87 & 0.87 & 3.33 & 78.12 & 78.09 & 54.09 & 0.80 \\
					
					IPW(0.05) & Poor & 1 & -18.70 & 29.62 & 22.98 & 22.88 & 0.92 & -10.51 & 26.07 & 23.87 & 23.84 & 0.94 \\
					
					IPW(0.1) & Poor & 1 & -15.16 & 22.54 & 16.68 & 16.70 & 0.87 & -6.13 & 18.75 & 17.73 & 18.00 & 0.95 \\
					
					IPW(0.15) & Poor & 1 & -14.37 & 21.10 & 15.46 & 15.45 & 0.85 & -4.61 & 17.74 & 17.14 & 16.81 & 0.93 \\
					
					OW & Poor & 1 & -14.92 & 19.52 & 12.59 & 12.42 & 0.77 & -7.74 & 15.97 & 13.98 & 13.90 & 0.90 \\
					
					MW & Poor & 1 & -13.45 & 18.66 & 12.94 & 13.05 & 0.83 & -6.44 & 15.48 & 14.09 & 14.44 & 0.93 \\
					
					EW & Poor & 1 & -17.27 & 22.57 & 14.54 & 13.82 & 0.76 & -9.08 & 18.45 & 16.07 & 15.52 & 0.91 \\

					\bottomrule
				\end{tabular}
				\begin{tablenotes}
					\tiny
					\item  Mod: Moderate; IPW($\alpha$): trimmed IPW with $I_{\alpha}(x)=\mathds{1}({\{\alpha\leq e(x)\leq 1-\alpha\}})$, for  $\alpha=0.05, 0.10,$ and $0.15.$
					\item Bias: relative bias in percentage; RMSE: root mean-squared error in $10^{-2}$; SD: empirical standard deviation in $10^{-2}$; SE: average estimated standard error in $10^{-2}$; CP: coverage probability of $95\%$ confidence interval. The results are based on 1000 simulated data sets.
				\end{tablenotes}
			\end{threeparttable}
		\end{table}		
		
		\begin{table}[t]
			\centering
			\begin{threeparttable}[]
				\centering\scriptsize\sf
				\caption{Variable omission: heterogeneous treatment effect ($N=500$).\label{varomi_HE500} { }}
				\begin{tabular}{p{1cm}p{0.61cm}p{0.75cm}  p{0.8cm}p{0.8cm}p{0.8cm}p{0.7cm}p{0.7cm}   p{0.7cm}p{0.7cm}p{0.7cm}p{0.7cm}p{0.7cm}}
					\toprule
					&  &  & \multicolumn{10}{c}{Propensity score misspecification} \\\cmidrule(lr){4-13}
					&  &  & \multicolumn{5}{c}{None }& \multicolumn{5}{c}{Missing $X_2$} \\ \cmidrule(lr){4-8}\cmidrule(lr){9-13}
					Weight & Overlap & True & Bias & RMSE & SD & SE & CP & Bias & RMSE & SD & SE & CP \\
					\cmidrule(lr){1-13}

					IPW & Good & 1.52 & 0.34 & 17.25 & 17.25 & 14.75 & 0.92 & 12.62 & 24.75 & 15.59 & 13.96 & 0.65 \\
					
					IPW(0.05) & Good & 1.53 & 0.13 & 14.22 & 14.23 & 13.40 & 0.94 & 12.38 & 23.46 & 13.91 & 13.28 & 0.68 \\
					
					IPW(0.1) & Good & 1.53 & 0.07 & 12.36 & 12.36 & 12.10 & 0.95 & 11.91 & 22.22 & 12.67 & 12.56 & 0.68 \\
					
					IPW(0.15) & Good & 1.54 & 0.03 & 11.55 & 11.55 & 11.25 & 0.95 & 11.40 & 21.34 & 12.08 & 11.99 & 0.68 \\
					
					OW & Good & 1.55 & -0.31 & 10.40 & 10.40 & 9.63 & 0.93 & 11.64 & 21.36 & 11.45 & 10.86 & 0.63 \\
					
					MW & Good & 1.56 & -0.46 & 10.98 & 10.97 & 10.28 & 0.93 & 11.69 & 21.94 & 12.12 & 11.53 & 0.66 \\
					
					EW & Good & 1.54 & -0.23 & 10.60 & 10.60 & 9.84 & 0.94 & 11.76 & 21.52 & 11.54 & 10.95 & 0.62 \\ \addlinespace

					IPW & Mod & 1.32 & 6.48 & 42.94 & 42.10 & 31.01 & 0.81 & 28.08 & 49.14 & 32.18 & 26.40 & 0.52 \\
					
					IPW(0.05) & Mod & 1.36 & 2.03 & 20.91 & 20.73 & 20.15 & 0.93 & 23.84 & 37.82 & 19.33 & 19.71 & 0.56 \\
					
					IPW(0.1) & Mod & 1.42 & 0.72 & 15.49 & 15.46 & 15.56 & 0.94 & 21.84 & 34.56 & 15.45 & 15.71 & 0.49 \\
					
					IPW(0.15) & Mod & 1.47 & 0.39 & 13.58 & 13.57 & 13.78 & 0.95 & 20.91 & 33.99 & 14.32 & 14.31 & 0.43 \\
					
					OW & Mod & 1.44 & 0.20 & 11.13 & 11.13 & 11.09 & 0.95 & 23.65 & 36.06 & 12.05 & 12.11 & 0.19 \\
					
					MW & Mod & 1.47 & -0.07 & 11.63 & 11.64 & 11.92 & 0.96 & 23.76 & 37.18 & 12.69 & 12.97 & 0.22 \\
					
					EW & Mod & 1.42 & 0.47 & 12.01 & 12.00 & 11.76 & 0.95 & 23.81 & 35.93 & 12.41 & 12.44 & 0.22 \\ \addlinespace

					IPW & Poor & 1.17 & 20.80 & 76.81 & 72.89 & 48.34 & 0.70 & 45.83 & 78.22 & 56.97 & 42.09 & 0.52 \\
					
					IPW(0.05) & Poor & 1.29 & 1.16 & 22.52 & 22.48 & 22.63 & 0.93 & 30.33 & 45.15 & 22.30 & 22.20 & 0.53 \\
					
					IPW(0.1) & Poor & 1.39 & 0.78 & 16.57 & 16.54 & 17.16 & 0.95 & 29.24 & 43.98 & 16.53 & 16.79 & 0.32 \\
					
					IPW(0.15) & Poor & 1.47 & 0.10 & 16.93 & 16.93 & 16.17 & 0.93 & 27.85 & 43.92 & 16.05 & 15.84 & 0.25 \\
					
					OW & Poor & 1.38 & -0.14 & 12.80 & 12.80 & 12.80 & 0.95 & 32.21 & 46.38 & 13.29 & 13.36 & 0.08 \\
					
					MW & Poor & 1.42 & -0.24 & 13.60 & 13.61 & 13.75 & 0.96 & 31.89 & 47.49 & 13.87 & 14.22 & 0.10 \\
					
					EW & Poor & 1.34 & 0.14 & 13.99 & 14.00 & 13.90 & 0.94 & 32.50 & 45.91 & 14.11 & 14.01 & 0.13 \\ \addlinespace

					&  &  &  \multicolumn{5}{c}{Missing $X_1^2$} &   \multicolumn{5}{c}{Missing $X_2X_4$}  \\ \cmidrule(lr){4-8}\cmidrule(lr){9-13}
					Weight & Overlap & True & Bias & RMSE & SD & SE & CP & Bias & RMSE & SD & SE & CP\\
					\cmidrule(lr){1-13}	
					
					IPW & Good & 1.52 & -4.17 & 15.95 & 14.64 & 13.85 & 0.94 & -2.31 & 17.69 & 17.35 & 15.56 & 0.95 \\
					
					IPW(0.05) & Good & 1.53 & -4.34 & 15.65 & 14.18 & 13.58 & 0.94 & -2.66 & 15.94 & 15.42 & 14.70 & 0.95 \\
					
					IPW(0.1) & Good & 1.53 & -4.66 & 14.71 & 12.87 & 12.62 & 0.92 & -2.97 & 14.64 & 13.92 & 13.76 & 0.94 \\
					
					IPW(0.15) & Good & 1.54 & -4.60 & 13.83 & 11.88 & 11.59 & 0.89 & -3.12 & 13.95 & 13.10 & 13.06 & 0.93 \\
					
					OW & Good & 1.55 & -4.27 & 12.45 & 10.55 & 9.83 & 0.87 & -3.06 & 12.94 & 12.05 & 11.35 & 0.92 \\
					
					MW & Good & 1.56 & -4.10 & 12.72 & 10.99 & 10.30 & 0.88 & -2.93 & 13.20 & 12.38 & 11.65 & 0.92 \\
					
					EW & Good & 1.54 & -4.31 & 12.62 & 10.73 & 10.04 & 0.87 & -3.01 & 13.07 & 12.22 & 11.55 & 0.92 \\ \addlinespace

					IPW & Mod & 1.32 & -10.08 & 44.15 & 42.12 & 33.87 & 0.93 & -1.04 & 42.46 & 42.46 & 33.26 & 0.90 \\
					
					IPW(0.05) & Mod & 1.36 & -9.77 & 26.43 & 22.84 & 21.90 & 0.93 & -5.54 & 24.40 & 23.21 & 22.82 & 0.94 \\
					
					IPW(0.1) & Mod & 1.42 & -8.41 & 19.64 & 15.63 & 15.70 & 0.90 & -3.98 & 18.20 & 17.32 & 17.08 & 0.94 \\
					
					IPW(0.15) & Mod & 1.47 & -7.96 & 18.28 & 14.02 & 13.75 & 0.86 & -3.41 & 16.30 & 15.51 & 15.02 & 0.93 \\
					
					OW & Mod & 1.44 & -7.34 & 15.31 & 11.10 & 11.16 & 0.84 & -4.18 & 13.91 & 12.55 & 12.55 & 0.92 \\
					
					MW & Mod & 1.47 & -6.80 & 15.13 & 11.35 & 11.69 & 0.87 & -3.68 & 13.69 & 12.58 & 12.88 & 0.94 \\
					
					EW & Mod & 1.42 & -8.04 & 16.75 & 12.29 & 12.03 & 0.84 & -4.50 & 15.14 & 13.74 & 13.51 & 0.92 \\ \addlinespace

					IPW & Poor & 1.17 & -16.75 & 92.84 & 90.79 & 61.58 & 0.85 & 5.54 & 80.00 & 79.78 & 55.34 & 0.80 \\
					
					IPW(0.05) & Poor & 1.29 & -13.61 & 29.07 & 23.13 & 23.42 & 0.92 & -6.78 & 25.96 & 24.44 & 24.34 & 0.94 \\
					
					IPW(0.1) & Poor & 1.39 & -11.75 & 23.72 & 17.16 & 17.08 & 0.85 & -4.06 & 19.46 & 18.63 & 18.28 & 0.94 \\
					
					IPW(0.15) & Poor & 1.47 & -11.40 & 23.19 & 16.07 & 15.59 & 0.80 & -3.27 & 17.98 & 17.34 & 16.87 & 0.93 \\
					
					OW & Poor & 1.38 & -11.26 & 20.02 & 12.65 & 12.68 & 0.76 & -5.24 & 15.73 & 13.99 & 14.10 & 0.91 \\
					
					MW & Poor & 1.42 & -10.56 & 19.92 & 13.08 & 13.20 & 0.79 & -4.45 & 15.62 & 14.28 & 14.58 & 0.93 \\
					
					EW & Poor & 1.34 & -12.73 & 22.32 & 14.33 & 14.23 & 0.78 & -6.19 & 17.89 & 15.85 & 15.76 & 0.92 \\

					\bottomrule
				\end{tabular}
				\begin{tablenotes}
					\tiny
					\item  Mod: Moderate; IPW($\alpha$): trimmed IPW with $I_{\alpha}(x)=\mathds{1}({\{\alpha\leq e(x)\leq 1-\alpha\}})$, for   $\alpha=0.05, 0.10,$ and $0.15.$
					\item Bias: relative bias in percentage; RMSE: root mean-squared error in $10^{-2}$; SD: empirical standard deviation in $10^{-2}$; SE: average estimated standard error in $10^{-2}$; CP: coverage probability of $95\%$ confidence interval. The results are based on 1000 simulated data sets.
				\end{tablenotes}
			\end{threeparttable}
		\end{table}

		\begin{table}[]
			\centering
			\begin{threeparttable}[]
				\centering\scriptsize\sf
				\caption{Variable omission: homogeneous treatment effect ($N=1000$).\label{varomi1000} { }}
				\begin{tabular}{p{1cm}p{0.61cm}p{0.75cm}  p{0.8cm}p{0.8cm}p{0.8cm}p{0.7cm}p{0.7cm}   p{0.7cm}p{0.7cm}p{0.7cm}p{0.7cm}p{0.7cm}}
					\toprule
					&  &  & \multicolumn{10}{c}{Propensity score misspecification} \\\cmidrule(lr){4-13}
					&  &  & \multicolumn{5}{c}{None }& \multicolumn{5}{c}{Missing $X_2$} \\ \cmidrule(lr){4-8}\cmidrule(lr){9-13}
					Weight & Overlap & True & Bias & RMSE & SD & SE & CP & Bias & RMSE & SD & SE & CP \\
					\cmidrule(lr){1-13}	
					IPW & Good & 1 & 0.68 & 10.39 & 10.38 & 9.79 & 0.93 & 17.48 & 20.02 & 9.75 & 9.51 & 0.50 \\
					
					IPW(0.05) & Good & 1 & 0.25 & 8.94 & 8.94 & 9.01 & 0.95 & 17.38 & 19.51 & 8.87 & 9.08 & 0.51 \\
					
					IPW(0.1) & Good & 1 & 0.09 & 8.23 & 8.24 & 8.23 & 0.95 & 17.30 & 19.31 & 8.59 & 8.58 & 0.47 \\
					
					IPW(0.15) & Good & 1 & 0.02 & 7.81 & 7.81 & 7.77 & 0.95 & 17.14 & 19.09 & 8.41 & 8.29 & 0.46 \\
					
					OW & Good & 1 & 0.18 & 6.84 & 6.84 & 6.79 & 0.96 & 18.18 & 19.72 & 7.67 & 7.59 & 0.34 \\
					
					MW & Good & 1 & 0.27 & 7.16 & 7.16 & 7.19 & 0.96 & 19.26 & 20.86 & 8.02 & 8.02 & 0.34 \\
					
					EW & Good & 1 & 0.19 & 6.96 & 6.96 & 6.89 & 0.95 & 17.95 & 19.53 & 7.70 & 7.62 & 0.36 \\ \addlinespace

					IPW & Mod & 1 & 1.33 & 35.07 & 35.06 & 24.03 & 0.85 & 28.71 & 39.25 & 26.77 & 19.29 & 0.50 \\
					
					IPW(0.05) & Mod & 1 & -0.38 & 14.42 & 14.42 & 14.31 & 0.94 & 26.93 & 30.59 & 14.52 & 14.14 & 0.51 \\
					
					IPW(0.1) & Mod & 1 & -0.13 & 10.23 & 10.24 & 10.70 & 0.96 & 28.55 & 30.52 & 10.79 & 10.85 & 0.25 \\
					
					IPW(0.15) & Mod & 1 & 0.05 & 9.55 & 9.55 & 9.49 & 0.96 & 31.24 & 32.74 & 9.82 & 9.80 & 0.12 \\
					
					OW & Mod & 1 & -0.27 & 7.82 & 7.82 & 7.76 & 0.95 & 33.23 & 34.30 & 8.48 & 8.41 & 0.03 \\
					
					MW & Mod & 1 & -0.23 & 8.32 & 8.32 & 8.30 & 0.94 & 35.86 & 37.00 & 9.11 & 9.00 & 0.02 \\
					
					EW & Mod & 1 & -0.40 & 8.36 & 8.36 & 8.20 & 0.95 & 31.80 & 32.98 & 8.74 & 8.63 & 0.05 \\ \addlinespace

					IPW & Poor & 1 & 13.31 & 58.97 & 57.48 & 39.46 & 0.73 & 40.73 & 59.27 & 43.07 & 32.33 & 0.54 \\
					
					IPW(0.05) & Poor & 1 & 0.70 & 16.88 & 16.88 & 15.70 & 0.92 & 36.62 & 39.96 & 16.00 & 15.38 & 0.35 \\
					
					IPW(0.1) & Poor & 1 & 0.64 & 12.13 & 12.12 & 11.70 & 0.94 & 42.39 & 44.00 & 11.80 & 11.60 & 0.06 \\
					
					IPW(0.15) & Poor & 1 & 0.65 & 11.41 & 11.40 & 11.03 & 0.93 & 46.50 & 47.83 & 11.20 & 10.96 & 0.01 \\
					
					OW & Poor & 1 & 0.30 & 9.15 & 9.15 & 8.82 & 0.94 & 46.38 & 47.34 & 9.51 & 9.22 & 0.00 \\
					
					MW & Poor & 1 & 0.34 & 9.55 & 9.54 & 9.44 & 0.95 & 49.71 & 50.71 & 10.03 & 9.86 & 0.00 \\
					
					EW & Poor & 1 & 0.37 & 10.11 & 10.11 & 9.60 & 0.93 & 43.84 & 44.99 & 10.12 & 9.69 & 0.01 \\ \addlinespace

					&  &  &  \multicolumn{5}{c}{Missing $X_1^2$} &   \multicolumn{5}{c}{Missing $X_2X_4$}  \\ \cmidrule(lr){4-8}\cmidrule(lr){9-13}
					Weight & Overlap & True & Bias & RMSE & SD & SE & CP & Bias & RMSE & SD & SE & CP\\
					\cmidrule(lr){1-13}			
					IPW & Good & 1 & -7.15 & 12.06 & 9.71 & 9.40 & 0.90 & -3.54 & 11.72 & 11.18 & 10.67 & 0.95 \\
					
					IPW(0.05) & Good & 1 & -7.20 & 11.93 & 9.52 & 9.29 & 0.89 & -4.07 & 11.03 & 10.26 & 10.20 & 0.93 \\
					
					IPW(0.1) & Good & 1 & -7.54 & 11.65 & 8.88 & 8.80 & 0.88 & -4.39 & 10.75 & 9.81 & 9.68 & 0.93 \\
					
					IPW(0.15) & Good & 1 & -7.22 & 10.95 & 8.24 & 8.13 & 0.85 & -4.35 & 10.34 & 9.38 & 9.26 & 0.92 \\
					
					OW & Good & 1 & -6.25 & 9.42 & 7.05 & 6.98 & 0.84 & -3.92 & 9.10 & 8.22 & 8.14 & 0.92 \\
					
					MW & Good & 1 & -5.51 & 9.15 & 7.31 & 7.27 & 0.88 & -3.42 & 8.89 & 8.21 & 8.24 & 0.93 \\
					
					EW & Good & 1 & -6.51 & 9.71 & 7.20 & 7.10 & 0.84 & -3.99 & 9.28 & 8.39 & 8.26 & 0.92 \\ \addlinespace

					IPW & Mod & 1 & -21.39 & 37.80 & 31.18 & 25.38 & 0.93 & -7.42 & 33.46 & 32.64 & 24.79 & 0.92 \\
					
					IPW(0.05) & Mod & 1 & -17.46 & 23.75 & 16.11 & 15.57 & 0.83 & -9.96 & 19.35 & 16.60 & 16.33 & 0.93 \\
					
					IPW(0.1) & Mod & 1 & -12.97 & 16.73 & 10.57 & 10.88 & 0.80 & -7.36 & 13.86 & 11.75 & 12.01 & 0.92 \\
					
					IPW(0.15) & Mod & 1 & -11.54 & 15.06 & 9.69 & 9.56 & 0.77 & -5.48 & 11.99 & 10.67 & 10.54 & 0.92 \\
					
					OW & Mod & 1 & -11.31 & 13.77 & 7.86 & 7.86 & 0.69 & -6.74 & 11.22 & 8.98 & 8.92 & 0.88 \\
					
					MW & Mod & 1 & -9.98 & 12.87 & 8.14 & 8.22 & 0.76 & -5.70 & 10.75 & 9.12 & 9.09 & 0.90 \\
					
					EW & Mod & 1 & -13.00 & 15.66 & 8.73 & 8.48 & 0.66 & -7.59 & 12.36 & 9.75 & 9.60 & 0.87 \\ \addlinespace

					IPW & Poor & 1 & -32.88 & 74.86 & 67.28 & 49.90 & 0.93 & -5.22 & 62.15 & 61.96 & 45.06 & 0.84 \\
					
					IPW(0.05) & Poor & 1 & -19.76 & 26.28 & 17.33 & 16.38 & 0.78 & -9.45 & 20.18 & 17.84 & 16.89 & 0.92 \\
					
					IPW(0.1) & Poor & 1 & -16.14 & 20.17 & 12.10 & 11.84 & 0.71 & -6.30 & 14.05 & 12.56 & 12.65 & 0.93 \\
					
					IPW(0.15) & Poor & 1 & -14.14 & 17.84 & 10.89 & 10.88 & 0.74 & -4.09 & 12.54 & 11.86 & 11.76 & 0.94 \\
					
					OW & Poor & 1 & -14.47 & 17.06 & 9.03 & 8.81 & 0.62 & -7.19 & 12.42 & 10.13 & 9.84 & 0.88 \\
					
					MW & Poor & 1 & -12.75 & 15.78 & 9.30 & 9.18 & 0.71 & -5.98 & 11.78 & 10.15 & 10.10 & 0.91 \\
					
					EW & Poor & 1 & -17.20 & 19.96 & 10.12 & 9.91 & 0.60 & -8.70 & 14.50 & 11.61 & 11.07 & 0.87 \\

					\bottomrule
				\end{tabular}
				\begin{tablenotes}
					\tiny
					\item  Mod: Moderate; IPW($\alpha$): trimmed IPW with $I_{\alpha}(x)=\mathds{1}({\{\alpha\leq e(x)\leq 1-\alpha\}})$, for   $\alpha=0.05, 0.10,$ and $0.15.$
					\item Bias: relative bias in percentage; RMSE: root mean-squared error in $10^{-2}$; SD: empirical standard deviation in $10^{-2}$; SE: average estimated standard error in $10^{-2}$; CP: coverage probability of $95\%$ confidence interval. The results are based on 1000 simulated data sets.
				\end{tablenotes}
			\end{threeparttable}
		\end{table}

		\begin{table}[]
			\centering
			\begin{threeparttable}[]
				\centering\scriptsize\sf
				\caption{Variable omission: heterogeneous treatment effect ($N=1000$).\label{varomi_HE1000} { }}
				\begin{tabular}{p{1cm}p{0.61cm}p{0.75cm}  p{0.8cm}p{0.8cm}p{0.8cm}p{0.7cm}p{0.7cm}   p{0.7cm}p{0.7cm}p{0.7cm}p{0.7cm}p{0.7cm}}
					\toprule
					&  &  & \multicolumn{10}{c}{Propensity score misspecification} \\\cmidrule(lr){4-13}
					&  &  & \multicolumn{5}{c}{None }& \multicolumn{5}{c}{Missing $X_2$} \\ \cmidrule(lr){4-8}\cmidrule(lr){9-13}
					Weight & Overlap & True & Bias & RMSE & SD & SE & CP & Bias & RMSE & SD & SE & CP \\
					\cmidrule(lr){1-13}				
					
					IPW & Good & 1.52 & -0.02 & 11.12 & 11.12 & 10.42 & 0.95 & 11.77 & 20.60 & 10.14 & 9.96 & 0.50 \\
					
					IPW(0.05) & Good & 1.53 & -0.24 & 9.17 & 9.17 & 9.42 & 0.96 & 11.65 & 19.92 & 8.98 & 9.39 & 0.52 \\
					
					IPW(0.1) & Good & 1.53 & -0.33 & 8.32 & 8.31 & 8.46 & 0.96 & 11.35 & 19.39 & 8.56 & 8.78 & 0.49 \\
					
					IPW(0.15) & Good & 1.54 & -0.33 & 7.88 & 7.87 & 7.88 & 0.94 & 10.88 & 18.66 & 8.15 & 8.43 & 0.49 \\
					
					OW & Good & 1.55 & -0.31 & 6.91 & 6.90 & 6.83 & 0.94 & 11.45 & 19.29 & 7.56 & 7.68 & 0.36 \\
					
					MW & Good & 1.56 & -0.31 & 7.30 & 7.29 & 7.19 & 0.95 & 11.73 & 20.01 & 7.98 & 8.08 & 0.39 \\
					
					EW & Good & 1.54 & -0.31 & 7.01 & 7.00 & 6.98 & 0.94 & 11.47 & 19.27 & 7.60 & 7.75 & 0.37 \\ \addlinespace

					IPW & Mod & 1.32 & 1.50 & 37.34 & 37.31 & 25.95 & 0.82 & 25.32 & 43.34 & 27.51 & 20.47 & 0.44 \\
					
					IPW(0.05) & Mod & 1.36 & 0.29 & 14.70 & 14.70 & 14.80 & 0.94 & 21.73 & 33.02 & 14.59 & 14.50 & 0.45 \\
					
					IPW(0.1) & Mod & 1.42 & 0.34 & 10.77 & 10.76 & 10.88 & 0.94 & 20.87 & 31.51 & 10.96 & 11.06 & 0.24 \\
					
					IPW(0.15) & Mod & 1.47 & 0.16 & 9.37 & 9.38 & 9.58 & 0.96 & 20.33 & 31.60 & 10.02 & 9.95 & 0.14 \\
					
					OW & Mod & 1.44 & -0.00 & 7.63 & 7.64 & 7.87 & 0.95 & 23.22 & 34.43 & 8.44 & 8.59 & 0.02 \\
					
					MW & Mod & 1.47 & -0.05 & 7.96 & 7.97 & 8.36 & 0.96 & 23.52 & 35.74 & 8.94 & 9.13 & 0.03 \\
					
					EW & Mod & 1.42 & -0.06 & 8.30 & 8.30 & 8.45 & 0.95 & 23.16 & 33.94 & 8.74 & 8.87 & 0.04 \\ \addlinespace

					IPW & Poor & 1.17 & 12.38 & 62.87 & 61.21 & 41.96 & 0.73 & 41.33 & 64.80 & 43.14 & 33.48 & 0.50 \\
					
					IPW(0.05) & Poor & 1.29 & 0.90 & 16.34 & 16.30 & 16.24 & 0.94 & 29.58 & 41.55 & 16.11 & 15.81 & 0.34 \\
					
					IPW(0.1) & Poor & 1.39 & 0.62 & 12.15 & 12.12 & 11.92 & 0.94 & 28.98 & 42.11 & 11.87 & 11.78 & 0.09 \\
					
					IPW(0.15) & Poor & 1.47 & 0.24 & 11.60 & 11.60 & 11.10 & 0.94 & 27.28 & 41.62 & 11.36 & 11.04 & 0.05 \\
					
					OW & Poor & 1.38 & 0.14 & 9.09 & 9.09 & 9.03 & 0.95 & 32.07 & 45.26 & 9.55 & 9.43 & 0.00 \\
					
					MW & Poor & 1.42 & -0.02 & 9.71 & 9.71 & 9.56 & 0.95 & 31.73 & 46.30 & 10.10 & 9.97 & 0.01 \\
					
					EW & Poor & 1.34 & 0.29 & 10.08 & 10.07 & 9.96 & 0.95 & 32.27 & 44.56 & 10.19 & 9.97 & 0.01 \\ \addlinespace

					&  &  &  \multicolumn{5}{c}{Missing $X_1^2$} &   \multicolumn{5}{c}{Missing $X_2X_4$}  \\ \cmidrule(lr){4-8}\cmidrule(lr){9-13}
					Weight & Overlap & True & Bias & RMSE & SD & SE & CP & Bias & RMSE & SD & SE & CP\\
					\cmidrule(lr){1-13}	
					
					IPW & Good & 1.52 & -4.48 & 12.06 & 9.95 & 9.69 & 0.91 & -2.60 & 12.27 & 11.62 & 11.06 & 0.94 \\
					
					IPW(0.05) & Good & 1.53 & -4.65 & 12.02 & 9.70 & 9.55 & 0.90 & -2.93 & 11.36 & 10.45 & 10.43 & 0.93 \\
					
					IPW(0.1) & Good & 1.53 & -5.13 & 11.82 & 8.83 & 8.97 & 0.87 & -3.21 & 11.03 & 9.87 & 9.76 & 0.92 \\
					
					IPW(0.15) & Good & 1.54 & -5.02 & 11.23 & 8.13 & 8.20 & 0.85 & -3.11 & 10.69 & 9.55 & 9.25 & 0.90 \\
					
					OW & Good & 1.55 & -4.33 & 9.69 & 7.00 & 6.98 & 0.83 & -2.89 & 9.52 & 8.41 & 8.06 & 0.90 \\
					
					MW & Good & 1.56 & -3.99 & 9.57 & 7.26 & 7.25 & 0.86 & -2.60 & 9.42 & 8.50 & 8.17 & 0.91 \\
					
					EW & Good & 1.54 & -4.43 & 9.91 & 7.17 & 7.14 & 0.83 & -2.92 & 9.65 & 8.54 & 8.22 & 0.90 \\ \addlinespace

					IPW & Mod & 1.32 & -13.82 & 36.12 & 31.16 & 26.39 & 0.94 & -4.38 & 34.45 & 33.98 & 26.14 & 0.91 \\
					
					IPW(0.05) & Mod & 1.36 & -11.65 & 22.79 & 16.34 & 16.01 & 0.87 & -6.49 & 18.77 & 16.56 & 16.60 & 0.94 \\
					
					IPW(0.1) & Mod & 1.42 & -8.64 & 16.38 & 10.91 & 11.11 & 0.83 & -4.46 & 13.55 & 11.99 & 12.05 & 0.93 \\
					
					IPW(0.15) & Mod & 1.47 & -8.27 & 15.68 & 9.86 & 9.61 & 0.76 & -3.61 & 12.00 & 10.75 & 10.56 & 0.92 \\
					
					OW & Mod & 1.44 & -7.65 & 13.49 & 7.81 & 7.93 & 0.72 & -4.27 & 10.66 & 8.71 & 8.90 & 0.91 \\
					
					MW & Mod & 1.47 & -6.95 & 12.99 & 8.02 & 8.23 & 0.76 & -3.59 & 10.21 & 8.74 & 9.05 & 0.92 \\
					
					EW & Mod & 1.42 & -8.68 & 15.00 & 8.60 & 8.65 & 0.71 & -4.85 & 11.76 & 9.55 & 9.65 & 0.90 \\ \addlinespace

					IPW & Poor & 1.17 & -25.50 & 79.12 & 73.32 & 52.09 & 0.92 & -2.26 & 63.14 & 63.11 & 46.82 & 0.83 \\
					
					IPW(0.05) & Poor & 1.29 & -14.61 & 25.65 & 17.34 & 16.84 & 0.82 & -6.50 & 19.08 & 17.13 & 17.33 & 0.94 \\
					
					IPW(0.1) & Poor & 1.39 & -12.49 & 21.26 & 12.20 & 11.99 & 0.70 & -4.19 & 13.97 & 12.70 & 12.74 & 0.92 \\
					
					IPW(0.15) & Poor & 1.47 & -11.59 & 20.36 & 11.19 & 10.94 & 0.66 & -3.28 & 13.06 & 12.15 & 11.80 & 0.93 \\
					
					OW & Poor & 1.38 & -11.08 & 17.78 & 9.08 & 8.97 & 0.61 & -5.01 & 12.02 & 9.83 & 9.94 & 0.90 \\
					
					MW & Poor & 1.42 & -10.12 & 17.22 & 9.45 & 9.26 & 0.65 & -4.25 & 11.83 & 10.17 & 10.16 & 0.91 \\
					
					EW & Poor & 1.34 & -12.92 & 20.14 & 10.20 & 10.24 & 0.62 & -6.09 & 13.87 & 11.20 & 11.26 & 0.89 \\

					\bottomrule
				\end{tabular}
				\begin{tablenotes}
					\tiny
					\item  Mod: Moderate; IPW($\alpha$): trimmed IPW with $I_{\alpha}(x)=\mathds{1}({\{\alpha\leq e(x)\leq 1-\alpha\}})$, for  $\alpha=0.05, 0.10,$ and $0.15.$
					\item Bias: relative bias in percentage; RMSE: root mean-squared error in $10^{-2}$; SD: empirical standard deviation in $10^{-2}$; SE: average estimated standard error in $10^{-2}$; CP: coverage probability of $95\%$ confidence interval. The results are based on 1000 simulated data sets.
				\end{tablenotes}
			\end{threeparttable}
		\end{table}

		
		\subsection{Under variable transformation}
		Tables \ref{N500} and \ref{N1000} provide the results under variable transformation for $N=500$ and  $N=1000$, respectively.
		\begin{table}[]
			\centering
			\begin{threeparttable}[]
				\centering\tiny\sf
				\caption{Variable transformation ($N=500$).\label{N500} { }}
				\begin{tabular}{p{1.1cm}p{0.6cm}p{0.31cm}   p{0.6cm}p{0.4cm}p{0.3cm}p{0.3cm}p{0.4cm}  p{0.6cm}p{0.4cm}p{0.3cm}p{0.3cm}p{0.4cm}  p{0.6cm}p{0.4cm}p{0.3cm}p{0.3cm}p{0.4cm}}
					\toprule
					\multicolumn{18}{c}{\bf Homogeneous treatment effect} \\\cmidrule(lr){1-18}
					&  &  & \multicolumn{15}{c}{Propensity score misspecification} \\\cmidrule(lr){4-18}
					&  &  & \multicolumn{5}{c}{None }& \multicolumn{5}{c}{Mild} &   \multicolumn{5}{c}{Major}   \\ \cmidrule(lr){4-8}\cmidrule(lr){9-13}\cmidrule(lr){14-18}
					Weight & Overlap & True &Bias & RMSE & SD & SE & CP & Bias & RMSE & SD & SE  & CP & Bias & RMSE & SD & SE & CP\\
					\cmidrule(lr){1-18}

					IPW & Good & 1 & 0.05 & 11.38 & 11.38 & 10.83 & 0.95 & 0.30 & 10.80 & 10.80 & 10.69 & 0.95 & 0.53 & 12.00 & 11.99 & 11.07 & 0.96 \\
					
					IPW(0.05) & Good & 1 & -0.22 & 10.36 & 10.37 & 10.44 & 0.96 & 0.28 & 10.21 & 10.21 & 10.39 & 0.95 & 0.66 & 10.32 & 10.30 & 10.30 & 0.95 \\
					
					IPW(0.1) & Good & 1 & -0.05 & 9.93 & 9.93 & 10.32 & 0.96 & 0.40 & 10.09 & 10.08 & 10.27 & 0.95 & 0.83 & 10.08 & 10.05 & 10.18 & 0.95 \\
					
					IPW(0.15) & Good & 1 & 0.13 & 10.19 & 10.19 & 10.42 & 0.94 & 0.48 & 10.18 & 10.17 & 10.37 & 0.95 & 1.03 & 10.14 & 10.09 & 10.27 & 0.95 \\
					
					OW & Good & 1 & 0.03 & 9.75 & 9.75 & 10.02 & 0.96 & 0.38 & 9.66 & 9.66 & 9.96 & 0.95 & 1.13 & 9.69 & 9.63 & 9.88 & 0.95 \\
					
					MW & Good & 1 & 0.08 & 9.89 & 9.89 & 10.23 & 0.96 & 0.33 & 9.77 & 9.77 & 10.15 & 0.95 & 1.19 & 9.79 & 9.72 & 10.07 & 0.95 \\
					
					EW & Good & 1 & 0.00 & 9.78 & 9.79 & 10.01 & 0.95 & 0.37 & 9.70 & 9.70 & 9.96 & 0.95 & 1.06 & 9.73 & 9.67 & 9.88 & 0.95 \\ \addlinespace

					IPW & Mod & 1 & 1.00 & 20.34 & 20.33 & 14.69 & 0.89 & -0.01 & 23.43 & 23.44 & 17.37 & 0.91 & -1.20 & 26.10 & 26.08 & 19.68 & 0.91 \\
					
					IPW(0.05) & Mod & 1 & -0.13 & 14.13 & 14.14 & 12.75 & 0.92 & 0.55 & 14.00 & 13.99 & 12.82 & 0.91 & 0.86 & 14.02 & 14.00 & 12.63 & 0.91 \\
					
					IPW(0.1) & Mod & 1 & -0.60 & 13.55 & 13.55 & 12.59 & 0.92 & 0.24 & 13.50 & 13.50 & 12.61 & 0.92 & 1.03 & 13.22 & 13.19 & 12.34 & 0.93 \\
					
					IPW(0.15) & Mod & 1 & -0.73 & 14.06 & 14.05 & 12.99 & 0.92 & -0.09 & 13.83 & 13.84 & 12.99 & 0.93 & 0.79 & 13.34 & 13.32 & 12.59 & 0.92 \\
					
					OW & Mod & 1 & -0.49 & 12.59 & 12.59 & 11.78 & 0.93 & -0.12 & 12.37 & 12.38 & 11.61 & 0.93 & 1.30 & 12.21 & 12.15 & 11.35 & 0.93 \\
					
					MW & Mod & 1 & -0.63 & 12.72 & 12.71 & 12.08 & 0.94 & -0.31 & 12.48 & 12.48 & 11.91 & 0.94 & 1.44 & 12.37 & 12.29 & 11.64 & 0.94 \\
					
					EW & Mod & 1 & -0.35 & 12.72 & 12.72 & 11.77 & 0.92 & -0.04 & 12.63 & 12.63 & 11.68 & 0.92 & 1.09 & 12.39 & 12.35 & 11.42 & 0.92 \\ \addlinespace

					IPW & Poor & 1 & 2.00 & 23.37 & 23.29 & 16.61 & 0.88 & 0.79 & 30.90 & 30.91 & 22.20 & 0.92 & -2.04 & 37.43 & 37.39 & 29.17 & 0.92 \\
					
					IPW(0.05) & Poor & 1 & 0.24 & 15.43 & 15.43 & 14.80 & 0.94 & 0.70 & 16.11 & 16.10 & 15.00 & 0.94 & 2.10 & 15.33 & 15.19 & 14.49 & 0.93 \\
					
					IPW(0.1) & Poor & 1 & 0.61 & 15.42 & 15.41 & 14.72 & 0.95 & 1.42 & 15.19 & 15.13 & 14.71 & 0.95 & 2.54 & 14.60 & 14.38 & 14.09 & 0.94 \\
					
					IPW(0.15) & Poor & 1 & 0.55 & 15.54 & 15.54 & 15.41 & 0.95 & 1.40 & 15.35 & 15.29 & 15.32 & 0.95 & 2.74 & 14.91 & 14.66 & 14.63 & 0.95 \\
					
					OW & Poor & 1 & 0.54 & 13.39 & 13.38 & 13.44 & 0.96 & 1.09 & 13.21 & 13.18 & 13.16 & 0.95 & 3.25 & 12.94 & 12.54 & 12.62 & 0.95 \\
					
					MW & Poor & 1 & 0.50 & 13.67 & 13.67 & 13.85 & 0.95 & 1.04 & 13.48 & 13.45 & 13.59 & 0.96 & 3.62 & 13.28 & 12.78 & 13.02 & 0.95 \\
					
					EW & Poor & 1 & 0.55 & 13.51 & 13.50 & 13.43 & 0.95 & 1.05 & 13.43 & 13.39 & 13.27 & 0.95 & 2.81 & 13.08 & 12.78 & 12.73 & 0.94 \\

					\bottomrule
					%
					%
					%
					\addlinespace
					&  &  & \multicolumn{15}{c}{Propensity score misspecification} \\\cmidrule(lr){4-18}
					&  &  & \multicolumn{5}{c}{None }& \multicolumn{5}{c}{Mild} &   \multicolumn{5}{c}{Major}   \\ \cmidrule(lr){4-8}\cmidrule(lr){9-13}\cmidrule(lr){14-18}
					Weight & Overlap & True &Bias & RMSE & SD & SE & CP & Bias & RMSE & SD & SE  & CP & Bias & RMSE & SD & SE & CP\\
					\cmidrule(lr){1-18}					
					
					IPW & Good & 1.46 & 0.34 & 12.65 & 12.65 & 11.14 & 0.92 & 2.15 & 12.44 & 12.05 & 10.97 & 0.92 & 2.60 & 13.13 & 12.58 & 11.33 & 0.92 \\
					
					IPW(0.05) & Good & 1.47 & 0.22 & 11.65 & 11.65 & 10.58 & 0.93 & 1.41 & 11.55 & 11.37 & 10.58 & 0.92 & 1.89 & 11.71 & 11.38 & 10.48 & 0.92 \\
					
					IPW(0.1) & Good & 1.50 & 0.14 & 11.25 & 11.25 & 10.38 & 0.93 & 0.67 & 11.17 & 11.13 & 10.37 & 0.93 & 1.09 & 11.19 & 11.07 & 10.31 & 0.93 \\
					
					IPW(0.15) & Good & 1.53 & 0.17 & 11.26 & 11.26 & 10.44 & 0.93 & 0.36 & 11.34 & 11.34 & 10.43 & 0.93 & 0.63 & 11.19 & 11.15 & 10.34 & 0.93 \\
					
					OW & Good & 1.52 & -0.02 & 10.92 & 10.92 & 10.09 & 0.93 & 0.68 & 10.91 & 10.86 & 10.04 & 0.93 & 1.23 & 10.94 & 10.78 & 9.97 & 0.93 \\
					
					MW & Good & 1.54 & -0.16 & 11.13 & 11.14 & 10.27 & 0.93 & 0.31 & 11.06 & 11.05 & 10.20 & 0.93 & 0.87 & 11.03 & 10.96 & 10.13 & 0.93 \\
					
					EW & Good & 1.51 & 0.03 & 10.94 & 10.95 & 10.09 & 0.93 & 0.91 & 10.96 & 10.88 & 10.06 & 0.93 & 1.46 & 11.01 & 10.79 & 9.98 & 0.92 \\ \addlinespace

					IPW & Mod & 1.24 & 2.20 & 21.04 & 20.88 & 15.40 & 0.87 & 4.70 & 23.83 & 23.11 & 17.92 & 0.86 & 5.54 & 28.36 & 27.53 & 20.55 & 0.86 \\
					
					IPW(0.05) & Mod & 1.35 & 0.66 & 13.37 & 13.35 & 12.82 & 0.93 & 2.25 & 14.12 & 13.80 & 13.08 & 0.93 & 2.53 & 13.92 & 13.50 & 12.77 & 0.92 \\
					
					IPW(0.1) & Mod & 1.42 & 0.31 & 13.11 & 13.11 & 12.62 & 0.94 & 1.08 & 13.09 & 13.00 & 12.74 & 0.95 & 1.33 & 13.02 & 12.88 & 12.44 & 0.95 \\
					
					IPW(0.15) & Mod & 1.48 & -0.01 & 13.39 & 13.40 & 13.03 & 0.94 & 0.38 & 13.68 & 13.68 & 13.03 & 0.94 & 0.55 & 13.07 & 13.05 & 12.66 & 0.94 \\
					
					OW & Mod & 1.43 & 0.04 & 11.87 & 11.88 & 11.93 & 0.95 & 1.32 & 11.88 & 11.74 & 11.78 & 0.94 & 2.06 & 11.92 & 11.55 & 11.51 & 0.94 \\
					
					MW & Mod & 1.47 & -0.11 & 12.03 & 12.03 & 12.21 & 0.95 & 0.77 & 11.93 & 11.89 & 12.04 & 0.95 & 1.45 & 11.93 & 11.75 & 11.78 & 0.94 \\
					
					EW & Mod & 1.40 & 0.13 & 11.96 & 11.97 & 11.95 & 0.94 & 1.71 & 12.15 & 11.92 & 11.86 & 0.94 & 2.51 & 12.20 & 11.69 & 11.59 & 0.93 \\ \addlinespace

					IPW & Poor & 1.10 & 7.24 & 26.43 & 25.20 & 17.20 & 0.79 & 9.28 & 35.54 & 34.05 & 23.48 & 0.83 & 9.68 & 37.94 & 36.43 & 28.25 & 0.81 \\
					
					IPW(0.05) & Poor & 1.31 & 1.21 & 16.36 & 16.29 & 15.11 & 0.92 & 2.53 & 16.86 & 16.54 & 15.33 & 0.92 & 2.58 & 16.19 & 15.85 & 14.76 & 0.92 \\
					
					IPW(0.1) & Poor & 1.40 & 0.65 & 16.44 & 16.42 & 15.05 & 0.92 & 0.94 & 16.26 & 16.21 & 14.98 & 0.93 & 0.55 & 15.18 & 15.17 & 14.38 & 0.93 \\
					
					IPW(0.15) & Poor & 1.47 & 0.08 & 16.51 & 16.51 & 15.45 & 0.92 & -0.22 & 16.52 & 16.53 & 15.43 & 0.93 & -0.57 & 15.21 & 15.20 & 14.68 & 0.93 \\
					
					OW & Poor & 1.38 & 0.59 & 14.04 & 14.02 & 13.78 & 0.94 & 1.76 & 13.98 & 13.77 & 13.49 & 0.94 & 2.27 & 13.37 & 13.01 & 12.91 & 0.94 \\
					
					MW & Poor & 1.43 & 0.39 & 14.26 & 14.26 & 14.13 & 0.94 & 1.05 & 14.14 & 14.07 & 13.85 & 0.94 & 1.37 & 13.37 & 13.23 & 13.28 & 0.94 \\
					
					EW & Poor & 1.35 & 0.78 & 14.11 & 14.08 & 13.82 & 0.94 & 2.43 & 14.29 & 13.91 & 13.66 & 0.94 & 3.06 & 13.77 & 13.15 & 13.05 & 0.94 \\

					\bottomrule
				\end{tabular}
				\begin{tablenotes}
					\tiny
					\item  Mod: Moderate; IPW($\alpha$): trimmed IPW with $I_{\alpha}(x)=\mathds{1}({\{\alpha\leq e(x)\leq 1-\alpha\}})$, for   $\alpha=0.05, 0.10,$ and $0.15.$
					\item Bias: relative bias in percentage; RMSE: root mean-squared error in $10^{-2}$; SD: empirical standard deviation in $10^{-2}$; SE: average estimated standard error in $10^{-2}$; CP: coverage probability of $95\%$ confidence interval. The results are based on 1000 simulated data sets.
				\end{tablenotes}
			\end{threeparttable}
		\end{table}

		\begin{table}[]
			\centering
			\begin{threeparttable}[]
				\centering\tiny\sf
				\caption{Variable transformation ($N=1000$).\label{N1000} { }}
				\begin{tabular}{p{1.1cm}p{0.6cm}p{0.3cm}   p{0.6cm}p{0.4cm}p{0.3cm}p{0.3cm}p{0.4cm}  p{0.6cm}p{0.4cm}p{0.3cm}p{0.3cm}p{0.4cm}  p{0.6cm}p{0.4cm}p{0.3cm}p{0.3cm}p{0.4cm}}
					\toprule
					\multicolumn{18}{c}{\bf Homogeneous treatment effect} \\\cmidrule(lr){1-18}
					&  &  & \multicolumn{15}{c}{Propensity score misspecification} \\\cmidrule(lr){4-18}
					&  &  & \multicolumn{5}{c}{None }& \multicolumn{5}{c}{Mild} &   \multicolumn{5}{c}{Major}   \\ \cmidrule(lr){4-8}\cmidrule(lr){9-13}\cmidrule(lr){14-18}
					Weight & Overlap & True &Bias & RMSE & SD & SE & CP & Bias & RMSE & SD & SE  & CP & Bias & RMSE & SD & SE & CP\\
					\cmidrule(lr){1-18}	
					
					IPW & Good & 1 & 0.33 & 8.17 & 8.17 & 7.79 & 0.95 & 0.46 & 7.96 & 7.96 & 7.78 & 0.96 & 0.68 & 8.41 & 8.38 & 7.95 & 0.96 \\
					
					IPW(0.05) & Good & 1 & 0.21 & 7.42 & 7.42 & 7.45 & 0.95 & 0.59 & 7.50 & 7.48 & 7.41 & 0.95 & 0.99 & 7.43 & 7.37 & 7.35 & 0.96 \\
					
					IPW(0.1) & Good & 1 & 0.25 & 7.32 & 7.32 & 7.33 & 0.95 & 0.57 & 7.33 & 7.31 & 7.29 & 0.95 & 1.11 & 7.36 & 7.28 & 7.24 & 0.95 \\
					
					IPW(0.15) & Good & 1 & 0.32 & 7.42 & 7.42 & 7.38 & 0.95 & 0.54 & 7.40 & 7.38 & 7.34 & 0.95 & 1.15 & 7.48 & 7.39 & 7.28 & 0.95 \\
					
					OW & Good & 1 & 0.21 & 7.15 & 7.15 & 7.09 & 0.95 & 0.53 & 7.10 & 7.08 & 7.05 & 0.95 & 1.29 & 7.15 & 7.03 & 7.00 & 0.95 \\
					
					MW & Good & 1 & 0.17 & 7.30 & 7.30 & 7.20 & 0.95 & 0.44 & 7.22 & 7.21 & 7.15 & 0.95 & 1.36 & 7.27 & 7.14 & 7.10 & 0.95 \\
					
					EW & Good & 1 & 0.23 & 7.14 & 7.14 & 7.10 & 0.95 & 0.54 & 7.09 & 7.08 & 7.06 & 0.95 & 1.23 & 7.14 & 7.04 & 7.01 & 0.95 \\ \addlinespace

					IPW & Mod & 1 & 0.03 & 15.32 & 15.33 & 11.85 & 0.92 & -1.91 & 21.04 & 20.97 & 14.63 & 0.94 & -3.41 & 28.17 & 27.98 & 18.16 & 0.94 \\
					
					IPW(0.05) & Mod & 1 & 0.20 & 9.34 & 9.34 & 9.17 & 0.94 & 0.51 & 9.48 & 9.47 & 9.28 & 0.94 & 1.25 & 9.18 & 9.09 & 9.02 & 0.94 \\
					
					IPW(0.1) & Mod & 1 & -0.11 & 9.13 & 9.13 & 8.97 & 0.94 & 0.32 & 9.00 & 9.00 & 9.02 & 0.94 & 1.28 & 8.72 & 8.63 & 8.76 & 0.95 \\
					
					IPW(0.15) & Mod & 1 & -0.60 & 9.23 & 9.22 & 9.23 & 0.95 & 0.10 & 9.17 & 9.17 & 9.22 & 0.95 & 1.27 & 9.05 & 8.96 & 8.92 & 0.94 \\
					
					OW & Mod & 1 & -0.31 & 8.43 & 8.43 & 8.32 & 0.95 & 0.07 & 8.27 & 8.28 & 8.20 & 0.95 & 1.59 & 8.21 & 8.06 & 8.01 & 0.94 \\
					
					MW & Mod & 1 & -0.44 & 8.64 & 8.63 & 8.50 & 0.95 & -0.12 & 8.49 & 8.49 & 8.38 & 0.95 & 1.68 & 8.38 & 8.21 & 8.19 & 0.94 \\
					
					EW & Mod & 1 & -0.25 & 8.49 & 8.49 & 8.36 & 0.95 & 0.06 & 8.36 & 8.36 & 8.28 & 0.94 & 1.33 & 8.28 & 8.18 & 8.09 & 0.94 \\ \addlinespace

					IPW & Poor & 1 & 0.19 & 20.92 & 20.93 & 14.00 & 0.89 & -3.31 & 29.23 & 29.06 & 21.53 & 0.93 & -7.12 & 38.64 & 38.00 & 31.89 & 0.94 \\
					
					IPW(0.05) & Poor & 1 & -0.36 & 11.19 & 11.19 & 10.72 & 0.94 & 0.57 & 11.51 & 11.50 & 10.89 & 0.93 & 1.73 & 10.86 & 10.72 & 10.42 & 0.94 \\
					
					IPW(0.1) & Poor & 1 & -0.06 & 11.21 & 11.21 & 10.61 & 0.94 & 0.45 & 11.27 & 11.26 & 10.61 & 0.93 & 2.00 & 10.54 & 10.35 & 10.10 & 0.94 \\
					
					IPW(0.15) & Poor & 1 & 0.36 & 11.55 & 11.55 & 10.96 & 0.94 & 0.43 & 11.43 & 11.43 & 10.90 & 0.94 & 2.05 & 10.70 & 10.50 & 10.34 & 0.94 \\
					
					OW & Poor & 1 & -0.13 & 9.90 & 9.90 & 9.55 & 0.94 & 0.31 & 9.68 & 9.68 & 9.35 & 0.94 & 2.45 & 9.52 & 9.20 & 8.95 & 0.94 \\
					
					MW & Poor & 1 & -0.04 & 10.09 & 10.09 & 9.77 & 0.94 & 0.32 & 9.82 & 9.82 & 9.59 & 0.95 & 2.76 & 9.84 & 9.44 & 9.18 & 0.94 \\
					
					EW & Poor & 1 & -0.25 & 10.03 & 10.03 & 9.60 & 0.94 & 0.12 & 9.87 & 9.87 & 9.50 & 0.94 & 1.88 & 9.56 & 9.38 & 9.10 & 0.94 \\
					
					\bottomrule
					%
					%
					%
					%
					\addlinespace
					\multicolumn{18}{c}{\bf Heterogeneous treatment effect} \\\cmidrule(lr){1-18}
					&  &  & \multicolumn{15}{c}{Propensity score misspecification} \\\cmidrule(lr){4-18}
					&  &  & \multicolumn{5}{c}{None }& \multicolumn{5}{c}{Mild} &   \multicolumn{5}{c}{Major}   \\ \cmidrule(lr){4-8}\cmidrule(lr){9-13}\cmidrule(lr){14-18}
					Weight & Overlap & True &Bias & RMSE & SD & SE & CP & Bias & RMSE & SD & SE  & CP & Bias & RMSE & SD & SE & CP\\
					\cmidrule(lr){1-18}					
					
					IPW & Good & 1.46 & 0.32 & 8.60 & 8.59 & 8.01 & 0.94 & 2.16 & 8.82 & 8.24 & 7.85 & 0.91 & 2.73 & 9.68 & 8.83 & 8.07 & 0.90 \\
					
					IPW(0.05) & Good & 1.47 & 0.32 & 7.96 & 7.95 & 7.52 & 0.93 & 1.47 & 8.08 & 7.78 & 7.51 & 0.93 & 1.97 & 8.23 & 7.71 & 7.43 & 0.92 \\
					
					IPW(0.1) & Good & 1.50 & 0.21 & 7.74 & 7.73 & 7.38 & 0.94 & 0.81 & 7.74 & 7.65 & 7.36 & 0.93 & 1.19 & 7.81 & 7.61 & 7.31 & 0.93 \\
					
					IPW(0.15) & Good & 1.53 & 0.23 & 7.73 & 7.72 & 7.41 & 0.94 & 0.34 & 7.57 & 7.56 & 7.40 & 0.95 & 0.75 & 7.66 & 7.58 & 7.33 & 0.94 \\
					
					OW & Good & 1.52 & 0.15 & 7.39 & 7.39 & 7.13 & 0.94 & 0.84 & 7.42 & 7.31 & 7.10 & 0.94 & 1.37 & 7.56 & 7.27 & 7.05 & 0.93 \\
					
					MW & Good & 1.54 & 0.08 & 7.49 & 7.50 & 7.23 & 0.94 & 0.54 & 7.46 & 7.42 & 7.19 & 0.94 & 1.06 & 7.56 & 7.39 & 7.14 & 0.93 \\
					
					EW & Good & 1.51 & 0.18 & 7.42 & 7.41 & 7.14 & 0.94 & 1.04 & 7.50 & 7.33 & 7.12 & 0.94 & 1.58 & 7.68 & 7.30 & 7.07 & 0.93 \\ \addlinespace

					IPW & Mod & 1.24 & 1.15 & 17.47 & 17.42 & 12.64 & 0.88 & 3.79 & 20.46 & 19.92 & 15.13 & 0.86 & 4.59 & 28.18 & 27.61 & 18.97 & 0.84 \\
					
					IPW(0.05) & Mod & 1.35 & 0.43 & 9.90 & 9.89 & 9.36 & 0.93 & 1.98 & 10.57 & 10.23 & 9.52 & 0.92 & 2.47 & 10.37 & 9.82 & 9.27 & 0.91 \\
					
					IPW(0.1) & Mod & 1.42 & 0.16 & 9.49 & 9.49 & 9.10 & 0.95 & 1.00 & 9.53 & 9.42 & 9.15 & 0.94 & 1.14 & 9.38 & 9.25 & 8.92 & 0.94 \\
					
					IPW(0.15) & Mod & 1.48 & -0.07 & 9.50 & 9.50 & 9.30 & 0.94 & 0.17 & 9.36 & 9.36 & 9.30 & 0.95 & 0.11 & 9.15 & 9.15 & 9.02 & 0.95 \\
					
					OW & Mod & 1.43 & -0.02 & 8.64 & 8.65 & 8.47 & 0.95 & 1.23 & 8.72 & 8.54 & 8.37 & 0.94 & 1.88 & 8.74 & 8.32 & 8.17 & 0.93 \\
					
					MW & Mod & 1.47 & -0.12 & 8.75 & 8.76 & 8.62 & 0.95 & 0.68 & 8.64 & 8.58 & 8.51 & 0.95 & 1.25 & 8.60 & 8.40 & 8.32 & 0.94 \\
					
					EW & Mod & 1.40 & 0.04 & 8.78 & 8.78 & 8.54 & 0.95 & 1.62 & 8.98 & 8.69 & 8.47 & 0.94 & 2.37 & 9.11 & 8.49 & 8.27 & 0.92 \\ \addlinespace

					IPW & Poor & 1.10 & 4.88 & 23.28 & 22.66 & 14.88 & 0.78 & 7.51 & 30.41 & 29.27 & 21.46 & 0.81 & 7.18 & 39.22 & 38.43 & 33.71 & 0.82 \\
					
					IPW(0.05) & Poor & 1.31 & 0.64 & 11.65 & 11.62 & 10.87 & 0.94 & 2.26 & 12.36 & 12.00 & 11.01 & 0.92 & 2.24 & 11.54 & 11.16 & 10.54 & 0.92 \\
					
					IPW(0.1) & Poor & 1.40 & 0.41 & 10.92 & 10.91 & 10.65 & 0.94 & 0.71 & 11.16 & 11.12 & 10.70 & 0.94 & 0.48 & 10.32 & 10.30 & 10.22 & 0.95 \\
					
					IPW(0.15) & Poor & 1.47 & 0.07 & 11.56 & 11.57 & 11.02 & 0.95 & -0.03 & 11.21 & 11.22 & 10.97 & 0.94 & -0.64 & 10.57 & 10.53 & 10.42 & 0.94 \\
					
					OW & Poor & 1.38 & 0.31 & 10.00 & 10.00 & 9.75 & 0.94 & 1.51 & 10.04 & 9.82 & 9.55 & 0.93 & 2.03 & 9.68 & 9.26 & 9.15 & 0.94 \\
					
					MW & Poor & 1.43 & 0.20 & 10.09 & 10.09 & 9.95 & 0.95 & 0.86 & 9.96 & 9.89 & 9.77 & 0.94 & 1.17 & 9.48 & 9.34 & 9.36 & 0.94 \\
					
					EW & Poor & 1.35 & 0.42 & 10.27 & 10.26 & 9.84 & 0.94 & 2.09 & 10.51 & 10.13 & 9.71 & 0.93 & 2.77 & 10.24 & 9.54 & 9.31 & 0.93 \\

					\bottomrule
				\end{tabular}
				\begin{tablenotes}
					\tiny
					\item  Mod: Moderate; IPW($\alpha$): trimmed IPW with $I_{\alpha}(x)=\mathds{1}({\{\alpha\leq e(x)\leq 1-\alpha\}})$, for   $\alpha=0.05, 0.10,$ and $0.15.$
					\item Bias: relative bias in percentage; RMSE: root mean-squared error in $10^{-2}$; SD: empirical standard deviation in $10^{-2}$; SE: average estimated standard error in $10^{-2}$; CP: coverage probability of $95\%$ confidence interval. The results are based on 1000 simulated data sets.
				\end{tablenotes}
			\end{threeparttable}
		\end{table}

		\section{Simulation  results, Part III: Low treatment prevalence 
		} \label{sec:lowpre}
\begin{figure*}[]
			\begin{center}
				\includegraphics[trim=2 23 20 35, clip, width=0.85\linewidth]{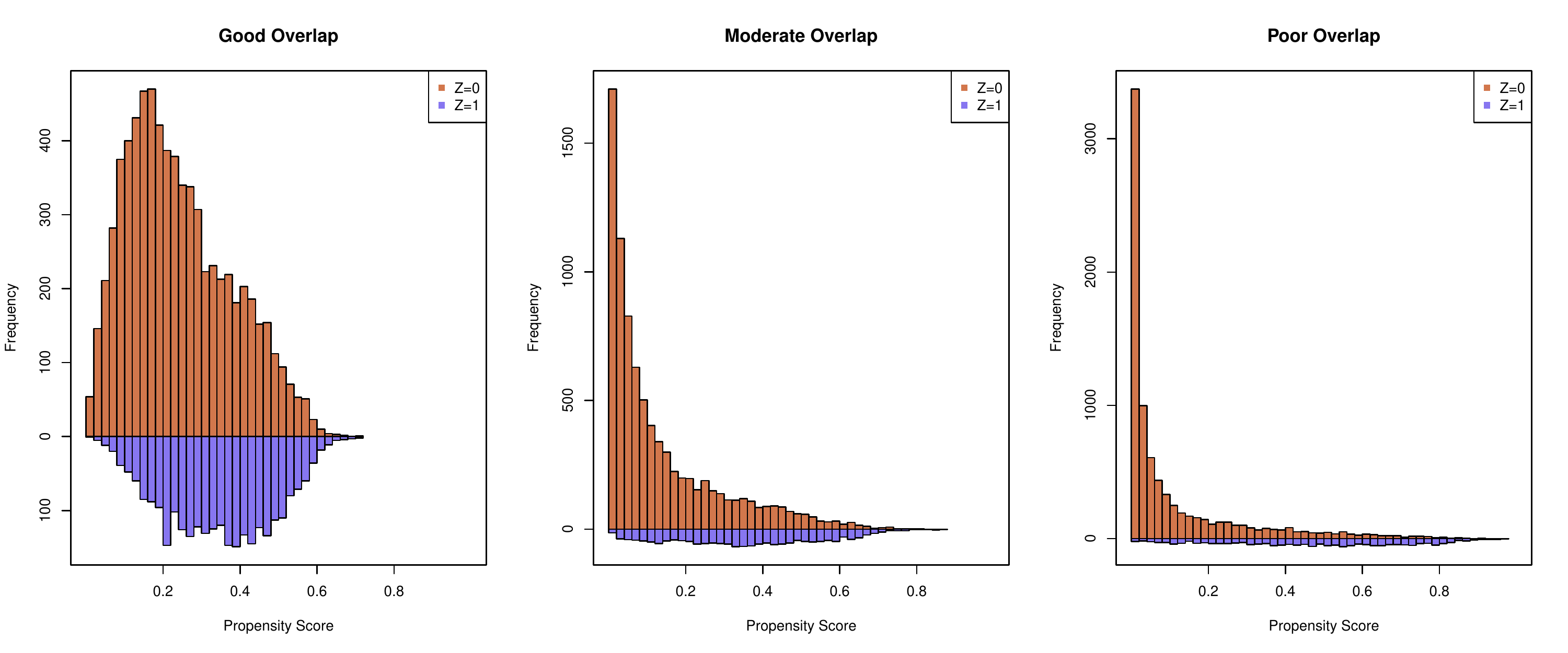}
				\includegraphics[trim=2 23 20 35, clip, width=0.85\linewidth]{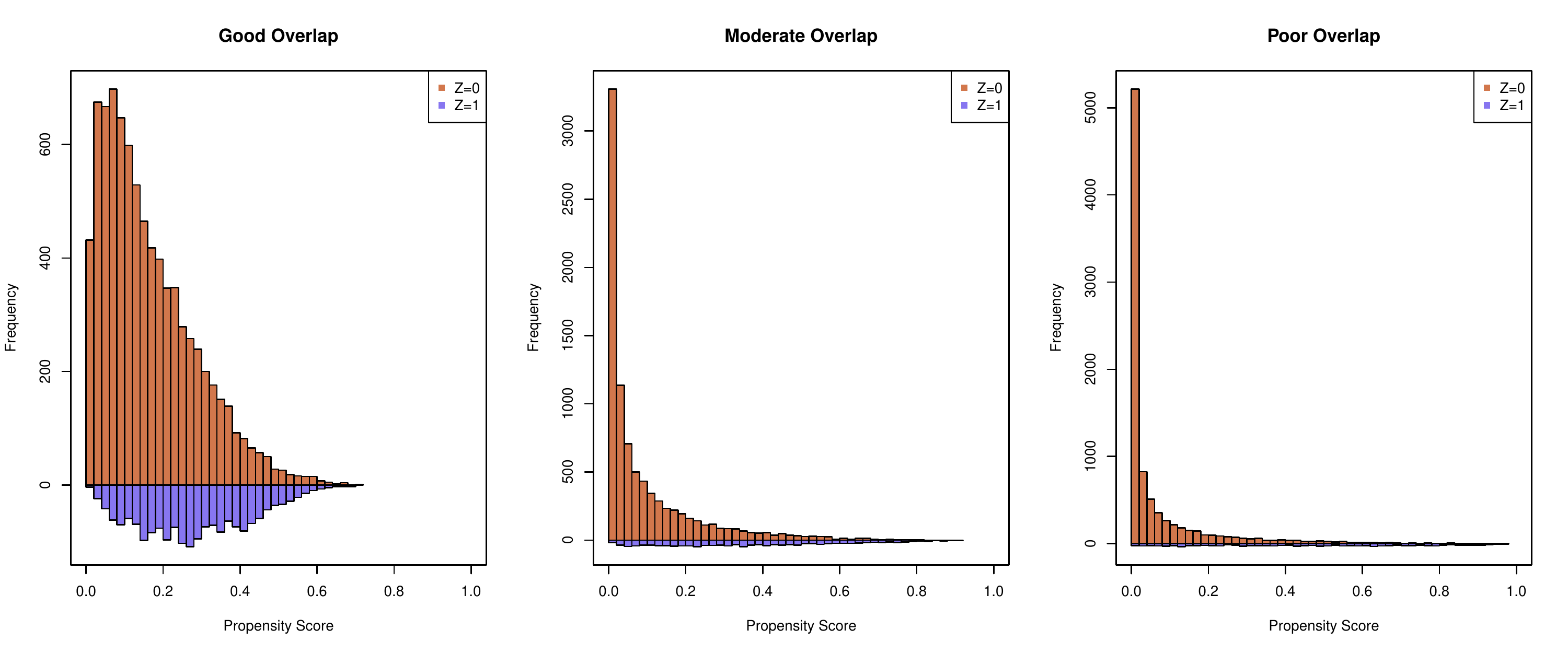}
			\end{center}
			\caption{Propensity scores (low prevalence of treatment), with good (left), moderate (middle), and poor (right) overlap. Variable omission (top panel) and variable transformation (bottom panel).
				\label{Mao_Overlap_Asy}}
		\end{figure*}

\begin{figure*}[h]
			\begin{center}
				\includegraphics[trim=5 20 10 5, clip, width=1\linewidth]{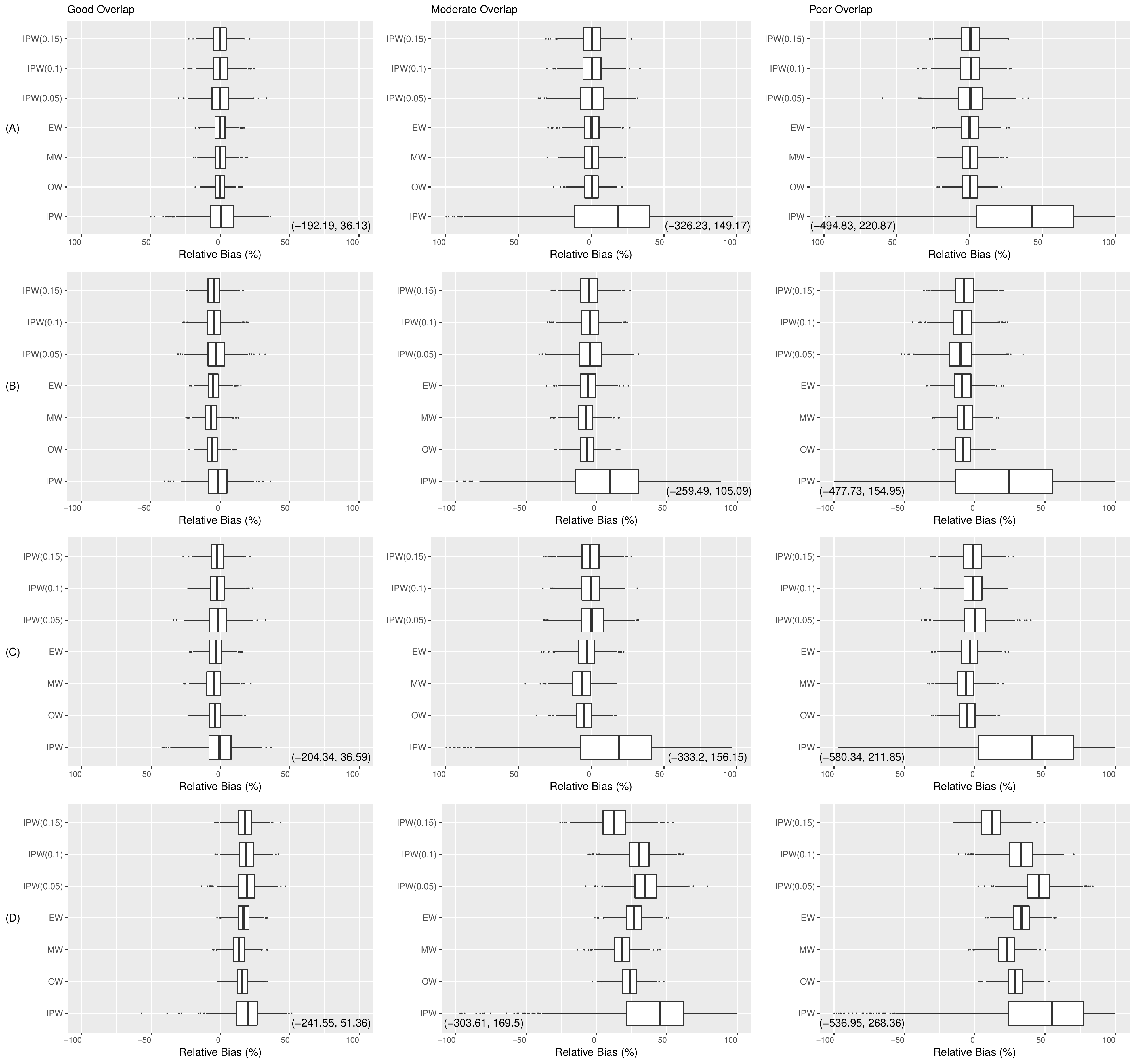}
			\end{center}
			\caption{Variable Omission: Relative bias   (homogeneous treatment effect and low prevalence of treatment), with $N=2,000$. \label{Boxplot_Mao_Asy} 
			}
			\subcaption*{{\bf Legend:} A: Correct PS Model; B: Missing $X_1^2$; C: Missing $X_2X_4$; D: Missing $X_2$.}
		\end{figure*}

\begin{figure*}[h]
			\begin{center}
				\includegraphics[trim=5 20 10 5, clip, width=1\linewidth]{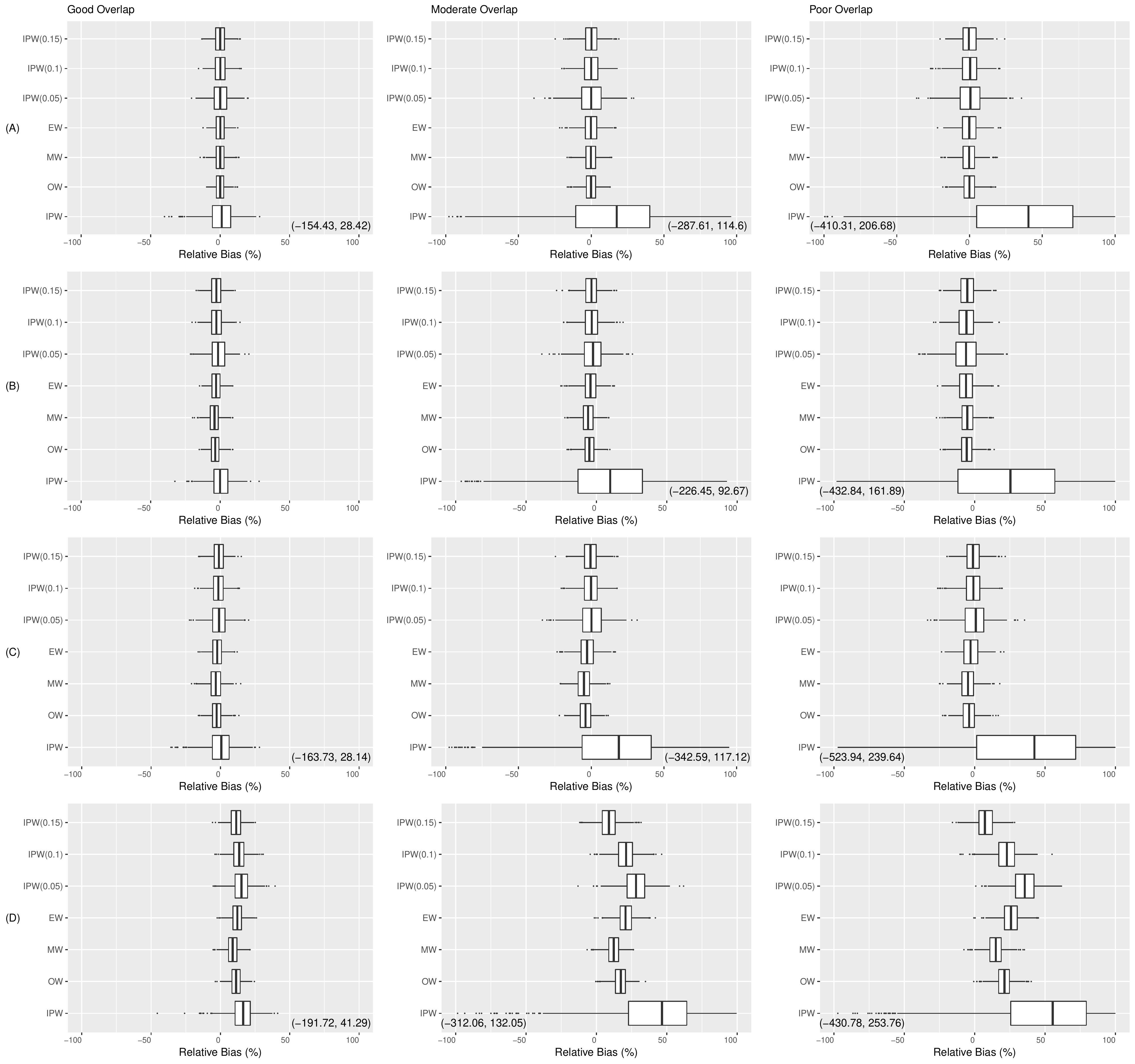}
			\end{center}
			\caption{Variable Omission: Relative bias   (heterogeneous treatment effect and low prevalence of treatment), with $N=2000$. \label{Boxplot_HE_Mao_Asy} 
			}
			\subcaption*{{\bf Legend:} A: Correct PS Model; B: Missing $X_1^2$; C: Missing $X_2X_4$; D: Missing $X_2$.}
		\end{figure*}

\begin{figure*}[h]
			\begin{center}
				\includegraphics[trim=5 25 5 5, clip, width=0.82\linewidth]{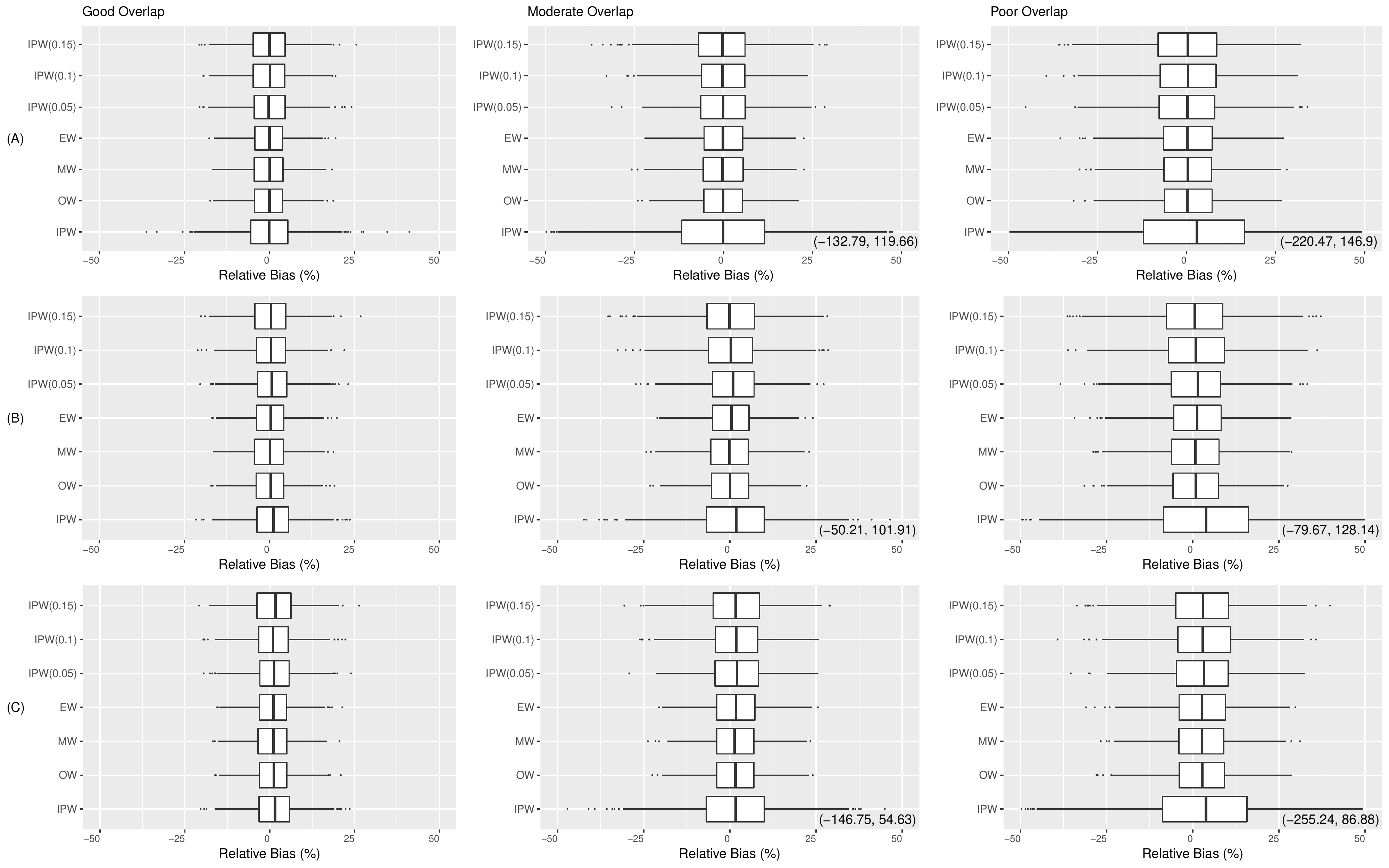}
			\end{center}
			\caption{Variable Transformation: Rel. bias (homogeneous trt. effect, low prevalence, and $N=2000$). \label{Boxplot_KS_Asy} 
			}
		\end{figure*}

\begin{figure*}[h]
			\begin{center}
				\includegraphics[trim=5 25 5 5, clip, width=0.82\linewidth]{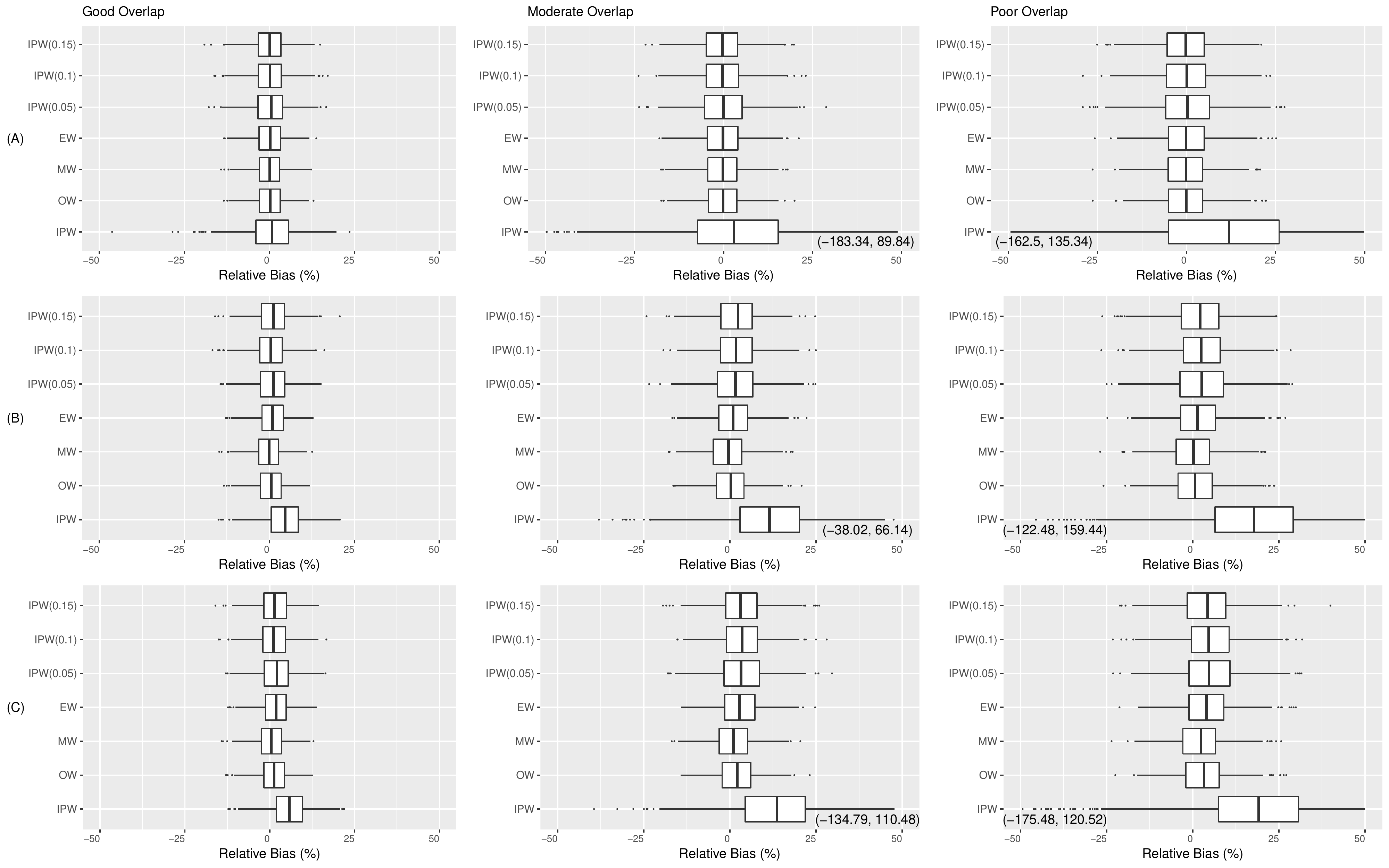}
			\end{center}
			\caption{Variable Transformation: Rel. bias  (heterogeneous trt. effect, low prevalence, and $N=2000$). \label{Boxplot_HE_KS_Asy} 
			}
			\subcaption*{{\bf Legend:} A: Correct PS Model; B: Mild PS Model Misspecification; C: Major PS Model Misspecification.}
		\end{figure*}

		In this section, we present the results under low prevalence of treatment (as indicated in Table \ref{trt_prevalence_appendix}),  respectively, under variable omission and under variable transformation (see Figure \ref{Mao_Overlap_Asy}) when  $N = 2000$. We consider both homogeneous and heterogeneous treatment effects.\\
		\begin{table}[]
			\centering
			\begin{threeparttable}[]
				\centering\small\sf
				\caption{Average treatment prevalence for different levels of PS overlap.\label{trt_prevalence_appendix} { }}
				\begin{tabular}{p{3.5cm}p{1cm}p{1cm}p{0.7cm}p{0.1cm}p{1cm}p{1cm}p{0.7cm}  }
					\toprule
					& \multicolumn{3}{c}{Medium prevalence} && \multicolumn{3}{c}{Low prevalence} \\ \cmidrule(lr){2-4}\cmidrule(lr){6-8}
					PS misspecification   & Good & Moderate & Poor  &&Good & Moderate & Poor  \\
					\cmidrule(lr){1-8}

					Variable omission & 47.42\% & 47.34\% & 47.12\% & & 	27.7\% & 	18.07\% & 	17.99\% \\
					Variable transformation  & 40.76\% & 37.63\% &	37.00\% &&  18.51\% & 	12.79\% & 	10.87\% \\
					\bottomrule
				\end{tabular}
			\end{threeparttable}
		\end{table}			
		Tables \ref{varomi_asy} and \ref{varomi_HE_asy} show the results for variable omission  whereas Table \ref{var_trans_Asy} shows the results for variable transformation under both homogeneous and heterogeneous treatment effects. The boxplots of relative bias percentages for variable omission are shown in Figures \ref{Boxplot_Mao_Asy} and \ref{Boxplot_HE_Mao_Asy}; those for  variable transformation are shown in Figures \ref{Boxplot_KS_Asy} and \ref{Boxplot_HE_KS_Asy}

		\begin{table}[h]
			\centering
			\begin{threeparttable}[]
				\centering\scriptsize\sf
				\caption{Variable omission: homogeneous treatment effect and  low prevalence of treatment ($N=2000$).\label{varomi_asy} { }}
				\begin{tabular}{p{1cm}p{0.6cm}p{0.7cm}  p{0.7cm}p{0.7cm}p{0.7cm}p{0.7cm}p{0.7cm}   p{0.7cm}p{0.7cm}p{0.7cm}p{0.7cm}p{0.7cm}}
					\toprule
					&  &  & \multicolumn{10}{c}{Propensity score misspecification} \\\cmidrule(lr){4-13}
					&  &  & \multicolumn{5}{c}{None }& \multicolumn{5}{c}{Missing $X_2$} \\ \cmidrule(lr){4-8}\cmidrule(lr){9-13}
					Weight & Overlap & True & Bias & RMSE & SD & SE & CP & Bias & RMSE & SD & SE & CP \\
					\cmidrule(lr){1-13}	
					
					IPW & Good & 1 & 0.40 & 13.88 & 13.88 & 11.47 & 0.92 & 18.57 & 23.29 & 14.07 & 10.52 & 0.48 \\
					IPW(0.05) & Good & 1 & 0.09 & 8.85 & 8.86 & 8.57 & 0.94 & 18.74 & 20.75 & 8.91 & 8.60 & 0.41 \\
					IPW(0.1) & Good & 1 & 0.07 & 7.19 & 7.20 & 6.96 & 0.94 & 18.48 & 19.94 & 7.50 & 7.38 & 0.30 \\
					IPW(0.15) & Good & 1 & -0.27 & 6.35 & 6.35 & 6.36 & 0.95 & 17.43 & 18.73 & 6.85 & 6.87 & 0.29 \\
					OW & Good & 1 & -0.14 & 5.26 & 5.26 & 5.30 & 0.95 & 15.83 & 16.84 & 5.76 & 5.86 & 0.22 \\
					MW & Good & 1 & -0.08 & 5.74 & 5.74 & 5.88 & 0.96 & 13.21 & 14.56 & 6.12 & 6.28 & 0.44 \\
					EW & Good & 1 & -0.10 & 5.56 & 5.56 & 5.52 & 0.94 & 16.64 & 17.69 & 6.00 & 6.05 & 0.20 \\  \addlinespace
					
					IPW & Mod & 1 & 8.65 & 49.49 & 48.75 & 34.61 & 0.77 & 37.73 & 56.27 & 41.76 & 29.77 & 0.50 \\
					IPW(0.05) & Mod & 1 & 0.43 & 11.49 & 11.48 & 11.28 & 0.95 & 34.85 & 36.82 & 11.91 & 11.38 & 0.16 \\
					IPW(0.1) & Mod & 1 & 0.36 & 9.38 & 9.37 & 8.93 & 0.94 & 30.09 & 31.95 & 10.73 & 9.60 & 0.14 \\
					IPW(0.15) & Mod & 1 & 0.60 & 8.86 & 8.85 & 8.55 & 0.94 & 12.73 & 17.34 & 11.78 & 9.16 & 0.67 \\
					OW & Mod & 1 & 0.15 & 6.95 & 6.96 & 6.68 & 0.94 & 23.35 & 24.51 & 7.48 & 7.23 & 0.11 \\
					MW & Mod & 1 & 0.12 & 7.55 & 7.56 & 7.41 & 0.95 & 17.90 & 19.56 & 7.89 & 7.77 & 0.35 \\
					EW & Mod & 1 & 0.21 & 7.83 & 7.83 & 7.52 & 0.94 & 26.54 & 27.75 & 8.10 & 7.93 & 0.09 \\  \addlinespace
					
					IPW & Poor & 1 & 32.37 & 84.20 & 77.77 & 45.69 & 0.56 & 53.43 & 89.20 & 71.47 & 43.92 & 0.47 \\
					IPW(0.05) & Poor & 1 & 0.37 & 12.15 & 12.16 & 11.98 & 0.94 & 45.21 & 46.83 & 12.21 & 12.08 & 0.06 \\
					IPW(0.1) & Poor & 1 & 0.22 & 9.72 & 9.72 & 9.86 & 0.96 & 32.32 & 34.62 & 12.41 & 10.44 & 0.17 \\
					IPW(0.15) & Poor & 1 & 0.37 & 9.12 & 9.12 & 9.34 & 0.96 & 12.01 & 15.80 & 10.27 & 9.76 & 0.76 \\
					OW & Poor & 1 & 0.14 & 7.19 & 7.19 & 7.33 & 0.96 & 28.83 & 29.81 & 7.59 & 7.83 & 0.04 \\
					MW & Poor & 1 & 0.16 & 7.73 & 7.73 & 7.92 & 0.96 & 22.34 & 23.72 & 8.00 & 8.24 & 0.22 \\
					EW & Poor & 1 & 0.11 & 8.40 & 8.40 & 8.35 & 0.94 & 33.13 & 34.24 & 8.63 & 8.61 & 0.03 \\  \addlinespace

					&  &  &  \multicolumn{5}{c}{Missing $X_1^2$} &   \multicolumn{5}{c}{Missing $X_2X_4$}  \\ \cmidrule(lr){4-8}\cmidrule(lr){9-13}
					Weight & Overlap & True & Bias & RMSE & SD & SE & CP & Bias & RMSE & SD & SE & CP\\
					\cmidrule(lr){1-13}	
					
					IPW & Good & 1 & -1.67 & 10.16 & 10.02 & 9.38 & 0.93 & -0.91 & 13.60 & 13.58 & 11.16 & 0.94 \\
					IPW(0.05) & Good & 1 & -2.81 & 9.17 & 8.73 & 8.47 & 0.93 & -1.75 & 9.34 & 9.18 & 8.78 & 0.94 \\
					IPW(0.1) & Good & 1 & -4.12 & 8.20 & 7.09 & 6.94 & 0.91 & -2.17 & 7.81 & 7.51 & 7.20 & 0.93 \\
					IPW(0.15) & Good & 1 & -4.43 & 7.75 & 6.36 & 6.34 & 0.89 & -2.03 & 6.94 & 6.64 & 6.59 & 0.94 \\
					OW & Good & 1 & -5.54 & 7.71 & 5.36 & 5.41 & 0.84 & -4.03 & 7.43 & 6.25 & 6.27 & 0.90 \\
					MW & Good & 1 & -6.38 & 8.65 & 5.84 & 5.99 & 0.82 & -4.97 & 8.70 & 7.14 & 7.23 & 0.89 \\
					EW & Good & 1 & -4.83 & 7.42 & 5.63 & 5.57 & 0.86 & -3.44 & 7.16 & 6.29 & 6.23 & 0.91 \\  \addlinespace
					
					IPW & Mod & 1 & 3.93 & 38.03 & 37.85 & 29.68 & 0.83 & 10.70 & 48.34 & 47.16 & 33.86 & 0.75 \\
					IPW(0.05) & Mod & 1 & -4.52 & 12.58 & 11.74 & 10.99 & 0.92 & 0.53 & 11.58 & 11.57 & 10.93 & 0.93 \\
					IPW(0.1) & Mod & 1 & -4.92 & 10.52 & 9.30 & 8.91 & 0.91 & -0.60 & 9.30 & 9.28 & 8.82 & 0.94 \\
					IPW(0.15) & Mod & 1 & -5.23 & 10.32 & 8.90 & 8.59 & 0.90 & -0.78 & 9.04 & 9.01 & 8.62 & 0.95 \\
					OW & Mod & 1 & -7.04 & 9.93 & 7.00 & 6.75 & 0.81 & -5.04 & 9.40 & 7.94 & 7.50 & 0.88 \\
					MW & Mod & 1 & -7.87 & 10.87 & 7.50 & 7.44 & 0.79 & -6.60 & 11.12 & 8.96 & 8.58 & 0.88 \\
					EW & Mod & 1 & -6.00 & 9.92 & 7.91 & 7.48 & 0.86 & -3.38 & 9.06 & 8.40 & 7.92 & 0.91 \\  \addlinespace
					
					IPW & Poor & 1 & 10.08 & 71.43 & 70.75 & 46.21 & 0.73 & 31.84 & 87.13 & 81.14 & 47.29 & 0.57 \\
					IPW(0.05) & Poor & 1 & -10.09 & 16.07 & 12.52 & 12.59 & 0.90 & -0.15 & 11.74 & 11.74 & 11.81 & 0.95 \\
					IPW(0.1) & Poor & 1 & -8.80 & 13.27 & 9.94 & 10.04 & 0.88 & -1.65 & 9.75 & 9.62 & 9.81 & 0.95 \\
					IPW(0.15) & Poor & 1 & -7.46 & 11.87 & 9.25 & 9.38 & 0.88 & -1.82 & 9.49 & 9.32 & 9.44 & 0.95 \\
					OW & Poor & 1 & -8.17 & 10.88 & 7.19 & 7.33 & 0.79 & -5.44 & 9.61 & 7.93 & 7.97 & 0.89 \\
					MW & Poor & 1 & -7.30 & 10.58 & 7.65 & 7.84 & 0.85 & -6.45 & 10.75 & 8.61 & 8.71 & 0.88 \\
					EW & Poor & 1 & -8.70 & 12.10 & 8.42 & 8.35 & 0.83 & -3.66 & 9.51 & 8.78 & 8.67 & 0.93 \\

					\bottomrule
				\end{tabular}
				\begin{tablenotes}
					\tiny
					\item  Mod: Moderate; IPW($\alpha$): trimmed IPW with $I_{\alpha}(x)=\mathds{1}({\{\alpha\leq e(x)\leq 1-\alpha\}})$, for  $\alpha=0.05, 0.10,$ and $0.15.$
					\item Bias: relative bias in percentage; RMSE: root mean-squared error in $10^{-2}$; SD: empirical standard deviation in $10^{-2}$; SE: average estimated standard error in $10^{-2}$; CP: coverage probability of $95\%$ confidence interval. The results are based on 1000 simulated data sets.
				\end{tablenotes}
			\end{threeparttable}
		\end{table}		
		
		\begin{table}[]
			\centering
			\begin{threeparttable}[]
				\centering\scriptsize\sf
				\caption{Variable omission: heterogeneous treatment effect and  low prevalence of treatment ($N=2000$).\label{varomi_HE_asy} { }}
				\begin{tabular}{p{1cm}p{0.6cm}p{0.7cm}  p{0.7cm}p{0.7cm}p{0.7cm}p{0.7cm}p{0.7cm}   p{0.7cm}p{0.7cm}p{0.7cm}p{0.7cm}p{0.7cm}}
					\toprule
					&  &  & \multicolumn{10}{c}{Propensity score misspecification} \\\cmidrule(lr){4-13}
					&  &  & \multicolumn{5}{c}{None }& \multicolumn{5}{c}{Missing $X_2$} \\ \cmidrule(lr){4-8}\cmidrule(lr){9-13}
					Weight & Overlap & True & Bias & RMSE & SD & SE & CP & Bias & RMSE & SD & SE & CP \\
					\cmidrule(lr){1-13}	
					
					IPW & Good & 1.38 & 0.45 & 15.09 & 15.08 & 12.25 & 0.90 & 15.69 & 26.36 & 15.05 & 11.29 & 0.42 \\
					IPW(0.05) & Good & 1.40 & 0.18 & 9.11 & 9.11 & 9.05 & 0.95 & 15.16 & 23.08 & 9.21 & 9.09 & 0.36 \\
					IPW(0.1) & Good & 1.44 & 0.11 & 7.39 & 7.39 & 7.22 & 0.94 & 13.36 & 20.69 & 7.79 & 7.79 & 0.31 \\
					IPW(0.15) & Good & 1.48 & -0.09 & 6.70 & 6.70 & 6.50 & 0.94 & 11.21 & 18.16 & 7.28 & 7.18 & 0.36 \\
					OW & Good & 1.46 & -0.04 & 5.58 & 5.59 & 5.35 & 0.94 & 11.24 & 17.58 & 6.24 & 6.05 & 0.24 \\
					MW & Good & 1.49 & -0.01 & 6.17 & 6.18 & 5.89 & 0.94 & 8.88 & 14.82 & 6.63 & 6.37 & 0.45 \\
					Entropy & Good & 1.45 & 0.00 & 5.87 & 5.87 & 5.64 & 0.95 & 12.25 & 18.84 & 6.43 & 6.31 & 0.20 \\  \addlinespace
					
					IPW & Mod & 1.12 & 7.87 & 53.98 & 53.28 & 37.52 & 0.76 & 39.83 & 64.88 & 46.98 & 32.39 & 0.48 \\
					IPW(0.05) & Mod & 1.30 & 0.15 & 12.39 & 12.39 & 12.02 & 0.94 & 27.88 & 38.32 & 12.51 & 12.27 & 0.19 \\
					IPW(0.1) & Mod & 1.42 & -0.02 & 9.49 & 9.49 & 9.25 & 0.94 & 20.72 & 31.18 & 10.61 & 10.18 & 0.20 \\
					IPW(0.15) & Mod & 1.49 & -0.03 & 8.76 & 8.76 & 8.71 & 0.95 & 8.84 & 16.78 & 10.39 & 9.43 & 0.70 \\
					OW & Mod & 1.38 & -0.17 & 6.71 & 6.71 & 6.83 & 0.95 & 17.17 & 24.79 & 7.37 & 7.51 & 0.11 \\
					MW & Mod & 1.43 & -0.28 & 7.16 & 7.15 & 7.48 & 0.96 & 12.21 & 19.02 & 7.59 & 7.90 & 0.38 \\
					EW & Mod & 1.34 & -0.09 & 7.89 & 7.90 & 7.84 & 0.95 & 20.78 & 29.01 & 8.44 & 8.41 & 0.09 \\  \addlinespace
					
					IPW & Poor & 1.05 & 33.33 & 89.26 & 82.17 & 49.62 & 0.59 & 61.78 & 99.92 & 76.09 & 48.09 & 0.46 \\
					IPW(0.05) & Poor & 1.33 & 0.34 & 13.52 & 13.52 & 12.74 & 0.94 & 35.45 & 48.87 & 13.05 & 12.85 & 0.06 \\
					IPW(0.1) & Poor & 1.43 & -0.05 & 10.49 & 10.49 & 10.16 & 0.94 & 22.37 & 34.32 & 12.47 & 10.90 & 0.18 \\
					IPW(0.15) & Poor & 1.49 & -0.11 & 9.56 & 9.56 & 9.47 & 0.95 & 7.66 & 15.49 & 10.45 & 9.92 & 0.77 \\
					OW & Poor & 1.39 & -0.19 & 7.66 & 7.66 & 7.50 & 0.95 & 20.86 & 30.14 & 8.20 & 8.09 & 0.05 \\
					MW & Poor & 1.44 & -0.28 & 8.29 & 8.28 & 8.02 & 0.95 & 14.81 & 22.99 & 8.63 & 8.39 & 0.29 \\
					EW & Poor & 1.34 & -0.15 & 8.89 & 8.89 & 8.81 & 0.95 & 25.68 & 35.73 & 9.26 & 9.16 & 0.04 \\   \addlinespace

					&  &  &  \multicolumn{5}{c}{Missing $X_1^2$} &   \multicolumn{5}{c}{Missing $X_2X_4$}  \\ \cmidrule(lr){4-8}\cmidrule(lr){9-13}
					Weight & Overlap & True & Bias & RMSE & SD & SE & CP & Bias & RMSE & SD & SE & CP\\
					\cmidrule(lr){1-13}

					IPW & Good & 1.38 & 0.04 & 10.72 & 10.73 & 9.91 & 0.94 & -0.19 & 14.56 & 14.56 & 11.85 & 0.92 \\
					IPW(0.05) & Good & 1.40 & -1.46 & 9.34 & 9.12 & 8.90 & 0.94 & -1.05 & 9.43 & 9.32 & 9.19 & 0.95 \\
					IPW(0.1) & Good & 1.44 & -2.66 & 8.20 & 7.25 & 7.18 & 0.92 & -1.55 & 7.95 & 7.64 & 7.39 & 0.92 \\
					IPW(0.15) & Good & 1.48 & -2.83 & 7.97 & 6.77 & 6.44 & 0.89 & -1.24 & 7.33 & 7.10 & 6.66 & 0.92 \\
					OW & Good & 1.46 & -3.56 & 7.70 & 5.68 & 5.43 & 0.83 & -2.70 & 7.57 & 6.46 & 6.23 & 0.90 \\
					MW & Good & 1.49 & -4.26 & 8.94 & 6.29 & 5.97 & 0.80 & -3.33 & 8.92 & 7.41 & 7.18 & 0.89 \\
					EW & Good & 1.45 & -2.97 & 7.29 & 5.89 & 5.64 & 0.87 & -2.27 & 7.25 & 6.47 & 6.25 & 0.91 \\  \addlinespace
					
					IPW & Mod & 1.12 & 5.96 & 40.88 & 40.35 & 32.26 & 0.80 & 10.75 & 53.19 & 51.83 & 36.50 & 0.74 \\
					IPW(0.05) & Mod & 1.30 & -2.57 & 12.35 & 11.89 & 11.65 & 0.94 & 0.31 & 12.08 & 12.08 & 11.64 & 0.94 \\
					IPW(0.1) & Mod & 1.42 & -3.57 & 10.68 & 9.41 & 9.17 & 0.90 & -0.40 & 9.28 & 9.27 & 9.08 & 0.95 \\
					IPW(0.15) & Mod & 1.49 & -3.82 & 10.40 & 8.72 & 8.68 & 0.90 & -0.71 & 8.82 & 8.76 & 8.73 & 0.94 \\
					OW & Mod & 1.38 & -5.04 & 9.71 & 6.79 & 6.86 & 0.83 & -3.96 & 9.21 & 7.42 & 7.56 & 0.89 \\
					MW & Mod & 1.43 & -5.86 & 11.15 & 7.36 & 7.51 & 0.81 & -5.06 & 10.98 & 8.27 & 8.60 & 0.87 \\
					EW & Mod & 1.34 & -4.18 & 9.62 & 7.84 & 7.76 & 0.90 & -2.74 & 9.06 & 8.29 & 8.17 & 0.93 \\  \addlinespace
					
					IPW & Poor & 1.05 & 13.04 & 75.66 & 74.46 & 51.00 & 0.71 & 33.92 & 92.54 & 85.47 & 51.41 & 0.57 \\
					IPW(0.05) & Poor & 1.33 & -6.44 & 16.11 & 13.65 & 13.15 & 0.92 & 0.01 & 13.13 & 13.14 & 12.47 & 0.93 \\
					IPW(0.1) & Poor & 1.43 & -5.99 & 13.55 & 10.51 & 10.25 & 0.87 & -1.22 & 10.42 & 10.28 & 10.01 & 0.94 \\
					IPW(0.15) & Poor & 1.49 & -5.18 & 12.24 & 9.50 & 9.45 & 0.87 & -1.17 & 9.86 & 9.71 & 9.50 & 0.94 \\
					OW & Poor & 1.39 & -5.69 & 11.02 & 7.67 & 7.47 & 0.80 & -4.12 & 9.93 & 8.12 & 8.05 & 0.88 \\
					MW & Poor & 1.44 & -5.30 & 11.20 & 8.21 & 7.94 & 0.82 & -4.85 & 11.26 & 8.84 & 8.75 & 0.86 \\
					EW & Poor & 1.34 & -6.06 & 12.02 & 8.85 & 8.76 & 0.86 & -2.81 & 9.82 & 9.07 & 9.05 & 0.95 \\

					\bottomrule
				\end{tabular}
				\begin{tablenotes}
					\tiny
					\item  Mod: Moderate; IPW($\alpha$): trimmed IPW with $I_{\alpha}(x)=\mathds{1}({\{\alpha\leq e(x)\leq 1-\alpha\}})$, for  $\alpha=0.05, 0.10,$ and $0.15.$
					\item Bias: relative bias in percentage; RMSE: root mean-squared error in $10^{-2}$; SD: empirical standard deviation in $10^{-2}$; SE: average estimated standard error in $10^{-2}$; CP: coverage probability of $95\%$ confidence interval. The results are based on 1000 simulated data sets.
				\end{tablenotes}
			\end{threeparttable}
		\end{table}

		%
		
		\begin{table}
			\centering
			\begin{threeparttable}
				\centering\scriptsize\sf
				\caption{Variable transformation: low prevalence of treatment ($N=2000$).\label{var_trans_Asy} { }}
				\begin{tabular}{p{1cm}p{0.65cm}p{0.3cm}   p{0.7cm}p{0.4cm}p{0.3cm}p{0.3cm}p{0.4cm}  p{0.7cm}p{0.4cm}p{0.3cm}p{0.3cm}p{0.4cm}  p{0.76cm}p{0.4cm}p{0.3cm}p{0.3cm}p{0.4cm}}
					\toprule
					\multicolumn{18}{c}{\bf Homogeneous treatment effect} \\\cmidrule(lr){1-18}
					&  &  & \multicolumn{15}{c}{Propensity score misspecification} \\\cmidrule(lr){4-18}
					&  &  & \multicolumn{5}{c}{None }& \multicolumn{5}{c}{Mild} &   \multicolumn{5}{c}{Major}   \\ \cmidrule(lr){4-8}\cmidrule(lr){9-13}\cmidrule(lr){14-18}
					Weight & Overlap & True &Bias & RMSE & SD & SE & CP & Bias & RMSE & SD & SE  & CP & Bias & RMSE & SD & SE & CP\\
					\cmidrule(lr){1-18}				
					
					IPW & Good & 1 & 0.04 & 8.61 & 8.61 & 8.06 & 0.95 & 1.13 & 7.11 & 7.02 & 6.82 & 0.94 & 1.42 & 7.02 & 6.88 & 6.70 & 0.94 \\
					IPW(0.05) & Good & 1 & 0.07 & 6.80 & 6.81 & 6.86 & 0.95 & 0.86 & 6.61 & 6.56 & 6.51 & 0.94 & 1.42 & 6.67 & 6.52 & 6.48 & 0.94 \\
					IPW(0.1) & Good & 1 & -0.00 & 6.71 & 6.72 & 6.80 & 0.96 & 0.55 & 6.64 & 6.62 & 6.60 & 0.94 & 1.16 & 6.70 & 6.60 & 6.55 & 0.94 \\
					IPW(0.15) & Good & 1 & -0.03 & 7.00 & 7.00 & 7.20 & 0.96 & 0.33 & 7.06 & 7.06 & 7.22 & 0.96 & 1.47 & 7.34 & 7.19 & 7.15 & 0.94 \\
					OW & Good & 1 & -0.00 & 5.99 & 5.99 & 6.13 & 0.96 & 0.31 & 5.99 & 5.99 & 6.16 & 0.96 & 1.08 & 6.07 & 5.98 & 6.12 & 0.95 \\
					MW & Good & 1 & -0.01 & 6.07 & 6.08 & 6.23 & 0.96 & 0.11 & 6.07 & 6.07 & 6.24 & 0.96 & 0.91 & 6.14 & 6.07 & 6.20 & 0.95 \\
					EW & Good & 1 & -0.01 & 6.05 & 6.05 & 6.17 & 0.96 & 0.48 & 6.04 & 6.02 & 6.14 & 0.95 & 1.17 & 6.12 & 6.01 & 6.11 & 0.94 \\  \addlinespace
					
					IPW & Mod & 1 & -0.87 & 20.66 & 20.65 & 14.66 & 0.87 & 1.61 & 13.75 & 13.66 & 10.26 & 0.86 & 1.30 & 14.29 & 14.24 & 10.54 & 0.88 \\
					IPW(0.05) & Mod & 1 & -0.03 & 9.29 & 9.30 & 9.13 & 0.94 & 0.79 & 8.89 & 8.86 & 8.56 & 0.94 & 2.08 & 9.06 & 8.83 & 8.47 & 0.93 \\
					IPW(0.1) & Mod & 1 & -0.37 & 9.23 & 9.23 & 9.12 & 0.95 & 0.04 & 9.29 & 9.30 & 9.01 & 0.94 & 1.85 & 9.20 & 9.02 & 8.89 & 0.93 \\
					IPW(0.15) & Mod & 1 & -0.40 & 9.98 & 9.97 & 9.60 & 0.95 & -0.12 & 10.25 & 10.25 & 9.96 & 0.94 & 1.78 & 10.14 & 9.99 & 9.81 & 0.94 \\
					OW & Mod & 1 & -0.31 & 7.90 & 7.90 & 7.78 & 0.95 & 0.05 & 7.86 & 7.87 & 7.90 & 0.95 & 1.52 & 7.93 & 7.78 & 7.77 & 0.94 \\
					MW & Mod & 1 & -0.32 & 8.04 & 8.04 & 7.97 & 0.95 & -0.22 & 8.00 & 8.01 & 7.99 & 0.95 & 1.34 & 8.04 & 7.93 & 7.86 & 0.94 \\
					EW & Mod & 1 & -0.35 & 8.10 & 8.09 & 7.89 & 0.94 & 0.39 & 8.02 & 8.01 & 7.84 & 0.94 & 1.64 & 8.06 & 7.90 & 7.73 & 0.94 \\  \addlinespace
					
					IPW & Poor & 1 & 2.01 & 29.86 & 29.81 & 15.98 & 0.73 & 4.08 & 21.91 & 21.53 & 12.18 & 0.75 & 2.03 & 25.09 & 25.02 & 15.56 & 0.79 \\
					IPW(0.05) & Poor & 1 & 0.26 & 11.50 & 11.51 & 10.77 & 0.93 & 1.29 & 11.07 & 11.00 & 10.31 & 0.93 & 3.00 & 11.31 & 10.91 & 10.09 & 0.91 \\
					IPW(0.1) & Poor & 1 & 0.40 & 11.67 & 11.67 & 10.74 & 0.93 & 0.88 & 11.74 & 11.71 & 10.76 & 0.93 & 2.94 & 11.92 & 11.56 & 10.48 & 0.90 \\
					IPW(0.15) & Poor & 1 & 0.30 & 12.13 & 12.14 & 11.27 & 0.92 & 0.59 & 12.65 & 12.64 & 11.66 & 0.92 & 2.74 & 12.34 & 12.04 & 11.41 & 0.93 \\
					OW & Poor & 1 & 0.28 & 9.68 & 9.68 & 9.16 & 0.93 & 0.75 & 9.76 & 9.73 & 9.32 & 0.94 & 2.66 & 9.89 & 9.53 & 9.05 & 0.93 \\
					MW & Poor & 1 & 0.30 & 9.88 & 9.88 & 9.39 & 0.93 & 0.51 & 9.93 & 9.92 & 9.42 & 0.93 & 2.56 & 10.02 & 9.69 & 9.16 & 0.93 \\
					EW & Poor & 1 & 0.27 & 9.91 & 9.91 & 9.30 & 0.93 & 1.15 & 9.99 & 9.93 & 9.25 & 0.93 & 2.74 & 10.11 & 9.73 & 9.01 & 0.91 \\

					\bottomrule
					%
					%
					%
					\addlinespace
					\multicolumn{18}{c}{\bf Heterogeneous treatment effect} \\\cmidrule(lr){1-18}
					&  &  & \multicolumn{5}{c}{None }& \multicolumn{5}{c}{Mild} &   \multicolumn{5}{c}{Major}   \\ \cmidrule(lr){4-8}\cmidrule(lr){9-13}\cmidrule(lr){14-18}
					Weight & Overlap & True &Bias & RMSE & SD & SE & CP & Bias & RMSE & SD & SE  & CP & Bias & RMSE & SD & SE & CP\\
					\cmidrule(lr){1-18}				
					
					IPW & Good & 1.21 & 0.59 & 8.86 & 8.83 & 8.24 & 0.93 & 4.48 & 8.94 & 7.13 & 6.96 & 0.86 & 5.72 & 9.83 & 6.99 & 6.82 & 0.81 \\
					IPW(0.05) & Good & 1.28 & 0.36 & 7.18 & 7.17 & 6.91 & 0.94 & 1.05 & 6.96 & 6.83 & 6.60 & 0.93 & 2.00 & 7.17 & 6.70 & 6.57 & 0.93 \\
					IPW(0.1) & Good & 1.36 & 0.15 & 6.85 & 6.85 & 6.87 & 0.94 & 0.37 & 6.78 & 6.76 & 6.68 & 0.94 & 1.25 & 6.95 & 6.74 & 6.65 & 0.94 \\
					IPW(0.15) & Good & 1.44 & 0.03 & 7.25 & 7.25 & 7.25 & 0.95 & 1.06 & 7.43 & 7.28 & 7.27 & 0.95 & 1.53 & 7.54 & 7.22 & 7.23 & 0.93 \\
					OW & Good & 1.36 & 0.13 & 6.09 & 6.09 & 6.22 & 0.96 & 0.33 & 6.08 & 6.07 & 6.25 & 0.95 & 1.20 & 6.23 & 6.02 & 6.23 & 0.94 \\
					MW & Good & 1.39 & 0.03 & 6.17 & 6.17 & 6.33 & 0.95 & -0.25 & 6.15 & 6.15 & 6.34 & 0.95 & 0.44 & 6.09 & 6.06 & 6.31 & 0.95 \\
					EW & Good & 1.33 & 0.19 & 6.14 & 6.14 & 6.25 & 0.95 & 0.88 & 6.21 & 6.10 & 6.23 & 0.94 & 1.81 & 6.51 & 6.05 & 6.21 & 0.94 \\  \addlinespace
					
					IPW & Mod & 1.01 & 2.62 & 22.28 & 22.14 & 15.49 & 0.82 & 11.41 & 18.01 & 13.85 & 10.65 & 0.67 & 13.08 & 19.81 & 14.78 & 11.05 & 0.64 \\
					IPW(0.05) & Mod & 1.26 & 0.05 & 9.61 & 9.61 & 9.29 & 0.95 & 1.58 & 9.50 & 9.29 & 8.78 & 0.93 & 3.40 & 10.23 & 9.30 & 8.69 & 0.90 \\
					IPW(0.1) & Mod & 1.38 & -0.17 & 9.42 & 9.42 & 9.21 & 0.94 & 1.85 & 9.62 & 9.28 & 9.14 & 0.94 & 3.35 & 10.18 & 9.08 & 9.05 & 0.92 \\
					IPW(0.15) & Mod & 1.46 & -0.40 & 9.43 & 9.42 & 9.68 & 0.96 & 2.08 & 10.25 & 9.79 & 10.03 & 0.94 & 3.37 & 11.01 & 9.85 & 9.93 & 0.92 \\
					OW & Mod & 1.32 & -0.21 & 7.79 & 7.79 & 7.97 & 0.96 & 0.16 & 7.85 & 7.85 & 8.10 & 0.96 & 1.95 & 8.19 & 7.78 & 8.00 & 0.95 \\
					MW & Mod & 1.37 & -0.31 & 7.92 & 7.91 & 8.13 & 0.96 & -0.53 & 7.96 & 7.93 & 8.17 & 0.96 & 0.99 & 8.03 & 7.92 & 8.07 & 0.96 \\
					EW & Mod & 1.27 & -0.15 & 8.00 & 8.01 & 8.10 & 0.95 & 0.98 & 8.10 & 8.01 & 8.06 & 0.95 & 2.85 & 8.66 & 7.87 & 7.97 & 0.93 \\  \addlinespace
					
					IPW & Poor & 0.92 & 8.96 & 30.35 & 29.23 & 16.68 & 0.68 & 18.06 & 27.31 & 21.70 & 12.51 & 0.57 & 19.01 & 29.36 & 23.61 & 14.77 & 0.56 \\
					IPW(0.05) & Poor & 1.27 & 0.21 & 11.72 & 11.72 & 10.99 & 0.93 & 2.41 & 11.90 & 11.51 & 10.54 & 0.92 & 4.77 & 12.92 & 11.42 & 10.34 & 0.88 \\
					IPW(0.1) & Poor & 1.39 & -0.08 & 11.41 & 11.42 & 10.80 & 0.93 & 2.57 & 11.82 & 11.28 & 10.85 & 0.93 & 4.73 & 13.12 & 11.36 & 10.65 & 0.90 \\
					IPW(0.15) & Poor & 1.47 & -0.29 & 11.63 & 11.63 & 11.29 & 0.94 & 2.05 & 12.74 & 12.39 & 11.70 & 0.92 & 3.91 & 13.53 & 12.26 & 11.47 & 0.91 \\
					OW & Poor & 1.32 & -0.17 & 9.59 & 9.59 & 9.38 & 0.94 & 0.65 & 9.72 & 9.69 & 9.56 & 0.94 & 2.97 & 10.31 & 9.54 & 9.32 & 0.93 \\
					MW & Poor & 1.38 & -0.30 & 9.77 & 9.77 & 9.58 & 0.94 & -0.01 & 9.84 & 9.84 & 9.62 & 0.94 & 1.92 & 10.02 & 9.66 & 9.38 & 0.94 \\
					EW & Poor & 1.27 & -0.06 & 9.79 & 9.79 & 9.56 & 0.94 & 1.48 & 10.03 & 9.86 & 9.51 & 0.93 & 3.91 & 10.89 & 9.70 & 9.31 & 0.91 \\

					\bottomrule
				\end{tabular}
				\begin{tablenotes}
					\tiny
					\item  Mod: Moderate; IPW($\alpha$): trimmed IPW with $I_{\alpha}(x)=\mathds{1}({\{\alpha\leq e(x)\leq 1-\alpha\}})$, for   $\alpha=0.05, 0.10,$ and $0.15.$
					\item Bias: relative bias in percentage; RMSE: root mean-squared error in $10^{-2}$; SD: empirical standard deviation in $10^{-2}$; SE: average estimated standard error in $10^{-2}$; CP: coverage probability of $95\%$ confidence interval. The results are based on 1000 simulated data sets.
				\end{tablenotes}
			\end{threeparttable}
		\end{table}

	\end{document}